\documentclass{aa}  

\usepackage{graphicx}
\usepackage{txfonts}
\usepackage{amsmath}
\usepackage{natbib}
\usepackage{lscape}
\usepackage{multicol}
\newcommand{\statcont} {\texttt{STATCONT}}

\newcommand{\phn} {$\phantom{0}$}

\newcommand{\hii} {H{\sc ii}}
\begin{document} 

   \title{\statcont: A statistical continuum level determination method for line-rich sources}

   \author{\'A.~S\'anchez-Monge\inst{1}
          \and
          P.~Schilke\inst{1}
          \and
          A.~Ginsburg\inst{2,3}
          \and
          R.~Cesaroni\inst{4}
          \and
          A.~Schmiedeke\inst{1,5}
          }

   \institute{I.\ Physikalisches Institut, Universit\"at zu K\"oln, Z\"ulpicher Str.\ 77, D-50937 K\"oln, Germany\\
              \email{sanchez@ph1.uni-koeln.de}
              \and
              National Radio Astronomy Observatory, 1003 Lopezville Road, Socorro, NM 87801, USA
              \and
              European Southern Observatory, Karl-Schwarzschild-Strasse 2, D-85748 Garching bei M\"unchen, Germany
              \and
              INAF-Osservatorio Astrofisico di Arcetri, Largo E.\ Fermi 5, I-50125 Firenze, Italy
              \and
              Max-Planck Institute for Extraterrestrial Physics, Giessenbachstrasse 1, D-85748 Garching bei M\"unchen, Germany
             }

   \date{Received ; accepted }

% \abstract{}{}{}{}{} 
% 5 {} token are mandatory
 
  \abstract{\statcont\ is a python-based tool designed to determine the continuum emission level in spectral data, in particular for sources with a line-rich spectrum. The tool inspects the intensity distribution of a given spectrum and automatically determines the continuum level by using different statistical approaches. The different methods included in \statcont\ are tested against synthetic data. We conclude that the sigma-clipping algorithm provides the most accurate continuum level determination, together with information on the uncertainty in its determination. This uncertainty can be used to correct the final continuum emission level, resulting in the here called `corrected sigma-clipping method' or c-SCM. The c-SCM has been tested against more than 750 different synthetic spectra reproducing typical conditions found towards astronomical sources. The continuum level is determined with a discrepancy of less than 1\% in 50\% of the cases, and less than 5\% in 90\% of the cases, provided at least 10\% of the channels are line free. The main products of \statcont\ are the continuum emission level, together with a conservative value of its uncertainty, and datacubes containing only spectral line emission, i.e., continuum-subtracted datacubes. \statcont\ also includes the option to estimate the spectral index, when different files covering different frequency ranges are provided.}
  

   \maketitle

%
%________________________________________________________________
\section{Introduction}\label{s:intro}

The continuum emission of astronomical objects provides important information on their physical properties. For example, in Galactic star-forming regions, the continuum emission at longer wavelengths in the centimeter domain (from 1~cm to 20~cm) can be used to characterize the properties of thermal ionized gas or synchrotron radiation (e.g., \citealt{WoodChurchwell1989, SanchezMonge2013a, SanchezMonge2013c, Beltran2016, Ramachandran2017}). In the millimeter/submillimeter domain (from 0.1~mm to 10~mm) the continuum emission likely traces a cold dust component, and therefore can be used to study the mass, structure and fragmentation level of cold and young sources (e.g., \citealt{Elia2010, Palau2014, Palau2017, Fontani2016, Schmiedeke2016, Koenig2017}). At even shorter wavelengths, in the infrared and visible, the continuum emission may trace hot/warm environments such as circumstellar disks around stars (e.g., \citealt{Aumann1984, Kraus2010, Kraus2017, Boley2016}) or directly the photosphere of the stars.

In addition to the continuum emission, many astronomical objects have spectral line features that contribute to their total emission. The determination of the continuum emission level requires, therefore, accurate handling of these line features. The classical approach is to identify and exclude the frequency intervals associated with spectral lines, in order to consider frequency ranges with only continuum emission. The identification of frequency ranges with (or without) spectral lines requires observations with high-enough spectral resolution (hereafter  `spectral line observing mode' observations) to resolve and sample with enough channel-bins the spectral line. Moreover, for astronomical sources with many spectral line features (e.g., hot molecular cores in high-mass star forming regions, \citealt{SanchezMonge2017, Allen2017}), a broad frequency coverage is necessary to ensure the presence of enough line-free frequency intervals to determine the continuum level. For this reason, spectral line surveys, covering large frequency ranges, constitute a unique tool to determine the continuum level in line-crowded sources.

Spectral line surveys have always had a special status among astronomical observations. They are extremely useful as a road map to researchers studying other sources with similar properties, often led to discoveries, are of great legacy value, and allow a very detailed modeling of the physical and chemical source structure thanks to the detection of hundreds or even thousands of transitions of different species. The development of millimeter astronomy together with the availability of new receivers covering broader frequency ranges allowed efficient spectral line surveys and revealed a population of objects, associated with the early stages of high-mass star formation, with line-rich spectra. These objects are commonly known as hot molecular cores (e.g., \citealt{Kurtz2000}). Two well-known objects of this type are the nearby Orion~KL region (e.g., \citealt{Ho1979, Sutton1985, Blake1986, Schilke1997, Schilke2001, Tercero2010, Crockett2014, Gong2015}) and the Sgr\,B2 star forming complex close to the Galactic center (e.g., \citealt{Cummins1986, Turner1989, Nummelin1998, Friedel2004, Belloche2013, Neill2014, Corby2015}). New receivers, with better sensitivities, proved that hot molecular cores are abundant throughout the Galaxy, and both observational and theoretical studies suggest that there can be about $10^3$--$10^4$ hot molecular cores (e.g., \citealt{Wilner2001, Furuya2005, Osorio2009}). For example, star-forming regions like G10.47$+$0.03, G29.96$-$0.02 or G31.41$+$0.31 contain sources with spectra as chemically rich as Orion~KL and Sgr\,B2 (e.g., \citealt{Cesaroni1994, Cesaroni2011, Walmsley1995, Wyrowski1999, Olmi2003, Rolffs2011}).

The advent of sensitive instruments like the Atacama Large Millimeter/submillimeter Array (ALMA, see \citealt{ALMA2015}) is proving that relatively modest dense cores (e.g., \citealt{SanchezMonge2013b, SanchezMonge2014, Beltran2014}) and even solar-type objects like IRAS~16293$-$2422 or NGC\,1333 (e.g., \citealt{Cazaux2003, Bottinelli2004, Jorgensen2016}) may show a rich chemistry, and therefore line-rich spectra. Similarly, a complex chemistry is also found towards disks (e.g., \citealt{Oberg2015, Guzman2015}), shocked gas (e.g., \citealt{Codella2010, Sugimura2011}) and photon-dominated regions (e.g., \citealt{TrevinoMorales2014, TrevinoMorales2016, Nagy2017}). In extragalactic astronomy, recent observations reach good enough sensitivities to detect an increasing number of spectral line features in the central regions of many galaxies (e.g., \citealt{GonzalezAlfonso2004, Martin2006, Martin2011, Aladro2011, Rangwala2011, Muller2011, Muller2014, Meier2015}), and it is expected that individual hot molecular cores will soon be detected in other galaxies (see e.g., \citealt{Shimonishi2016}). Finally, the existence of crowded, line-rich spectra is not only limited to the very early stages of star formation. Late evolutionary stages, like AGB (or asymptotic giant branch) stars, also show a rich and complex chemistry. IRC+10216 is one of the best well-known cases of a spectral-line rich AGB star (see e.g., \citealt{Cernicharo2000, Patel2011}).
In summary, current and future facilities like ALMA, NOEMA (NOrthern Extended Millimeter Array), the VLA (Karl G.\ Jansky Very Large Array), the SKA (Square Kilometer Array), the IRAM-30m telescope or APEX (Atacama Pathfinder Experiment) can go one step further in sensitivity allowing the detection of spectral line features that, with previous telescopes, were hindered by the noise level. Most of these line-rich sources are associated with detectable continuum emission that provides important information on the physical properties (e.g., mass, structure and fragmentation level) of the studied sources. The determination of this continuum emission becomes an essential, and sometimes tedious task in line-rich sources.

In this paper we present \statcont, a python-based tool, that automatically determines the continuum level of line-rich sources in a statistical way. The paper is organized as follows. In Sect.~\ref{s:cont} we list the challenges in the determination of the continuum level of line-rich sources and present our new method. We apply different statistical approaches to determine the continuum level in synthetic data (Sect.~\ref{s:contspectra}) and compare them in a statistical way (Sect.~\ref{s:statistics}).  We also study how noise may affect the continuum level determination (Sect.~\ref{s:contnoise}). Out of all the methods that we test, we show that the sigma-clipping method is the most accurate in many cases, and provides, besides the continuum level, an estimation of the uncertainty. A corrected sigma-clipping method is introduced and proved to be successful in more than 90\% of the cases, with discrepancies with the real continuum of less than 5\% (Sect.~\ref{s:correction}). In Sect.~\ref{s:realcases} we apply the new method to real-case examples, in particular to ALMA observational data, and compare with continuum images produced with the classical approach of identifying line-free channels by visual inspection. The conditions under which \statcont\ produces more accurate results are listed in Sect.~\ref{s:conditions}. The different procedures that compose \statcont, as well as the input and output files, are described in Sect.~\ref{s:procedures}. Finally, in Sect.~\ref{s:summary} we summarize the main aspects of the method and list some limitations as well as possible future improvements.

%----------------------------------------------------------------------
\begin{figure*}[t!]
\begin{center}
\begin{tabular}[b]{c c}
        \includegraphics[height=0.45\textheight]{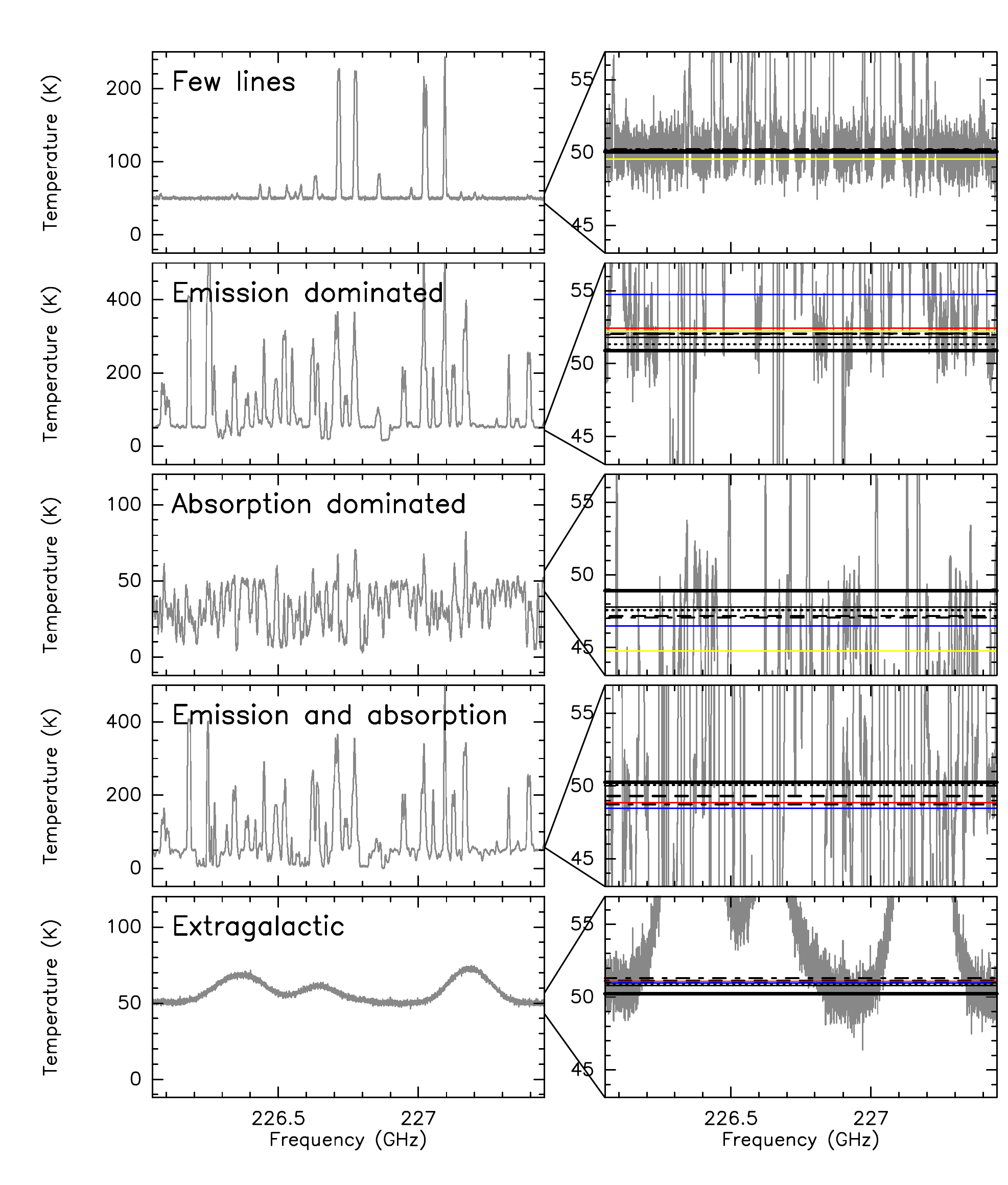} &
        \includegraphics[height=0.45\textheight]{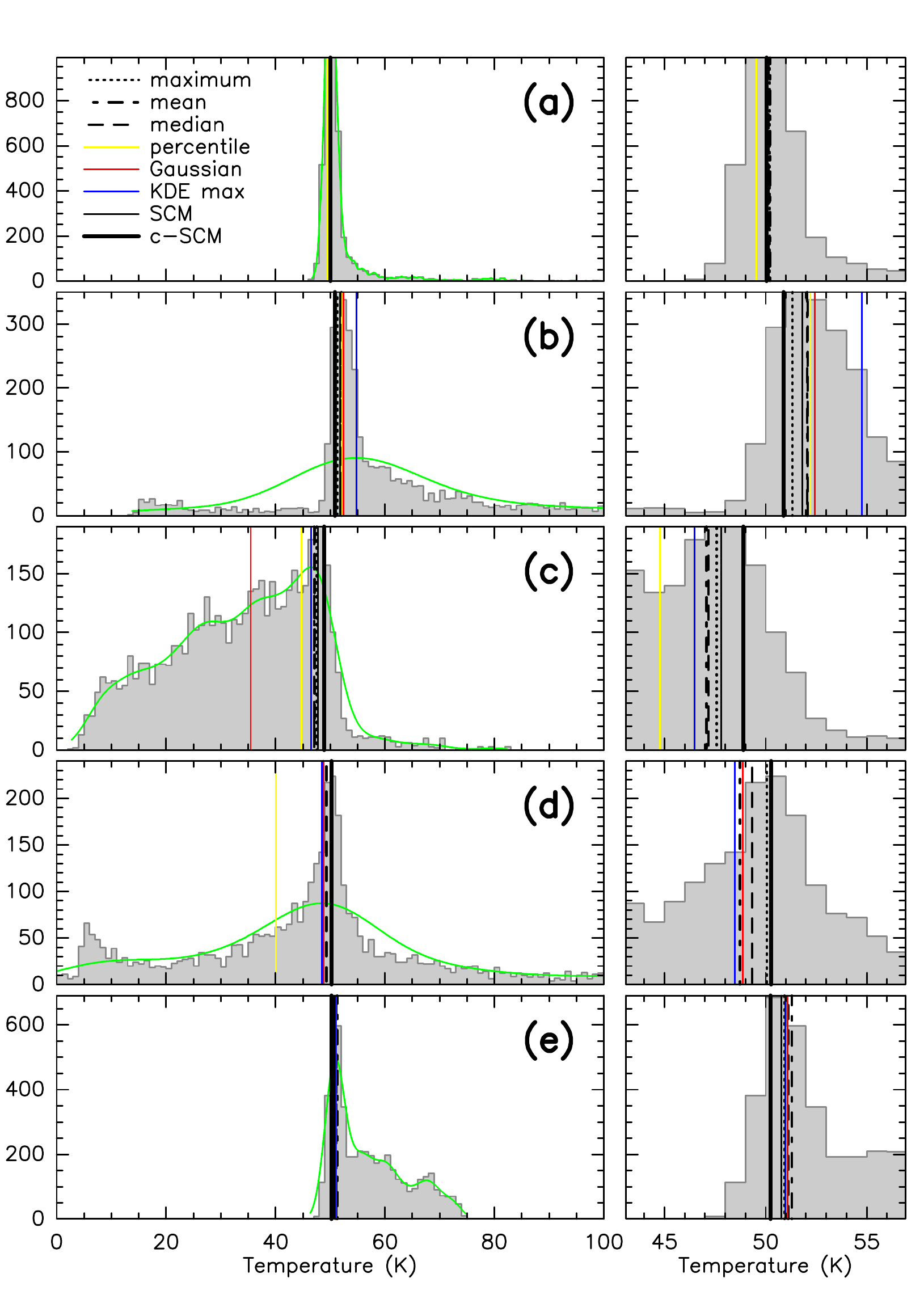} \\
\end{tabular}
\caption{Schematic description of the process of determination of the continuum level for six different types of spectra: (\textit{a}) spectrum with few lines, (\textit{b}) spectrum dominated by emission features, (\textit{c}) spectrum dominated by absorption features, (\textit{d}) spectrum with both bright emission and absorption features, and (\textit{e}) spectrum with broad lines, simulating an extragalactic source (see Sect.~\ref{s:cont}). From left to right, the different panels show: (\textit{left column}) synthetic spectra for each case, (\textit{middle-left column}) close-up view of the synthetic spectra with the continuum level indicated with horizontal colored lines, following the same scheme as in the right-column panels, (\textit{middle-right column}) histogram created from the synthetic spectra, with the different continuum level estimations indicated with vertical colored lines: `maximum continuum level' in black dotted, `mean continuum level' in black dot-dashed, `median continuum level' in black dotted, `percentile continuum level' in yellow, `Gaussian continuum level' in red, `KDE continuum level' in blue, `sigma-clip continuum level' or SCM in black, and `corrected sigma-clip continuum level' or cSCM in thick black (see Sect.~\ref{s:contspectra} for details of the different methods). The green solid line depicts the KDE, and (\textit{right column}) close-up view of the histogram around the 50~K value. The exact values of the continuum levels are listed in Table~\ref{t:contspectra}. The mean, median and Gaussian continuum levels correspond to the entire distribution (columns 3, 5 and 8 of Table~\ref{t:contspectra}, respectively). The 75th percentile is used for case (\textit{c}) and the 25th percentile for all the other cases.}
\label{f:contmethod}
\end{center}
\end{figure*}
%----------------------------------------------------------------------

%----------------------------------------------------------------------
\begin{table*}
\caption{\label{t:contspectra} Continuum level, in temperature units, for different synthetic spectra (see Fig.~\ref{f:contmethod}) for the methods described in Sect.~\ref{s:contspectra}}
\centering
\begin{tabular}{l c c c c c c c}
\hline\hline

\multicolumn{1}{c}{Type of spectra}
&maximum
&mean
&mean (sel.)
&median
&median (sel.)
\\

\multicolumn{1}{c}{(1)}
&(2)
&(3)
&(4)
&(5)
&(6)
\\
\hline
\textit{(a)} Few lines                &50.20  &\phn56.39  &50.22  &50.33  &50.14  \\
\textit{(b)} Emission dominated       &51.33  &106.71     &52.06  &60.43  &52.08  \\
\textit{(c)} Absorption dominated     &48.67  &\phn33.90  &45.69  &35.42  &45.95  \\
\textit{(d)} Emission and absorption  &50.06  &\phn83.62  &48.73  &51.02  &49.33  \\
\textit{(e)} Extragalactic            &50.93  &\phn56.90  &51.30  &54.88  &51.10  \\
\hline
\multicolumn{1}{c}{}
&percent 25$^\mathrm{th}$ / 75$^\mathrm{th}$
&Gaussian
&Gaussian (sel.)
&KDE max
&SCM
&c-SCM
\\
\multicolumn{1}{c}{}
&(7)
&(8)
&(9)
&(10)
&(11)
&(12)
\\
\hline
\textit{(a)} Few lines                &49.55~/~\phn51.45 &$50.08\pm1.11$  &$50.08\pm1.11$  &50.07  &$50.05\pm0.41$ &$50.05\pm0.41$\\
\textit{(b)} Emission dominated       &52.21~/~127.16    &$52.43\pm2.22$  &$52.25\pm1.97$  &54.76  &$51.82\pm0.94$ &$50.89\pm0.94$\\
\textit{(c)} Absorption dominated     &24.16~/~\phn44.79 &$35.47\pm14.7$  &$44.65\pm5.57$  &46.49  &$47.80\pm1.11$ &$48.91\pm1.11$\\
\textit{(d)} Emission and absorption  &40.08~/~\phn99.51 &$48.86\pm6.45$  &$49.67\pm3.93$  &48.47  &$50.27\pm0.83$ &$50.27\pm0.83$\\
\textit{(e)} Extragalactic            &51.10~/~\phn61.42 &$51.09\pm1.64$  &$51.05\pm1.56$  &50.98  &$50.79\pm0.54$ &$50.24\pm0.54$\\
\hline
\end{tabular}
\tablefoot{The real continuum level of each synthetic spectrum is 50~K.}
\end{table*}
%----------------------------------------------------------------------

%________________________________________________________________
\section{Determination of the continuum level}\label{s:cont}

The determination of the continuum emission level of astronomical sources observed in a spectral line observing mode is usually based on the identification of channels free of line emission, i.e., line-free channels. The procedure starts with a representative spectrum of the data, which is carefully inspected in the search for ranges of channels with emission not related to spectral lines. Then, the continuum level is obtained by fitting a function, usually a 0th or 1st-order polynomial function, to the line-free channels. This process is relatively simple and straightforward when analyzing line-poor astronomical sources, but becomes more challenging with increasing number of detected spectral lines. In this situation, the number of line-free channels is significantly reduced, their identification becomes tedious, and the final polynomial fitting is unreliable due to the limited number of selected channels.

A second problem arises when analyzing, for example, interferometric spectral datacubes, i.e., not a single spectrum but a 2-dimensional map with a third axis containing the spectral information (e.g., velocity, frequency). If the field of view contains more than one source with different systemic velocities (or line-of-sight velocities), it may happen that the line-free channels applicable for one source do not coincide with the line-free channels of another source in the field. In this scenario, there is no unique set of line-free channels that can be used throughout the whole map to determine the continuum emission level in the region. This overlap between lines at different positions often means that continuum subtraction in the observed \textit{uv}-domain (the raw visibilities measured by the interferometer) is impossible or inadvisable.

Here, we present an alternative method of determining the continuum emission level that can be applied to both single spectra and datacubes. With the aim of testing the method we have created synthetic spectra for different cases: (\textit{a}) line-poor spectrum, (\textit{b}) emission-line dominated spectrum, (\textit{c}) absorption dominated spectrum, (\textit{d}) spectrum with both emission and absorption features, and (\textit{e}) spectrum with broad lines simulating an extragalactic source. The left panels of Fig.~\ref{f:contmethod} show the different types of spectra. All these spectra have been produced with the XCLASS\footnote{The eXtended CASA Line Astronomy Software Suite (XCLASS) can be downloaded at https://www.astro.uni-koeln.de/projects/schilke/XCLASSInterface} software package \citep{Moeller2017}. In all cases we consider a fixed continuum level of 50~K and a number of emission and absorption components of different species (e.g., CH$_3$CN, CH$_3$OH, CH$_3$OCH$_3$). For cases (\textit{a}) to (\textit{d}) we assume a fixed line full width at half maximum (FWHM) of 8~km~s$^{-1}$ and a frequency range of 1.5~GHz, while for case (\textit{e}) we set the line FWHM to 300~km~s$^{-1}$. Different velocity components for each molecule are introduced in the simulations in order to increase the complexity of the spectra. Finally, we add a Gaussian noise level of $\sigma=1$~K to the synthetic spectra.

%________________________________________________________________
\subsection{Continuum level in spectral data: statistical methods}\label{s:contspectra}

In the process of determining the continuum level statistically, we first produce the spectra that we want to analyze. These spectra can be obtained from single-dish, single-pointing observations or they can correspond to the spectra of different pixels of a datacube. The spectra that we use to characterize the methods are shown in Fig.~\ref{f:contmethod} (left panels). The panels in the central column show a close-up view of each spectrum around the continuum level (i.e., 50~K in the examples).

The second step is the creation of a distribution (histogram) of the intensity measured in all channels of the spectra. The histograms of the selected cases are shown in the right panels of Fig.~\ref{f:contmethod}. We note that the shape of the histogram already contains information of the properties of the source we are analyzing. For example, in case of a spectrum with only noise around a certain continuum level, the histogram corresponds to the distribution of what is commonly known as white noise, i.e., a Gaussian distribution. In such a case, the width and location of the maximum indicate the rms noise level of the spectrum and the continuum emission level, respectively. For the cases shown in Fig.~\ref{f:contmethod}, the distribution of the histograms have some clear departures from a Gaussian profile. For example, in case (\textit{b}) we can identify a long tail towards positive intensities with respect to the peak of the histogram. This extension is due to the emission lines, above the continuum level, that dominate the spectrum. Similarly, case (\textit{d}) has a distribution that extends towards high intensity levels, but also towards low intensities, with a second peak at a level of 0~K corresponding to saturated absorption features. In case (\textit{c}), the presence of deep absorptions dominating more than half of the spectrum results in a histogram distribution that is skewed to low intensities with respect to its peak.

From the different cases shown in Fig.~\ref{f:contmethod} we can say that, in general, the maximum of the distribution is close to the continuum level of the spectra (i.e., 50~K). In addition, the histogram bins located around the maximum have in most of the cases a Gaussian-like shape similar to what is expected for a noise-only spectrum (see above). Therefore, the maximum of the distribution, together with the nearby histogram bins, likely describe the level of the continuum emission. In the following we describe different approaches to inspect this maximum and to determine the continuum level.

\paragraph{\textbf{Maximum continuum level}:}
In this approach, the continuum emission level is determined from the intensity of the histogram bin that corresponds to the maximum/peak of the distribution. For the examples shown in Fig.~\ref{f:contmethod} this continuum level is indicated by a black dotted line in the central and right panels. In Table~\ref{t:contspectra}, we list in column~2 the maximum continuum level determined for each of the spectra shown in the Figure. A first inspection reveals that this approach overestimates the continuum level when the spectrum is dominated by emission lines. Similarly, the continuum level is underestimated for absorption-dominated spectra. In general, the estimated continuum differs by $<5$\% with respect to the real one, however, the peak/maximum of the distribution, and therefore the estimated continuum level, strongly depends on the size of the histogram bins. In \statcont\ the size of the bins is defined by the rms noise level of the observations, which is 1~K in the synthetic spectra. Doubling the size of the histogram bins introduces variations in the continuum level determination corresponding to an additional 10\%. In general, wide bins may result in a too-smoothed and not accurate enough distribution, while narrow bins may result in very noisy distributions, and therefore, there is likely no optimal bin size.

\paragraph{\textbf{Mean continuum level}:}
In this method, the continuum emission level is determined from the mean value of the intensities of the spectrum. In this case, the histogram is irrelevant and, therefore, there is no dependence on the size of the histogram bins. However, the mean is not a proper determination of the continuum level if the distribution is skewed towards high or low intensities. This is clearly seen in cases (\textit{b}) and (\textit{d}), where the mean is determined to be about 100~K, which would correspond to an overestimation of the continuum level of about 100\% (see column~3 in Table~\ref{t:contspectra}, and the black dot-dashed lines in Fig.~\ref{f:contmethod}). A similar effect is seen for absorption-dominated spectra. Only in the cases of line-poor and extragalactic spectra, the continuum level differs by about 10\% with respect to the real one. The continuum level determination based on the mean is highly influenced by the long tails of the distribution, both towards high and low intensities. In order to avoid them, we automatically restrict the sample to consider or select only those intensities around the maximum/peak of the distribution, within e.g., 10 times the noise level of the observations. The continuum levels determined following this approach are listed as `mean (sel.)' in column~4 of Table~\ref{t:contspectra}. The discrepancy with respect to the correct value of the continuum level improves with respect to the standard mean continuum level, and is about 5--10\%. However, the selection of the range of intensities to be considered depends on the location of the maximum/peak of the distribution, and therefore on the size of the histogram bins.

\paragraph{\textbf{Median continuum level}:}
In this method, the continuum emission level is determined from the median value of the intensities of the spectrum. As in the previous case, the histogram and size of the bins are irrelevant. Different to the mean, the median value is less sensitive to outliers, which results in a better determination of the continuum level. For example, in case (\textit{b}) the continuum is estimated to be 60~K, compared to the value of 107~K obtained with the mean (cf.\ column~5 in Table~\ref{t:contspectra}, and black dashed lines in Fig.~\ref{f:contmethod}). We have applied the same selection criteria to avoid long tails towards high or low intensities, therefore considering only those intensities around the maximum/peak of the distribution. This choice results in a more precise determination of the continuum level (listed as `median (sel.)' in column~6 of Table~\ref{t:contspectra}), but depends on the histogram bins. The discrepancy with respect to the correct value of the continuum level is about 5--10\%.

%----------------------------------------------------------------------
\begin{figure}[t]
\begin{center}
\begin{tabular}[b]{c}
        \includegraphics[width=0.92\columnwidth]{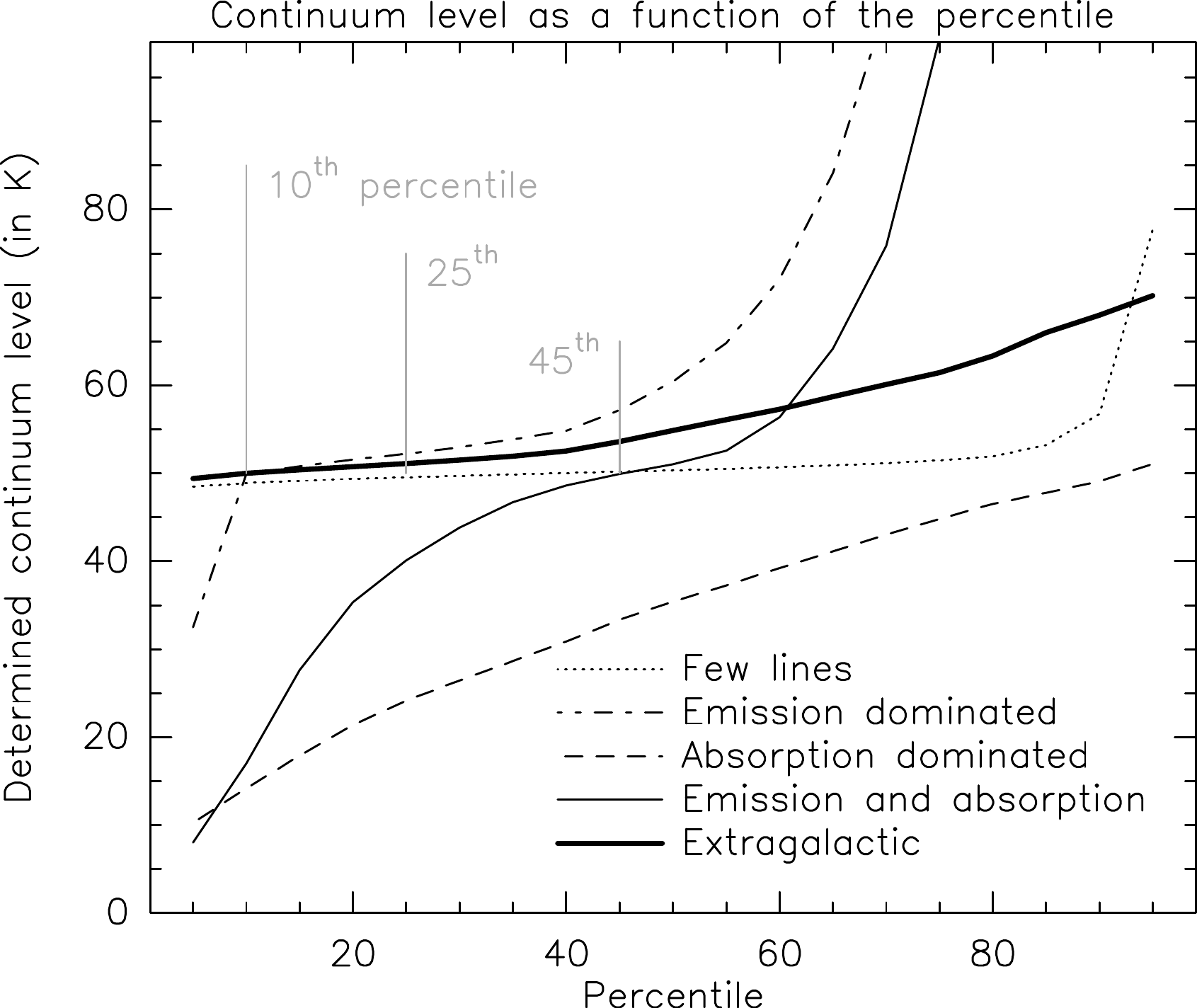} \\
\end{tabular}
\caption{Variation of the continuum emission level with the percentile used. Each line corresponds to the spectral test cases shown in Fig.~\ref{f:contmethod}. The real continuum level is at 50~K. The grey vertical lines show for which percentile the continuum emission level would match the real continuum level. Absorption-dominated spectra requires of higher percentiles, while a lower percentile works better for emission-dominated spectra.}
\label{f:percentiles}
\end{center}
\end{figure}
%----------------------------------------------------------------------

\paragraph{\textbf{Percentile continuum level}:}
A percentile is the value below which a given percentage of data points from a dataset fall. Commonly used are the 25th, the 50th and the 75th percentiles, which correspond to the value below which 25\%, 50\% and 75\% of the data points are found, respectively. With this definition the 50th percentile is directly the median of a dataset. We have used these three percentiles to determine the continuum level in the different test cases. In column~7 of Table~\ref{t:contspectra} we list the continuum levels determined using the 25th and 75th percentiles (see also yellow lines in Fig.~\ref{f:contmethod}), while the 50th percentile corresponds to the median (column~5). We find that the 25th percentile works better for emission-dominated spectra, like cases (\textit{b}), (\textit{d}), and (\textit{e}). For absorption-dominated spectra like case (\textit{c}), the 75th percentile is more accurate. The two percentiles give accurate results in line-poor spectra like case (\textit{a}). The discrepancy with respect to the correct value is in the range 5--20\%, if it is known in advance which percentile has to be used. In other words, it is necessary to establish which kind of spectra is analyzed (i.e., emission-dominated or absorption-dominated) in order to use the most accurate percentile. As an example, the 25th, 50th and 75th percentiles of the spectrum of case (\textit{d}) with emission and absorption lines are 40.1~K, 51.0~K and 99.5~K, respectively. The usage of a wrong percentile can result in uncertainties of about 100\% or more. This is clearly seen in Fig.~\ref{f:percentiles} where we show the variation of the continuum level determined using different percentiles. For most of the cases, the 25th percentile would result in an accurate determination of the continuum. However, absorption-dominated spectra are better described with percentiles above the 80th.

\paragraph{\textbf{Gaussian continuum level}:}
In this method we assume that the peak of the distribution resembles a Gaussian distribution with the location of the maximum and the width likely indicating the continuum level and the noise level, respectively. We fit a Gaussian function to the entire histogram distribution (column~8 in Table~\ref{t:contspectra}, and the red line in the examples of Fig.~\ref{f:contmethod}), and to those bins located around the peak of the distribution (column~9 in Table~\ref{t:contspectra}). The discrepancy between the `Gaussian continuum level' and the theoretical continuum level is in general around 5\% for most of the cases. The Gaussian fit to those bins located near the maximum of the distribution, improves the continuum level determination in particular for spectra such as that in case (\textit{c}) where most of the channels are affected by line features (either in emission or in absorption). This method is similar to the one used by \citet{Jorgensen2016} to determine the continuum emission level in ALMA observations of IRAS~16293$-$2422. A positive aspect of the Gaussian continuum level determination is that the width of the Gaussian function seems to be related to the discrepancy between the estimated and real continuum levels, and therefore can be used as an error in the determination of the continuum level. Similarly to the `maximum continuum level' determination, the size of the histogram bins may affect the shape of the distribution and therefore the Gaussian fit.

%----------------------------------------------------------------------
\begin{figure}[t]
\begin{center}
\begin{tabular}[b]{c}
        \includegraphics[width=0.95\columnwidth]{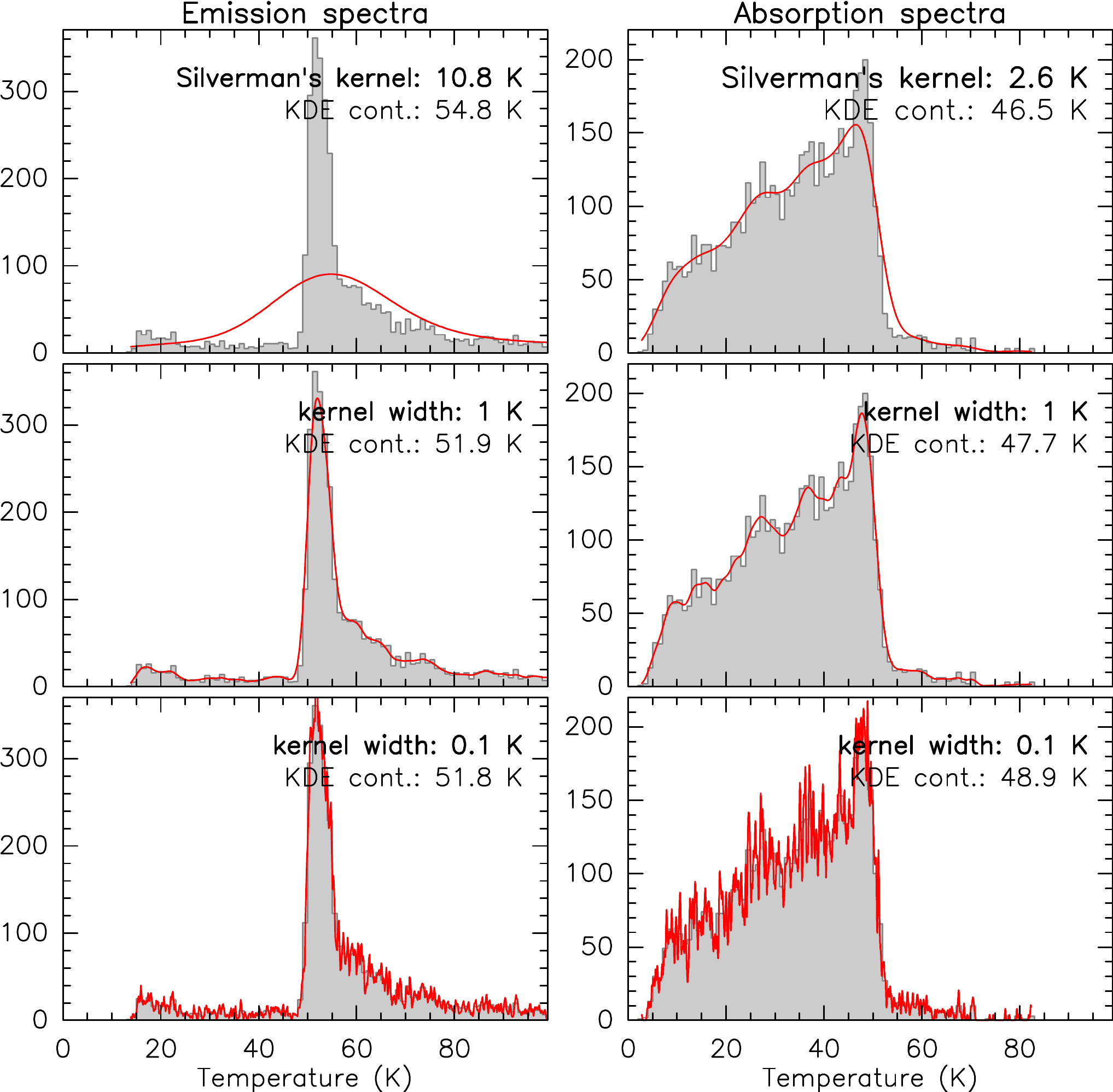} \\
\end{tabular}
\caption{Distribution of the intensities of emission-dominaed (\textit{left panels}) and absorption-dominated (\textit{right panels}) spectra cases of Fig.~\ref{f:contmethod}. The red solid lines show the KDE for different kernel widths: (\textit{top}) Silverman's rule, (\textit{middle}) kernel with of rms ($=1$~K), and (\textit{bottom}) kernel width one tenth of the rms ($=0.1$~K). The continuum level determined as the position of the maximum of the KDE is indicated in each panel.}
\label{f:KDEs}
\end{center}
\end{figure}
%----------------------------------------------------------------------

\paragraph{\textbf{KDE maximum continuum level}:}
The Kernel Density Estimation (KDE) method estimates the probability density function of a data set by applying a given kernel (function) that can be, for example, a Gaussian. There is a clear advantage of using KDEs instead of histograms when analyzing datasets with discrete data points. While the histogram shape is highly dependent on the selected bin size and the end points of the bins, the KDE removes this dependence by placing a kernel (or chosen function) at the value of each data point and then adding them up. In \statcont\ we use a Gaussian function with a kernel width determined following the \textit{Silverman's rule}\footnote{The kernel width for an univariate KDE in the Silverman's rule is determined as $(3n/4)^{-1/5}\times A$, where $n$ is the number of data points and $A$ is $\min({\sigma, IQR/1.34}$) with $\sigma$ the standard deviation and IQR the interquartile range (see Eqs.~3.28 and 3.30 in \citealt{Silverman1998}).}. The green solid line in Fig.~\ref{f:contmethod} shows the KDE. The location of the maximum of the KDE also targets the maximum of the distribution, and therefore the position of the continuum level, without being limited by the binning of the histogram. The blue line in Fig.~\ref{f:contmethod} indicates the `KDE maximum continuum level', which is also listed in column~10 of Table~\ref{t:contspectra}. The discrepancy with the theoretical value is typically about 5--10\%. The advantage of this method with respect to the Gaussian continuum level is that it does not depend on the ranges used to fit the Gaussian. The drawback is that it does not provide direct information on the width of the distribution and therefore on the accuracy of the continuum level, and it is still dependent on the kernel width considered, in a similar way to the width of the histogram bins. For example, while Silverman's rule works well for normally distributed populations, it may oversmooth if the population is multimodal \citep{Silverman1998}, as seems to happen in cases (\textit{b}) and (\textit{c}) in Fig.~\ref{f:contmethod}. In Fig.~\ref{f:KDEs}, we compare the KDE for two different spectra (emission and absorption-dominated cases) considering three different kernel widths: the Silverman kernel width, a width corresponding to the noise of the spectrum ($=1$~K), and a width one order of magnitude smaller than the noise ($=0.1$~K). While Silverman's width oversmooths the emission-dominated spectrum, the KDE with a width equal to the noise (smaller than the Silverman's width) seems to better describe multimodal populations. However, small kernel widths are more sensitive to fluctuations as shown in the bottom panels of Fig.~\ref{f:KDEs}, which might produce inaccurate continuum level determinations.

\paragraph{\textbf{Sigma-clipping continuum level}:}
In this last method we try to combine all the positive aspects of the previous methods while avoiding the drawbacks. It is based on the $\sigma$-clipping algorithm that determines the median and standard deviation of a distribution iteratively. In a first step the algorithm calculates the median ($\mu$) and standard deviation ($\sigma$) of the entire distribution. In a second step, it removes all the points that are smaller or larger than $\mu\pm\alpha\sigma$, where $\alpha$ is a parameter that can be provided by the user. The process is repeated until the stop criterion is reached: either a certain number of iterations, or when the new standard deviation is within a certain tolerance level of the previous iteration value. It is worth mentioning that the standard deviation is strongly dependent on outliers (or in our examples, on intensities far away from the continuum level). Therefore, in each iteration $\sigma$ will decrease or remain the same. If $\sigma$ is within the certain tolerance level, determined as $(\sigma_\mathrm{old}-\sigma_\mathrm{new})/\sigma_\mathrm{new}$, the process will stop because it has successfully removed all the outliers. The `sigma-clipping continuum level' of the examples in Fig.~\ref{f:contmethod} is shown with a thin black solid line, and the levels are listed in column~11 of Table~\ref{t:contspectra}. This method is by far the one with smallest discrepancies with respect to the theoretical level ($<5$\%). Furthermore, it does not depend on the size of the histogram bins, and the final standard deviation can be used as the uncertainty in the determination of the continuum level.

%----------------------------------------------------------------------
\begin{figure*}[t]
\begin{center}
\begin{tabular}[b]{c c c c}
        \includegraphics[height=0.115\textheight]{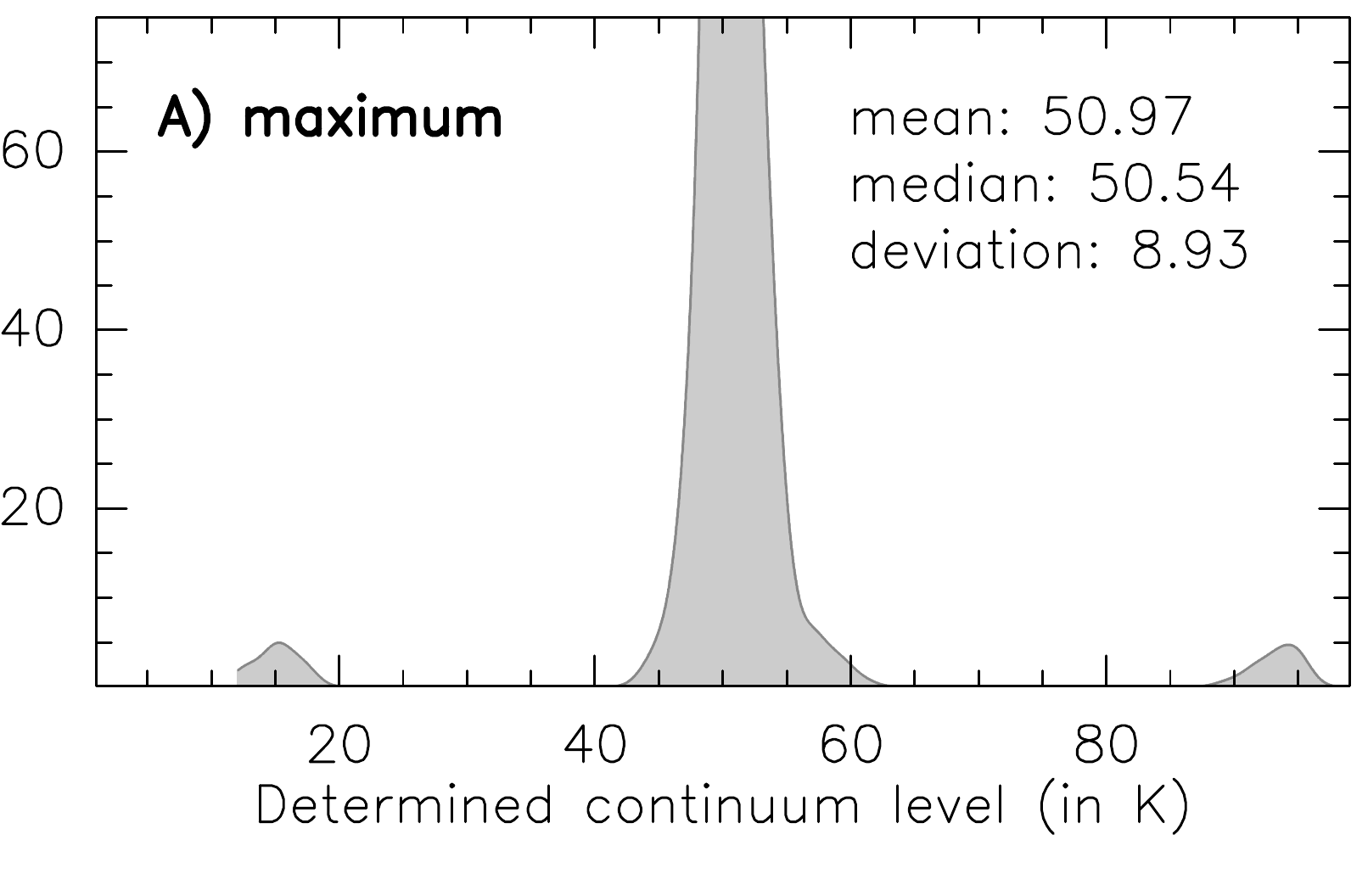} &
        \includegraphics[height=0.115\textheight]{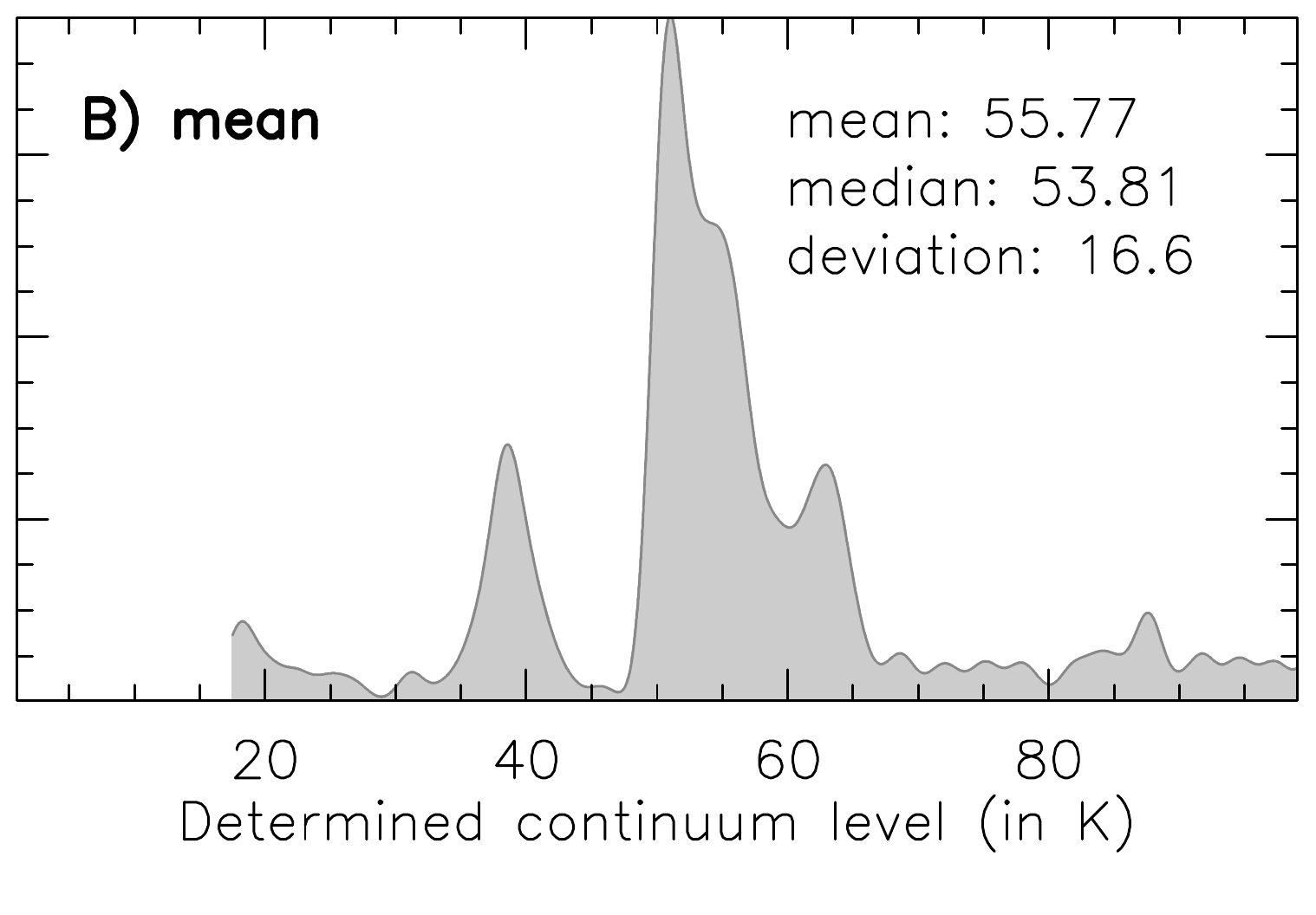} &
        \includegraphics[height=0.115\textheight]{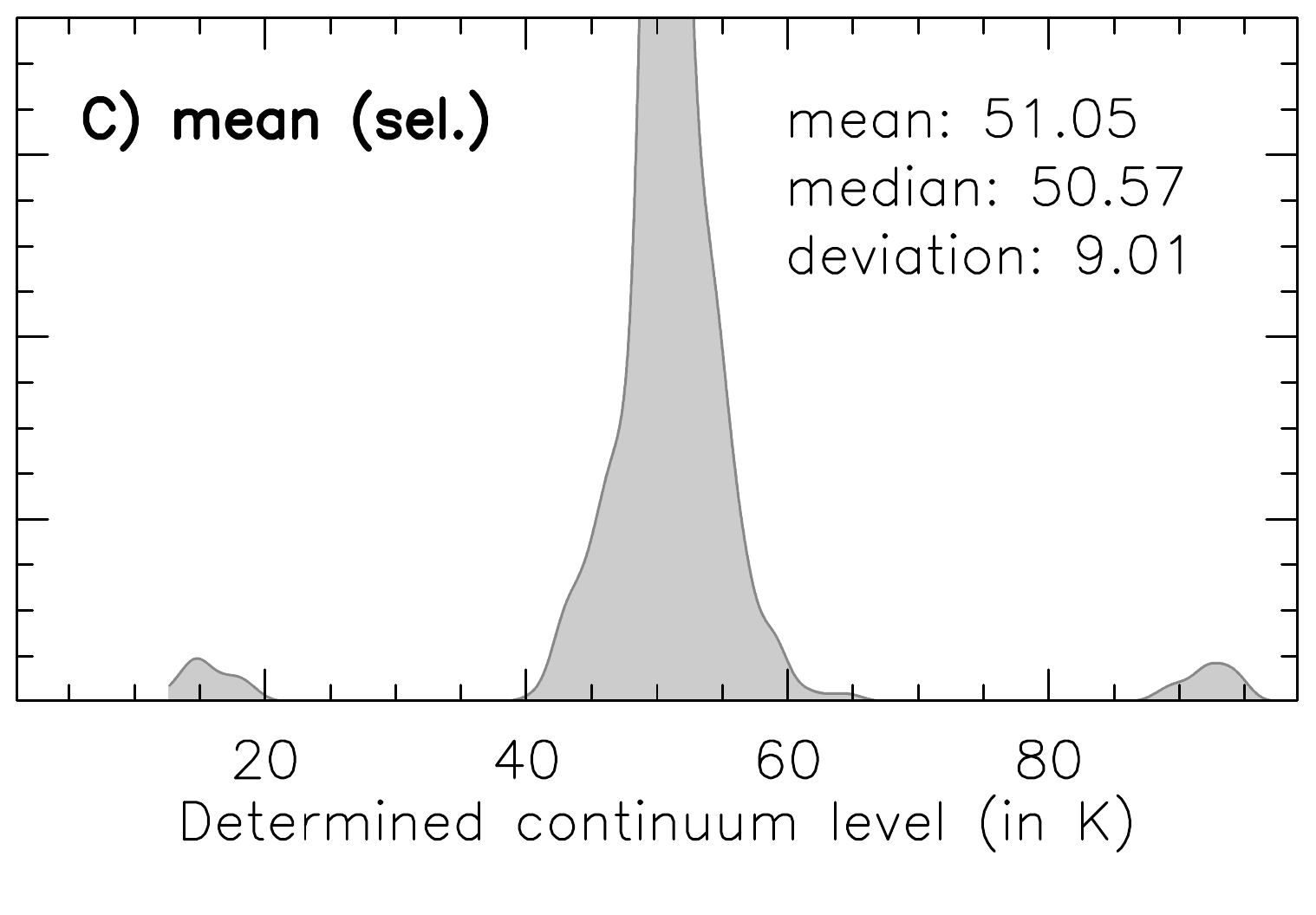} &
        \includegraphics[height=0.115\textheight]{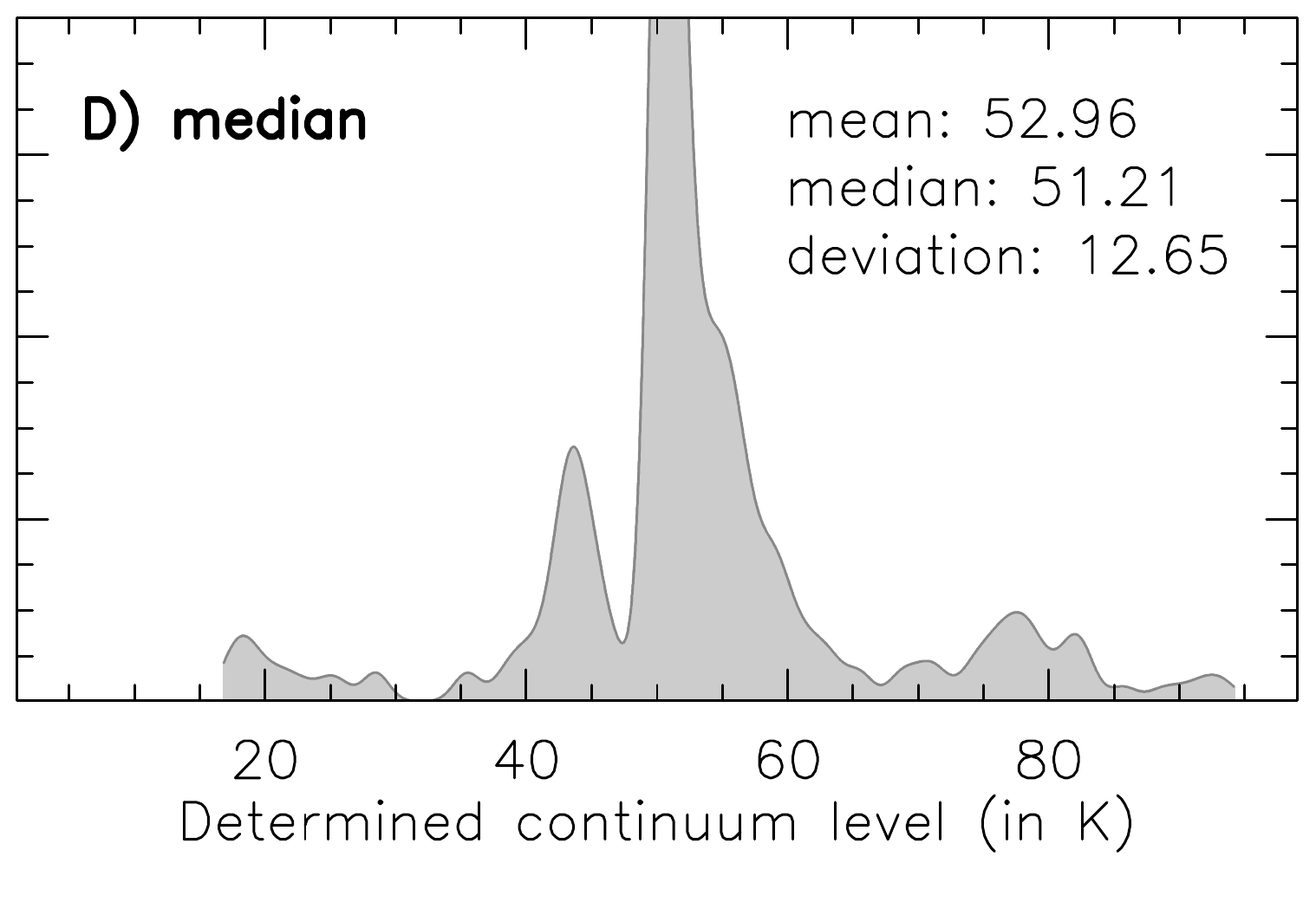} \\
        \includegraphics[height=0.115\textheight]{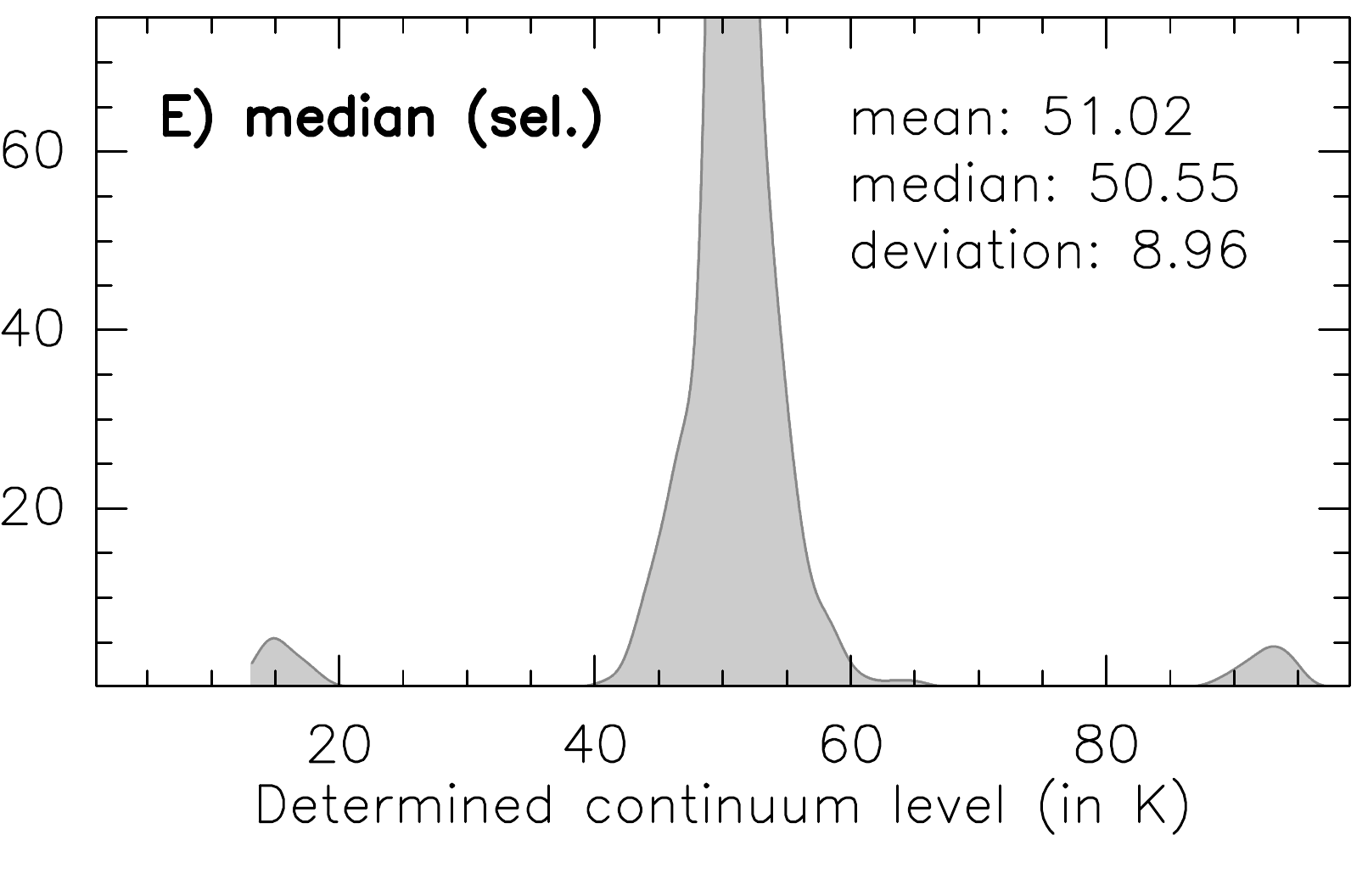} &
        \includegraphics[height=0.115\textheight]{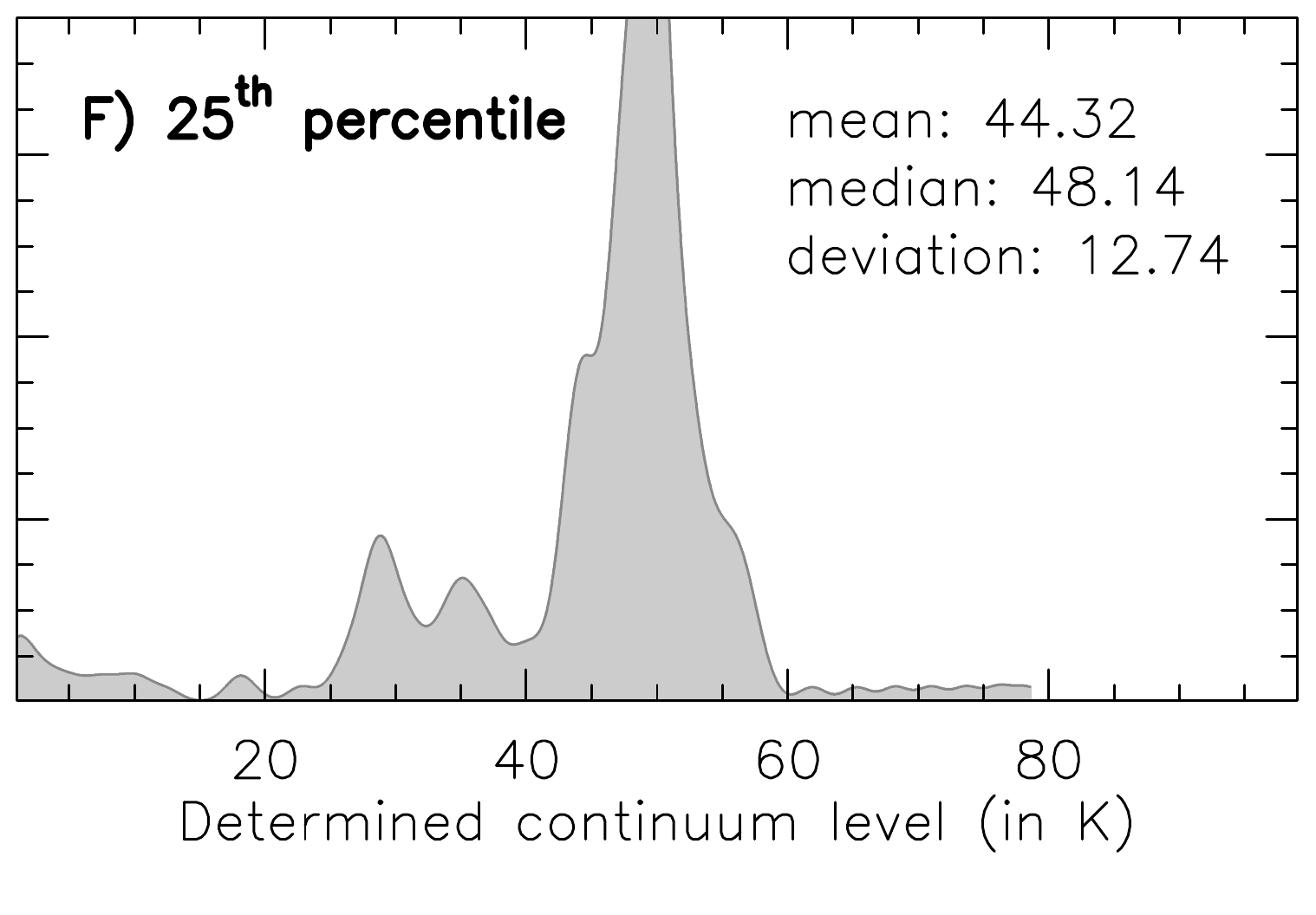} &
        \includegraphics[height=0.115\textheight]{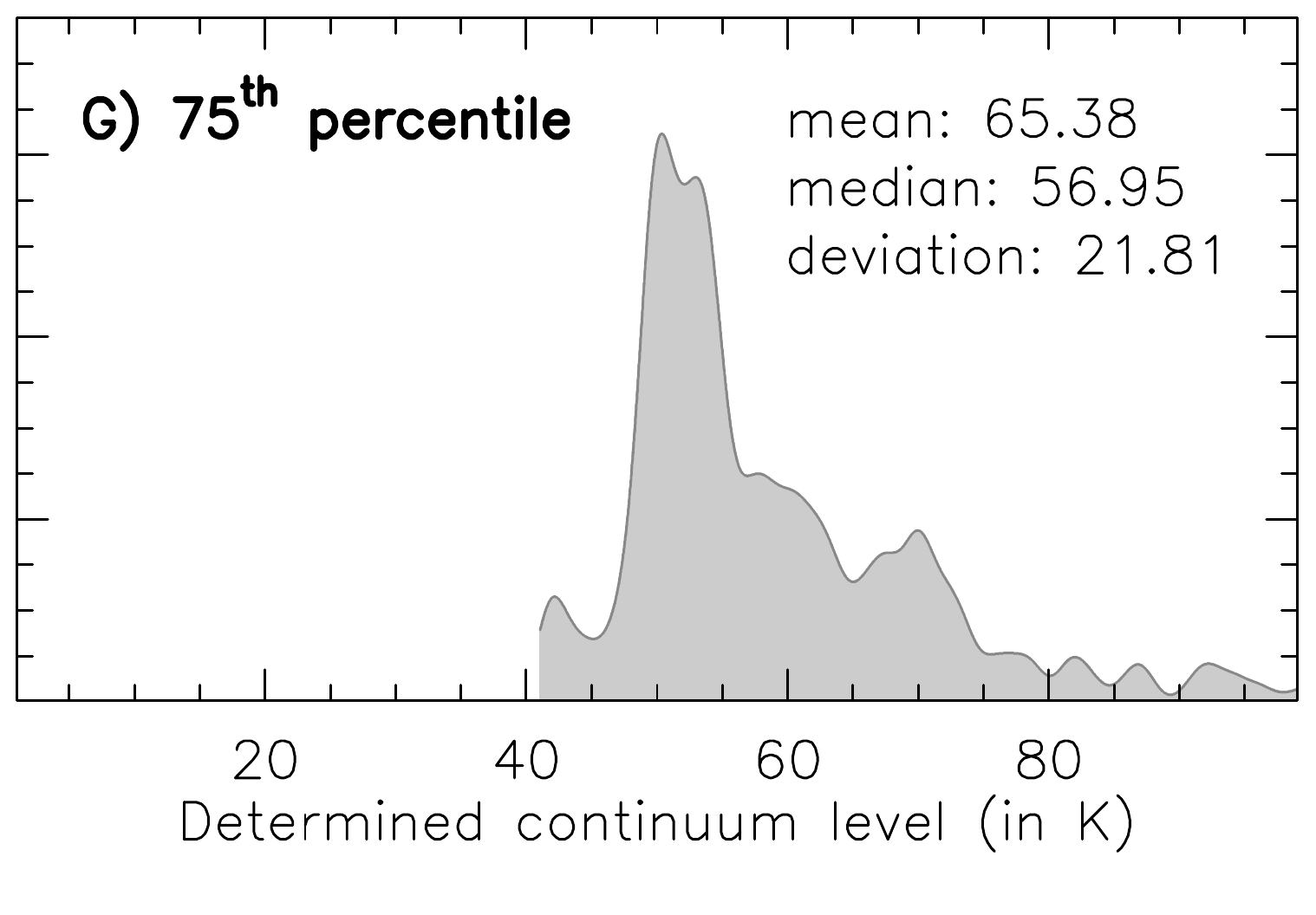} &
        \includegraphics[height=0.115\textheight]{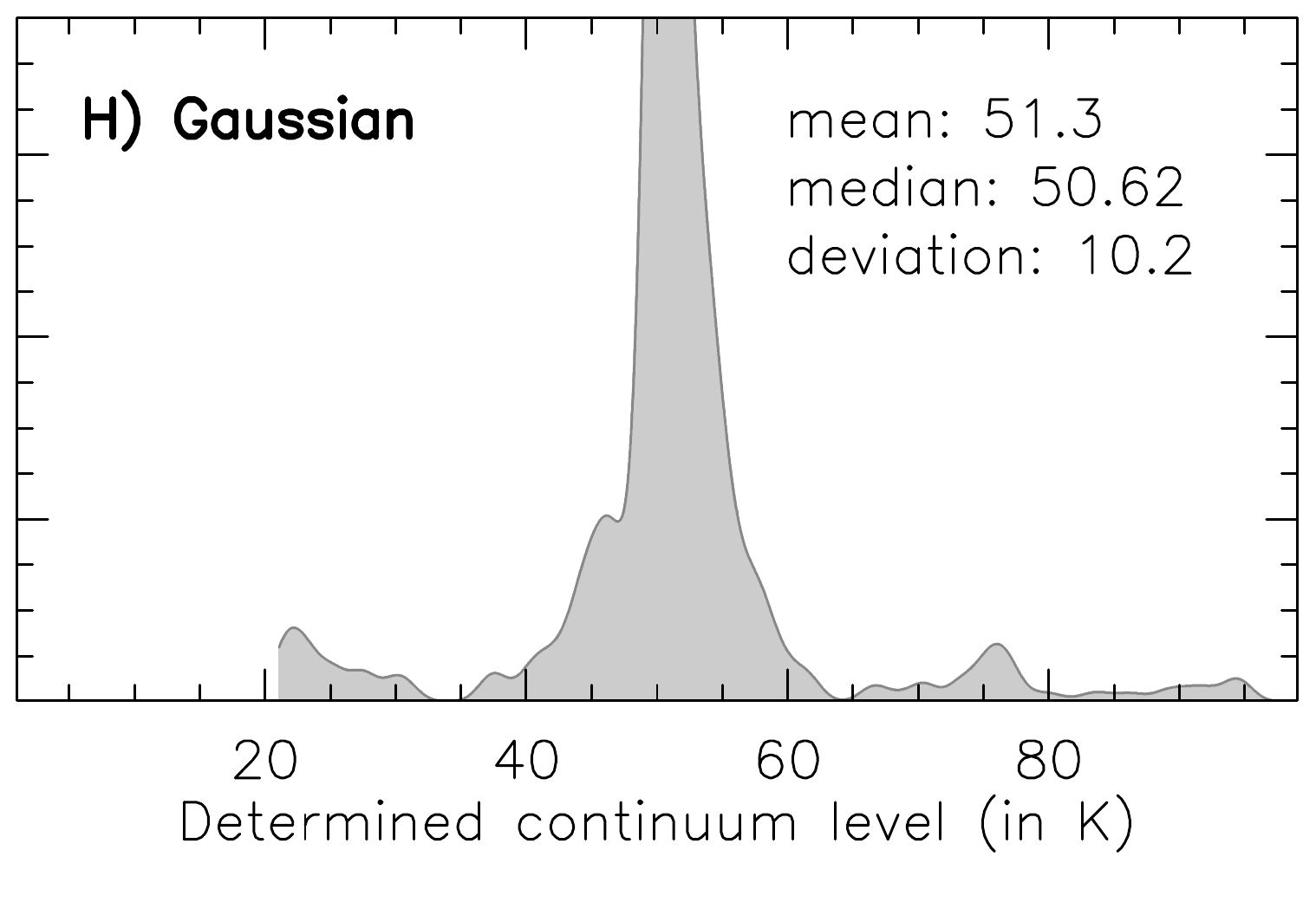} \\
        \includegraphics[height=0.115\textheight]{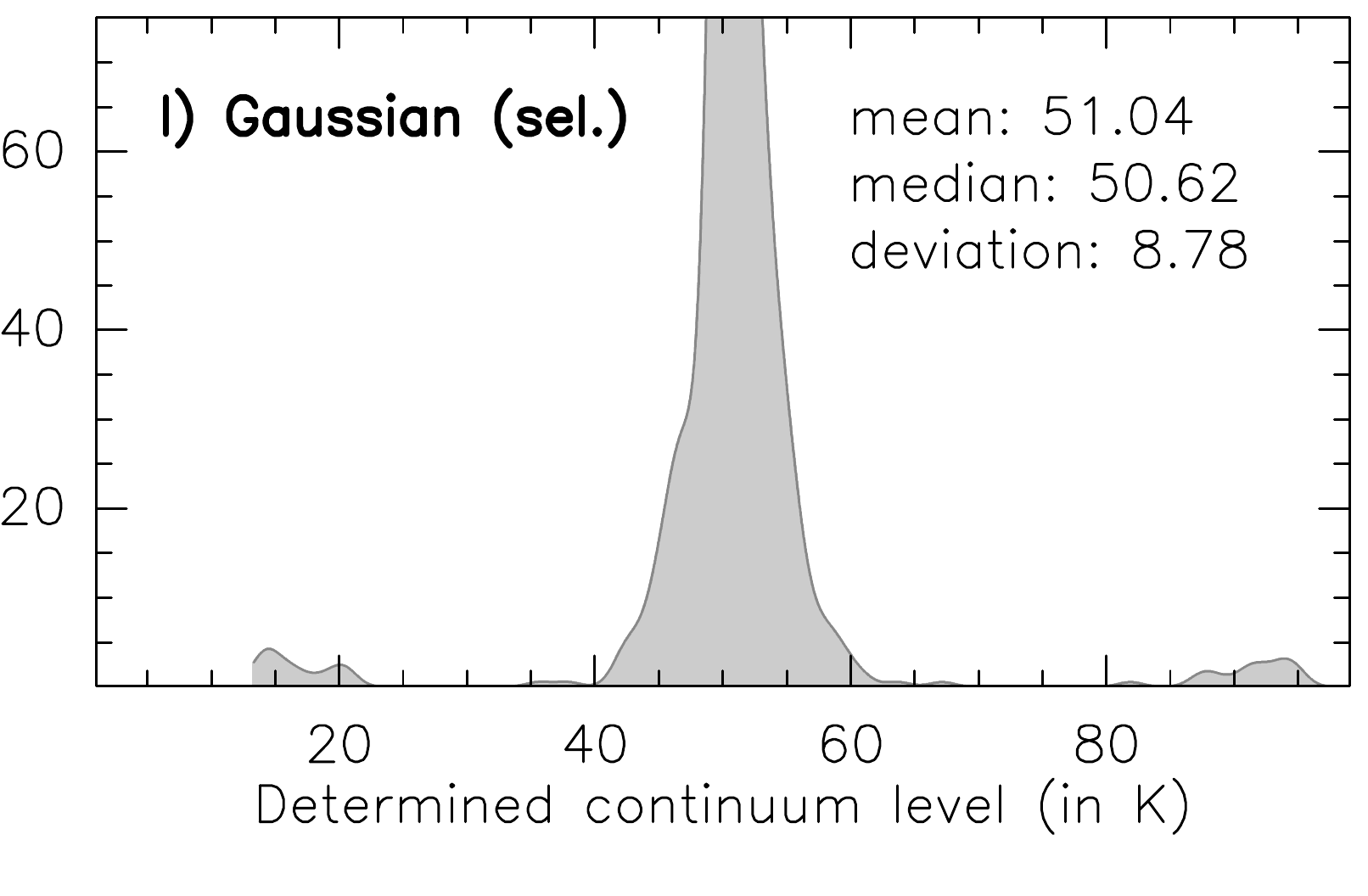} &
        \includegraphics[height=0.115\textheight]{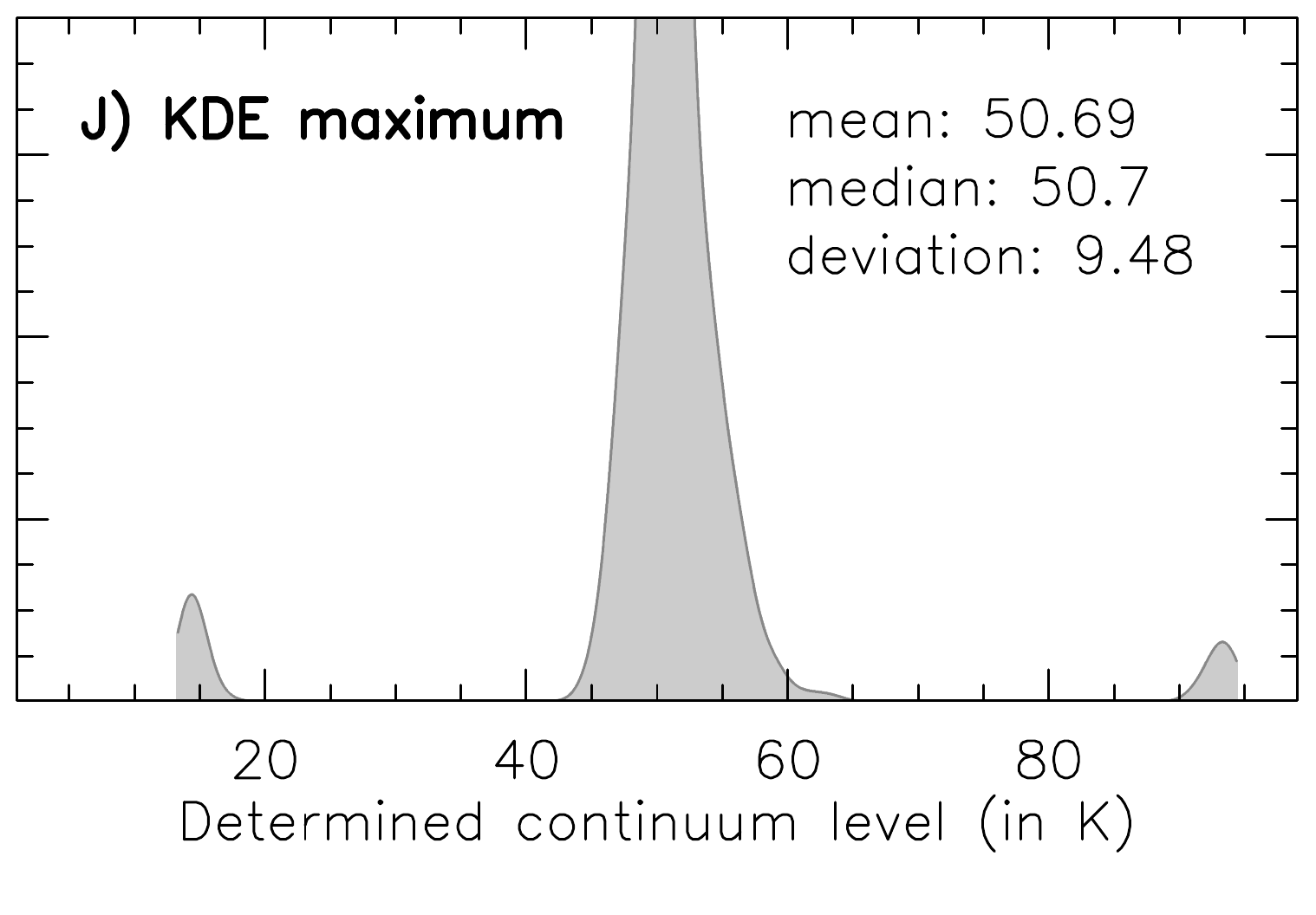} &
        \includegraphics[height=0.115\textheight]{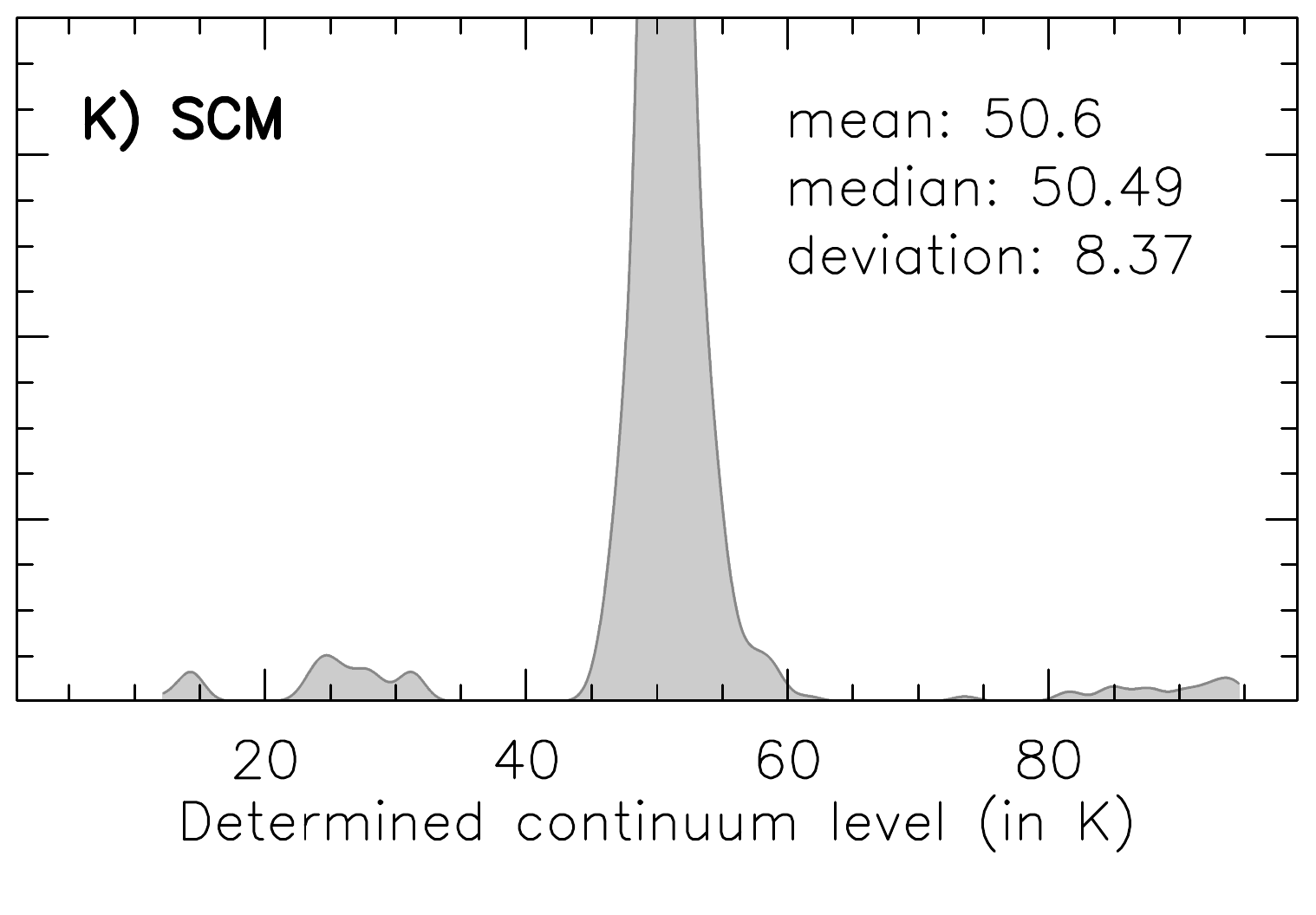} &
        \includegraphics[height=0.115\textheight]{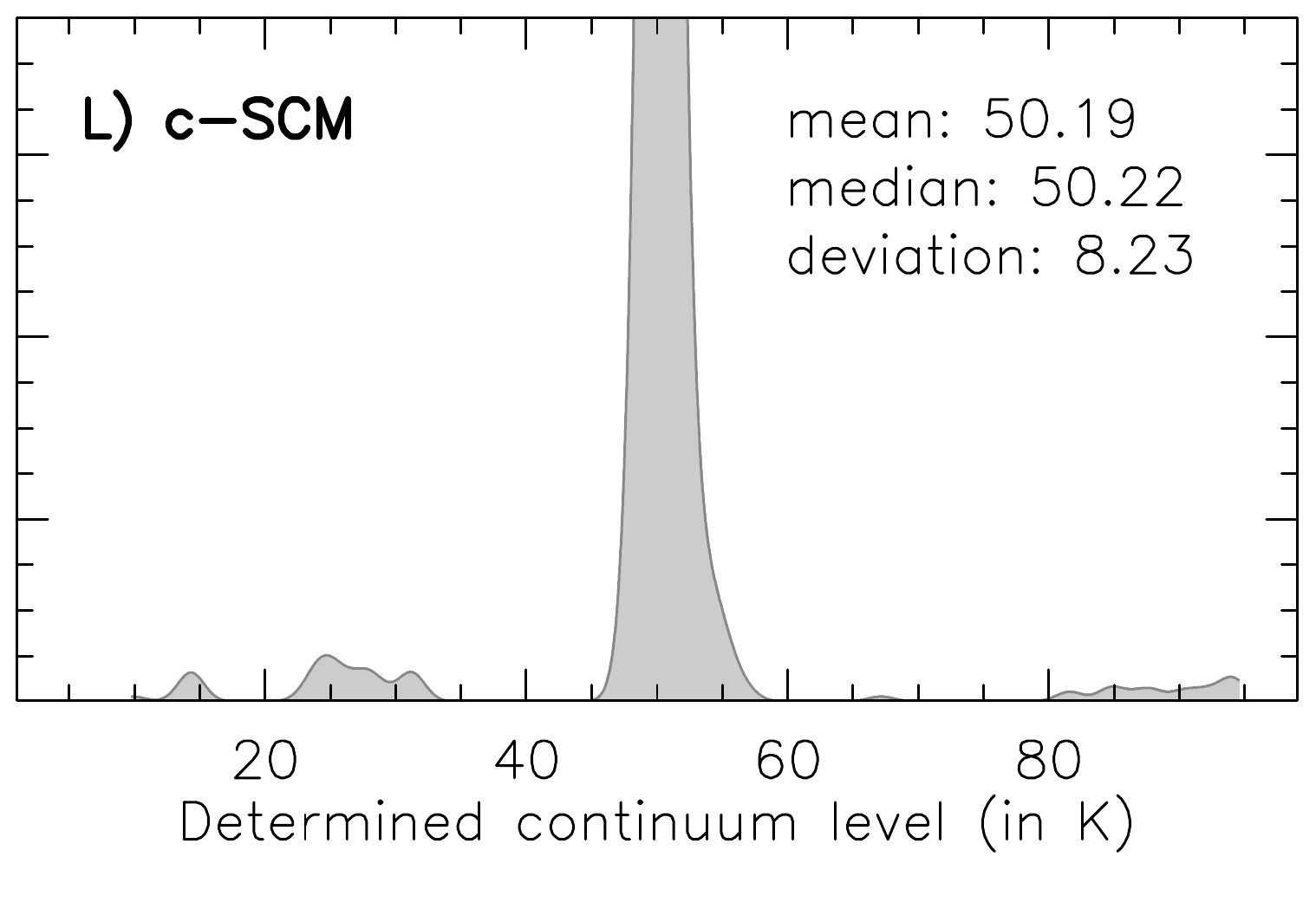} \\
\end{tabular}
\caption{Distribution of the continuum levels for 750 different synthetic spectra (see Sect.~\ref{s:statistics}) determined with the methods: (\textit{A}) maximum, (\textit{B}) mean, (\textit{C}) mean considering the bins around the maximum of the spectral histogram, (\textit{D}) median, (\textit{E}) median of the selected bins, (\textit{F}) the 25$^\mathrm{th}$ percentile, (\textit{G}) the 75$^\mathrm{th}$ percentile, (\textit{H}) Gaussian fit, (\textit{I}) Gaussian fit to the selected bins, (\textit{J}) the maximum of the KDE, (\textit{K}) the sigma-clipping method, and (\textit{L}) the corrected version of the sigma-clipping method as introduced in Sect.~\ref{s:correction}. For each panel, we indicate in the top-right corner the mean, median and standard deviation of the distributions (see also Table~\ref{t:statistics}).}
\label{f:statistics}
\end{center}
\end{figure*}
%----------------------------------------------------------------------

%----------------------------------------------------------------------
\begin{table*}
\caption{\label{t:statistics} Statistical comparison between the different continuum determination methods described in Sect.~\ref{s:contspectra}}
\centering
\begin{tabular}{l c c c c c c c c}
\hline\hline

&
&
&
&\multicolumn{5}{c}{Success rate (in \%) for different discrepancies}
\\
\cline{5-9}
\noalign{\smallskip}

\multicolumn{1}{c}{Continuum determination method}
&mean
&median
&deviation
&$<1$\%
&$<5$\%
&5--10\%
&10--25\%
&$>25$\%
\\
\multicolumn{1}{c}{(1)}
&(2)
&(3)
&(4)
&(5)
&(6)
&(7)
&(8)
&(9)
\\
\hline
\textit{(A)} maximum                  &50.97      &50.54      &\phn8.9   &26.4     &75.3     &16.5    &\phn3.3 &\phn4.9 \\
\textit{(B)} mean                     &55.77      &53.81      &16.6      &\phn7.8  &22.4     &13.6    &31.2    &32.8 \\
\textit{(C)} mean (sel.)              &51.05      &50.57      &\phn9.0   &25.3     &63.9     &20.0    &11.1    &\phn5.0 \\
\textit{(D)} median                   &52.96      &51.21      &12.7      &16.5     &41.5     &13.2    &25.9    &19.4 \\
\textit{(E)} median (sel.)            &51.02      &50.55      &\phn9.0   &27.6     &67.9     &18.4    &\phn8.7 &\phn5.0 \\
\textit{(F)} 25$^\mathrm{th}$ percent &44.32      &48.14      &12.7      &11.0     &41.5     &16.9    &18.1    &23.5 \\
\textit{(G)} 75$^\mathrm{th}$ percent &65.38      &56.95      &21.8      &\phn7.3  &25.9     &15.2    &22.0    &36.9 \\
\textit{(H)} Gaussian                 &51.31      &50.62      &10.2      &26.8     &57.2     &17.8    &13.1    &11.9 \\
\textit{(I)} Gaussian (sel.)          &51.04      &50.62      &\phn8.8   &25.6     &65.7     &19.9    &\phn9.2 &\phn5.2 \\
\textit{(J)} KDE max                  &50.69      &50.70      &\phn9.5   &21.6     &70.5     &16.4    &\phn7.3 &\phn5.8 \\
\textit{(K)} SCM                      &50.60      &50.49      &\phn8.4   &33.8     &77.1     &12.9    &\phn3.2 &\phn6.8 \\
\textit{(L)} c-SCM                    &50.19      &50.22      &\phn8.2   &48.8     &87.9     &\phn4.4 &\phn0.9 &\phn6.8 \\
\hline
\end{tabular}
\tablefoot{The real continuum level of each of the 750 synthetic spectra is 50~K. Therefore, the closer the mean and median values are to this value, the more accurate the continuum determination method is. The success rate indicates how many cases (in \%) are within a certain level of discrepancy: difference between determined continuum level and real value of 50~K.}
\end{table*}
%----------------------------------------------------------------------

%----------------------------------------------------------------------
\begin{figure*}[t]
\begin{center}
\begin{tabular}[b]{c c c c}
        \includegraphics[height=0.115\textheight]{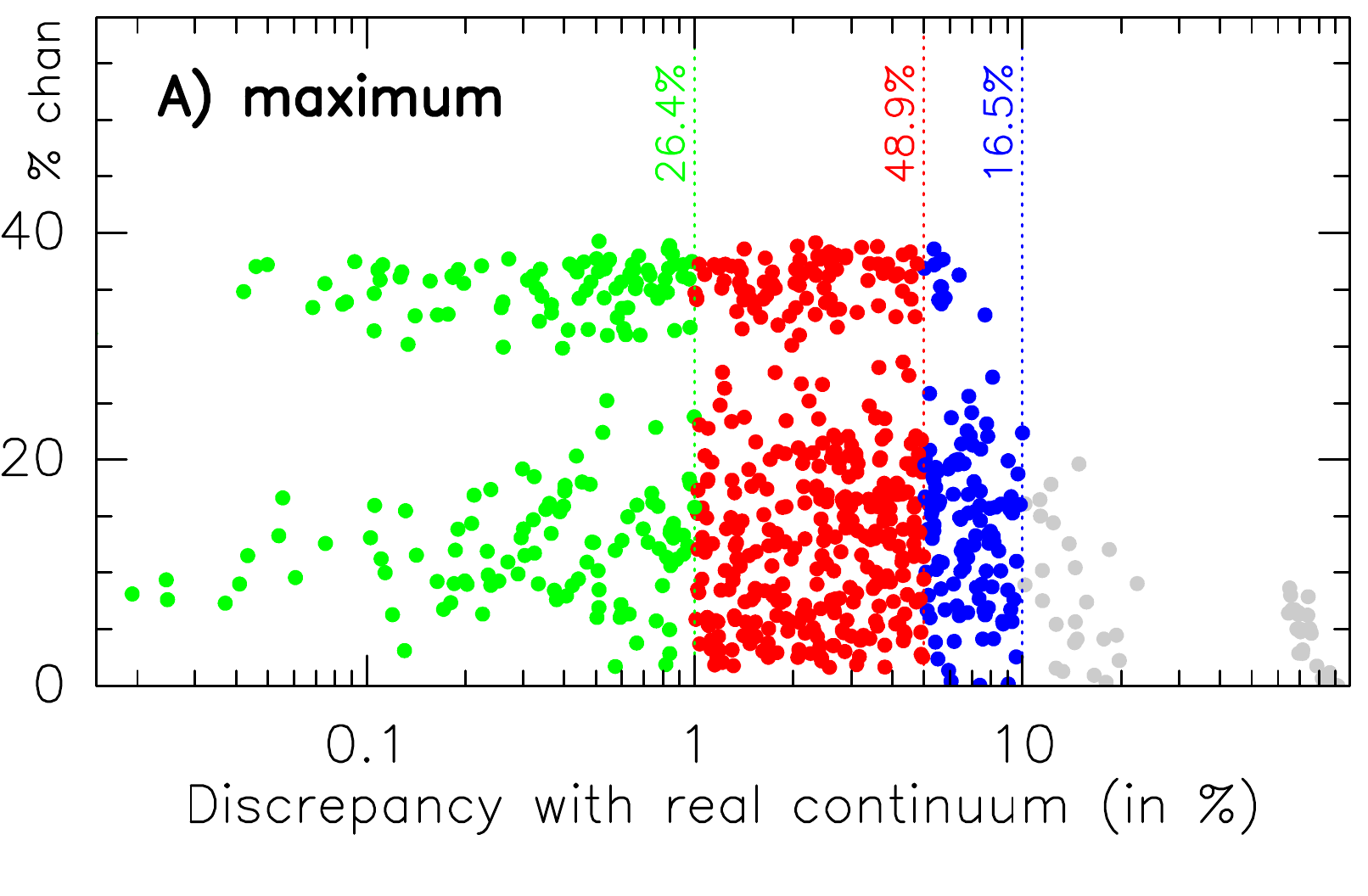} &
        \includegraphics[height=0.115\textheight]{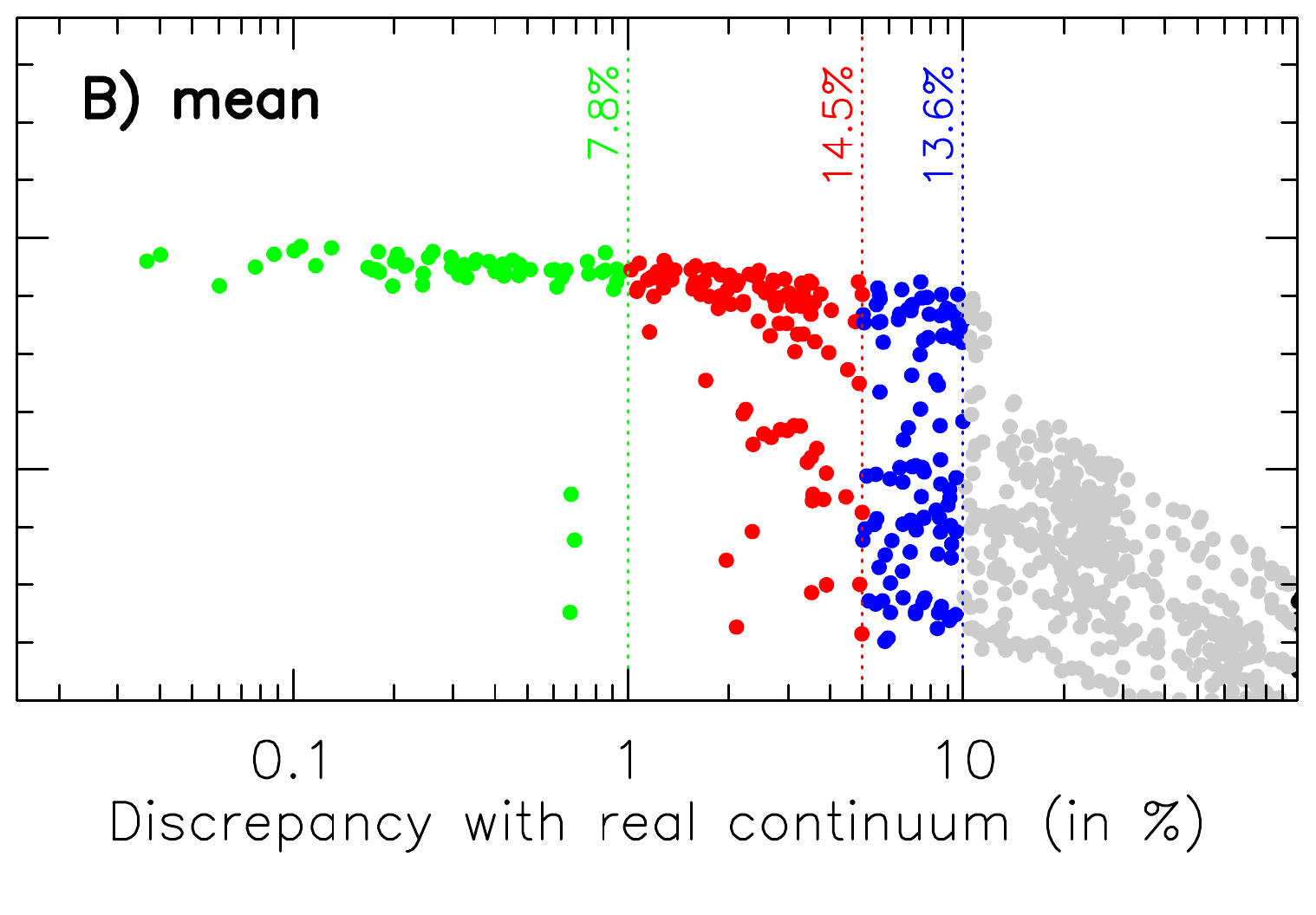} &
        \includegraphics[height=0.115\textheight]{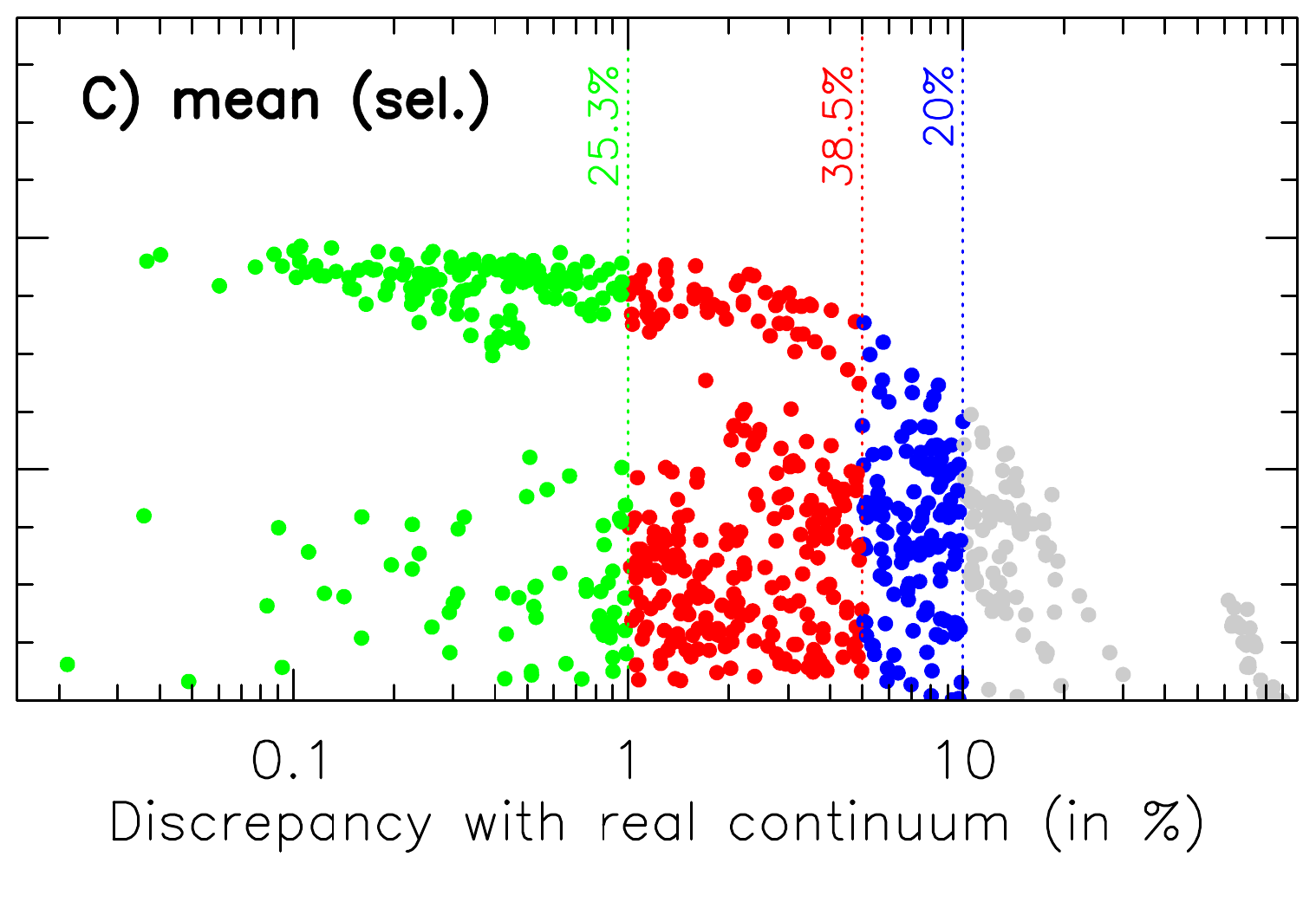} &
        \includegraphics[height=0.115\textheight]{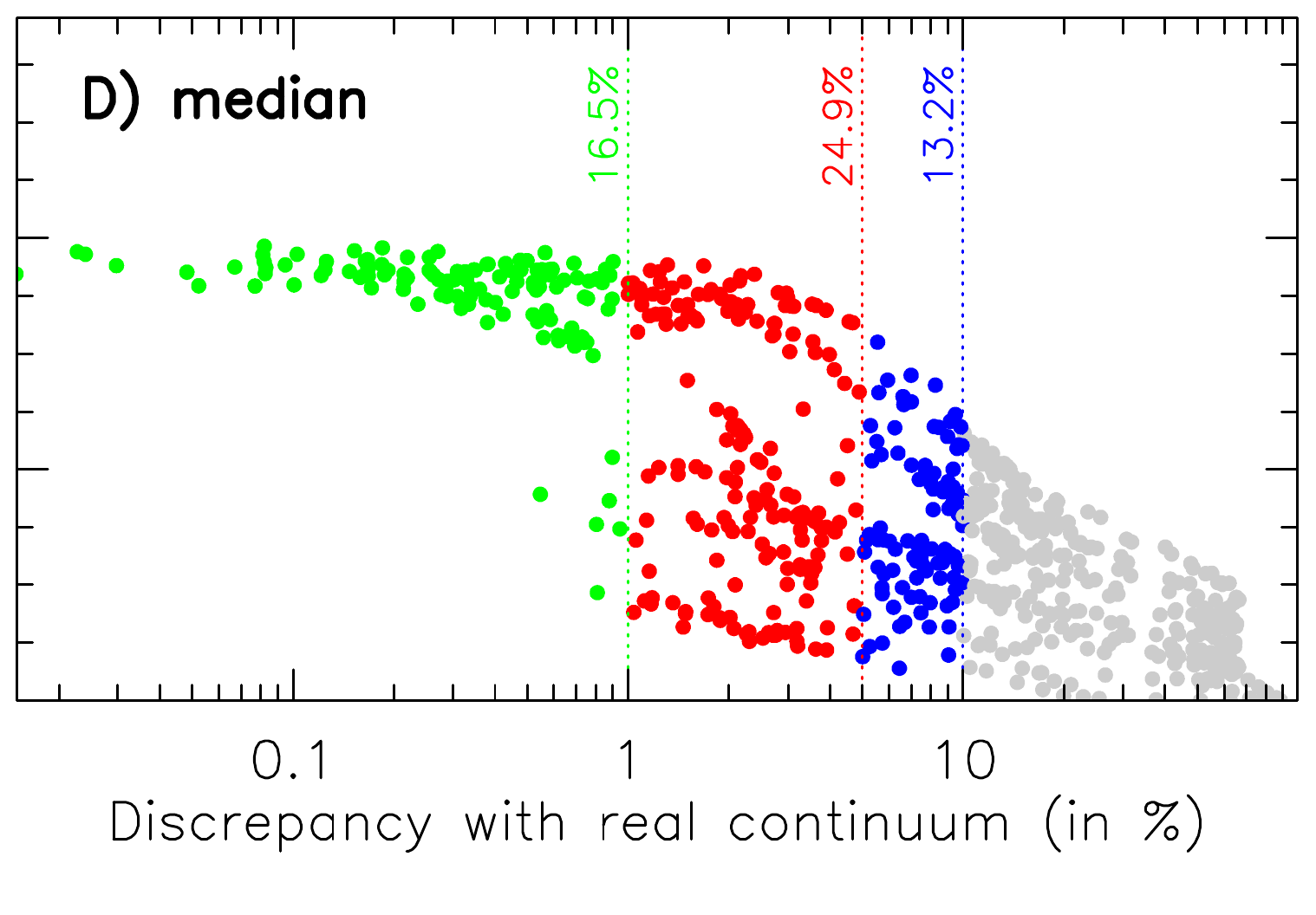} \\
        \includegraphics[height=0.115\textheight]{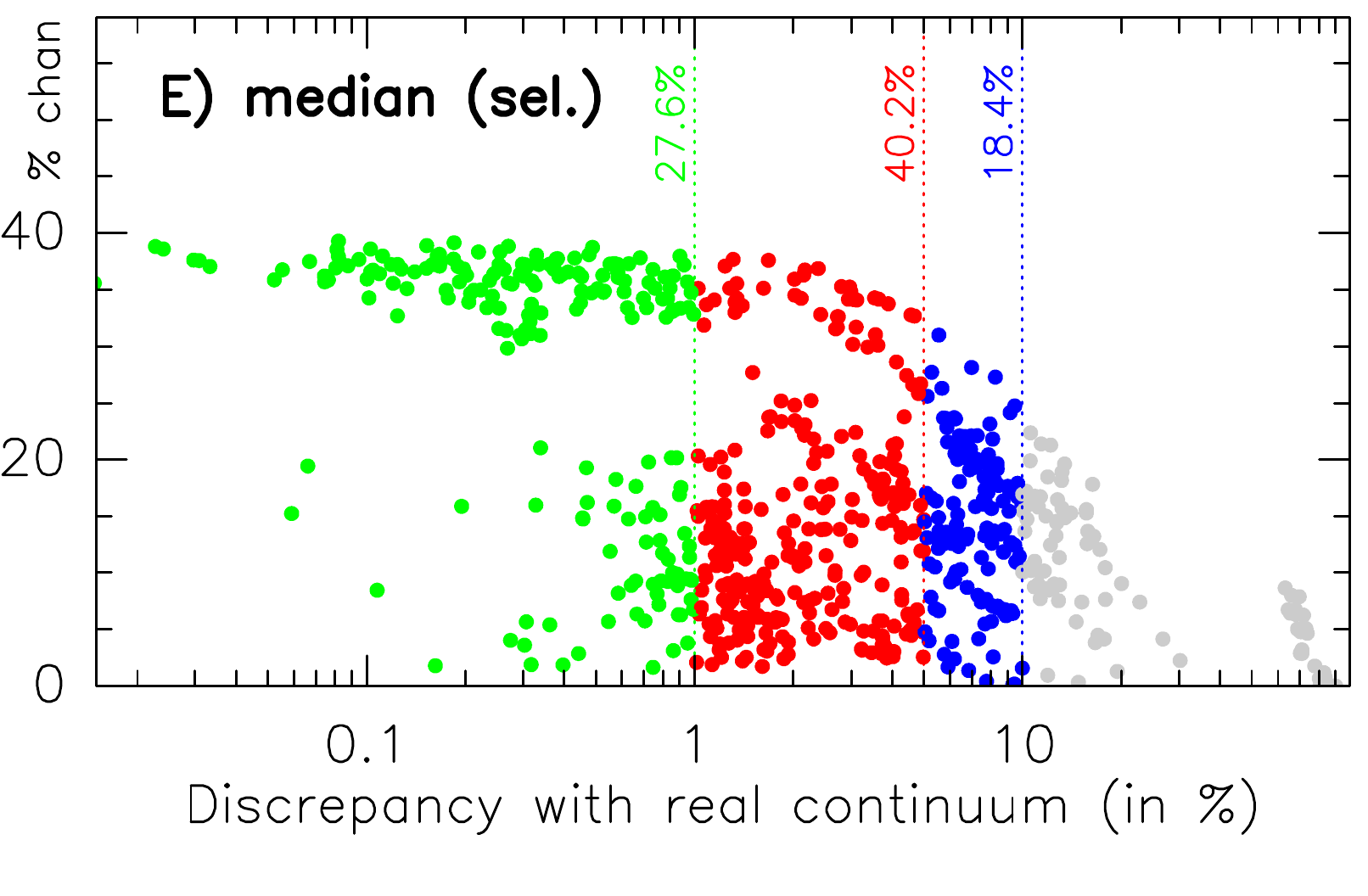} &
        \includegraphics[height=0.115\textheight]{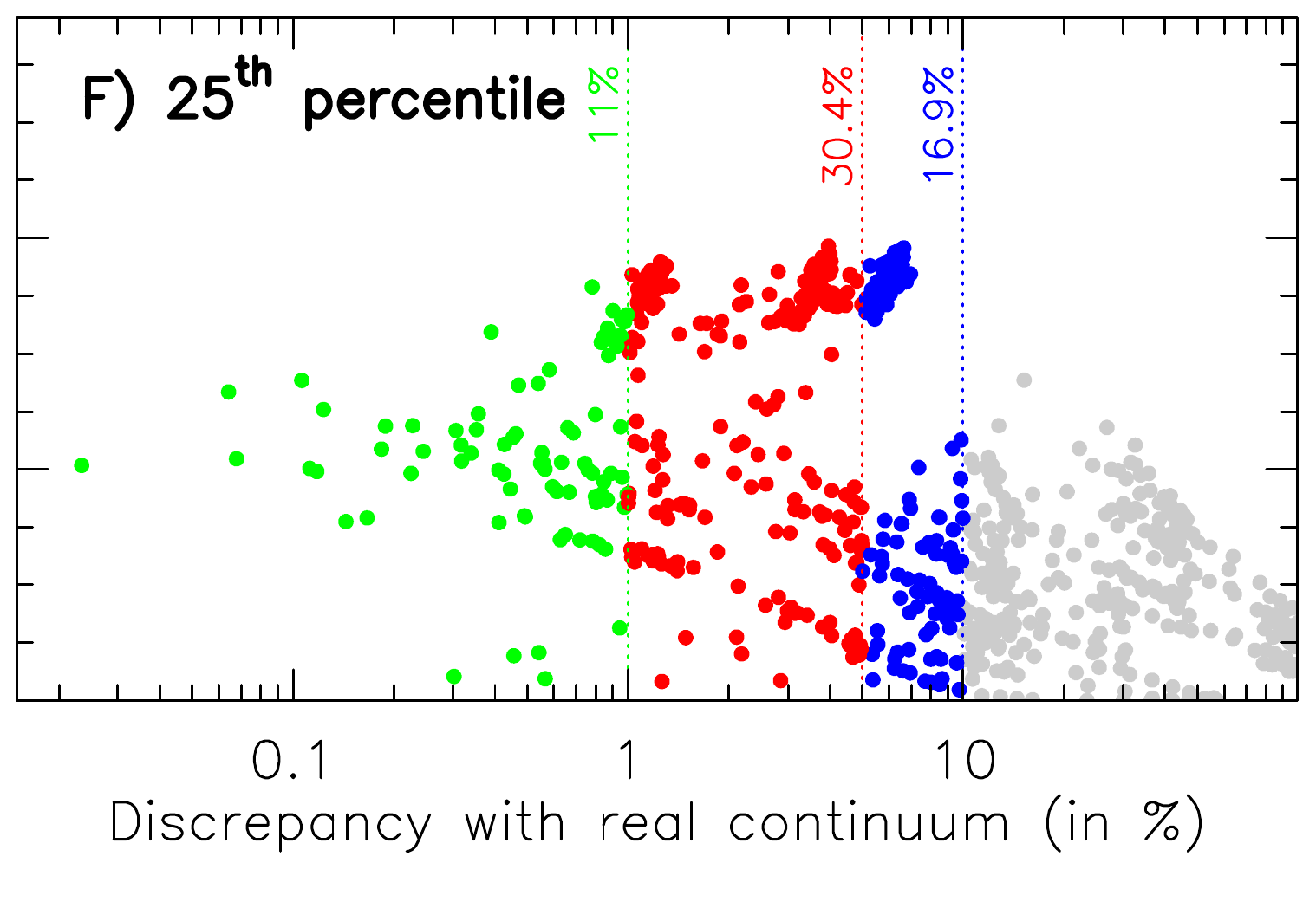} &
        \includegraphics[height=0.115\textheight]{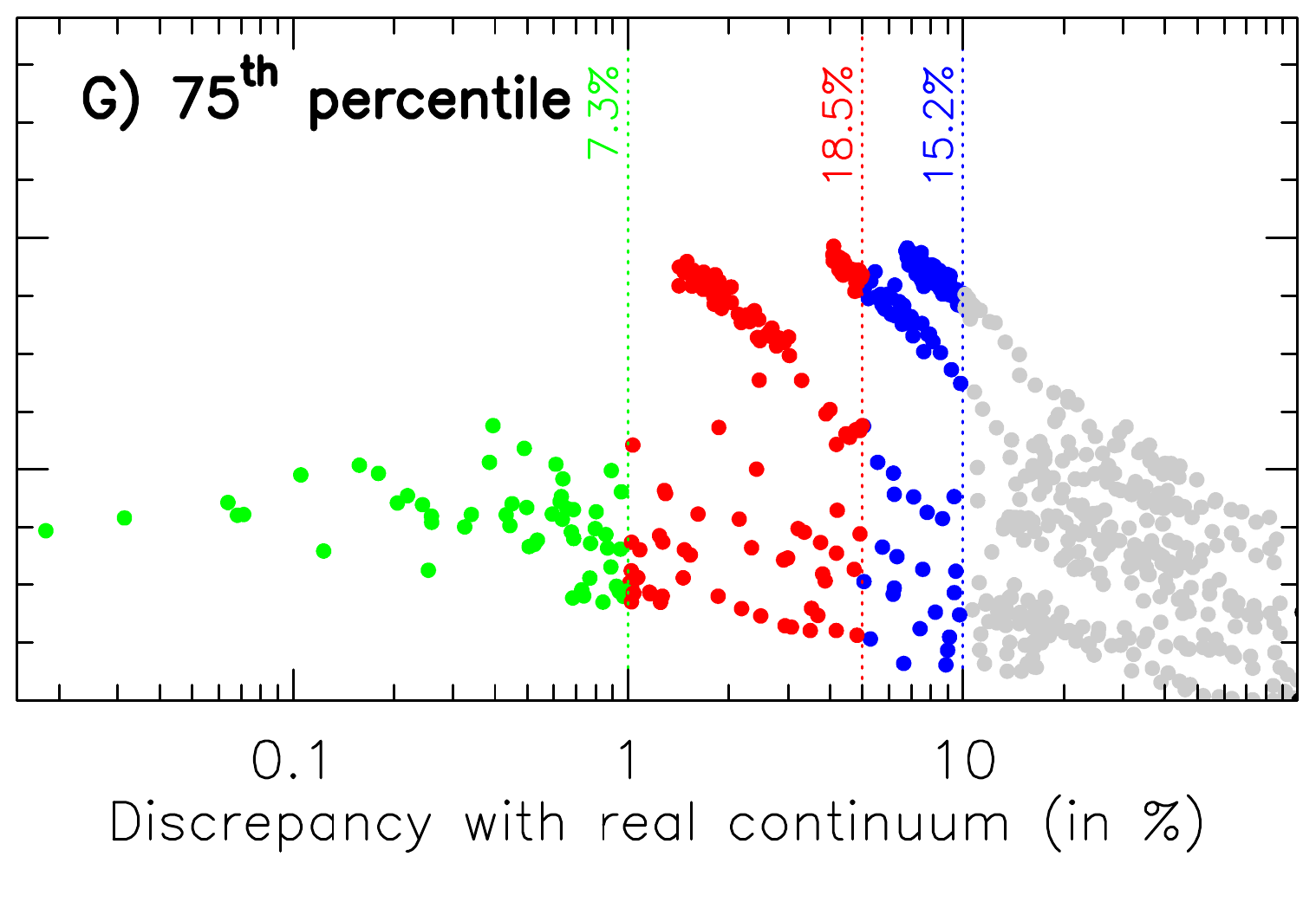} &
        \includegraphics[height=0.115\textheight]{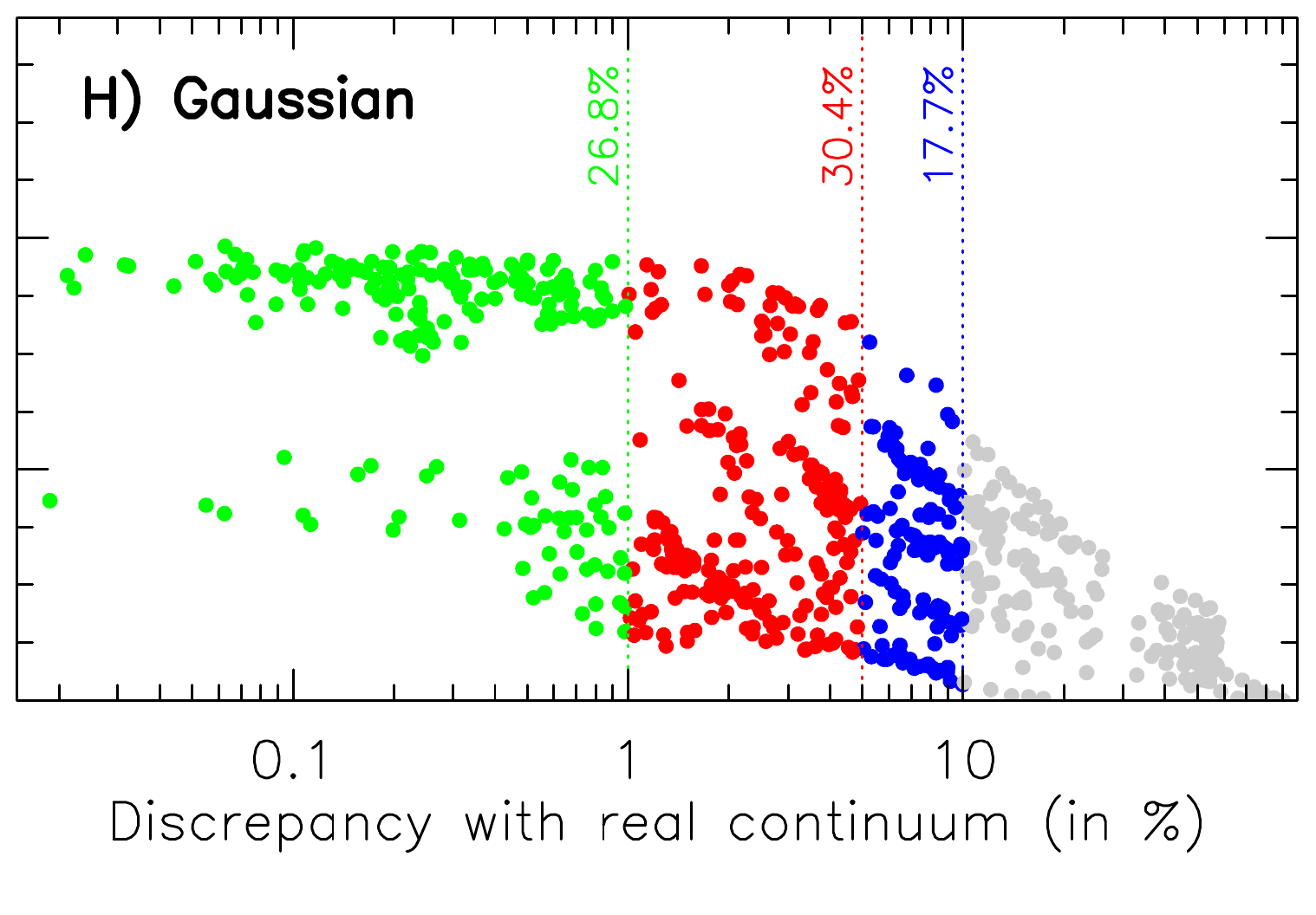} \\
        \includegraphics[height=0.115\textheight]{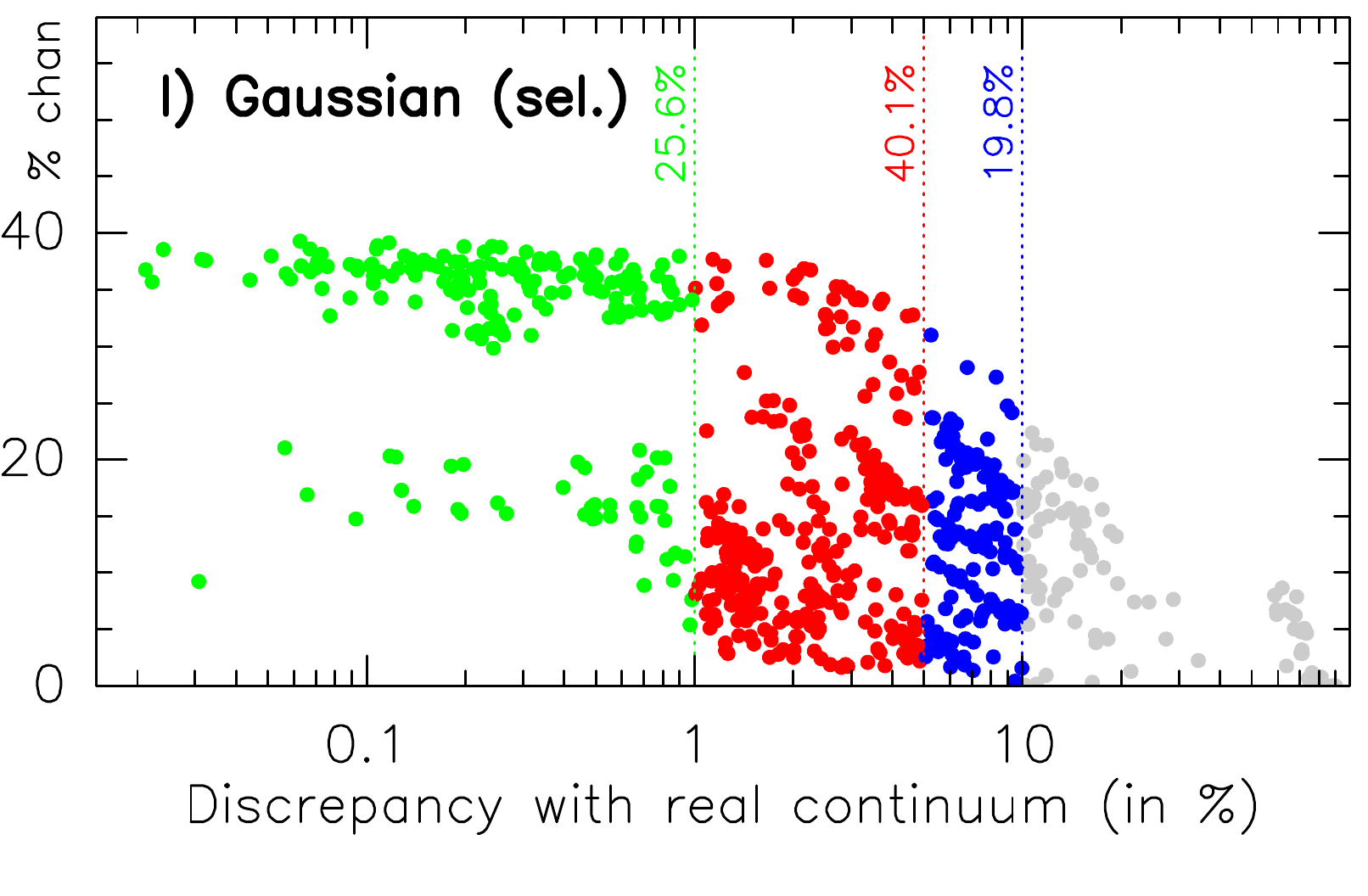} &
        \includegraphics[height=0.115\textheight]{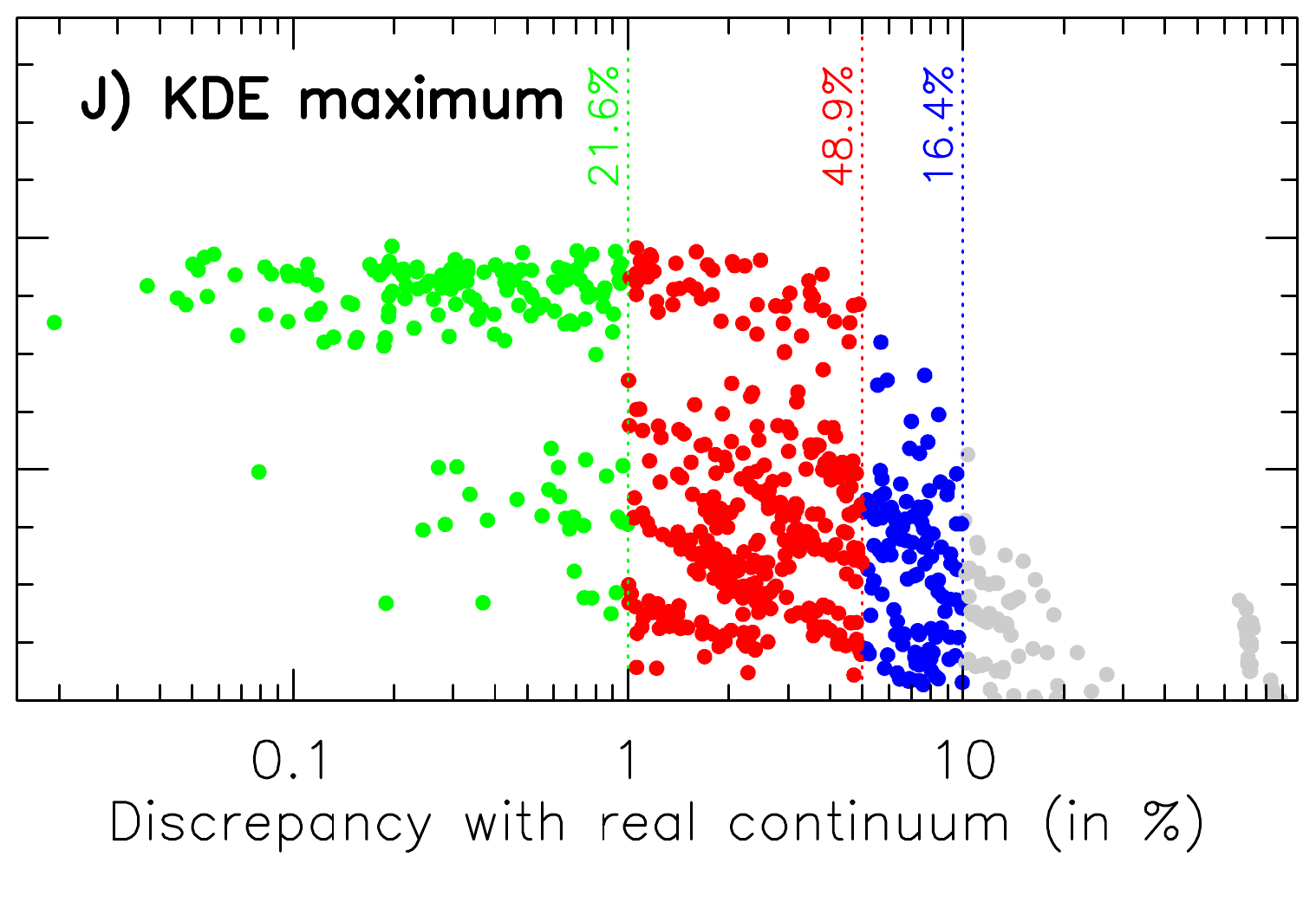} &
        \includegraphics[height=0.115\textheight]{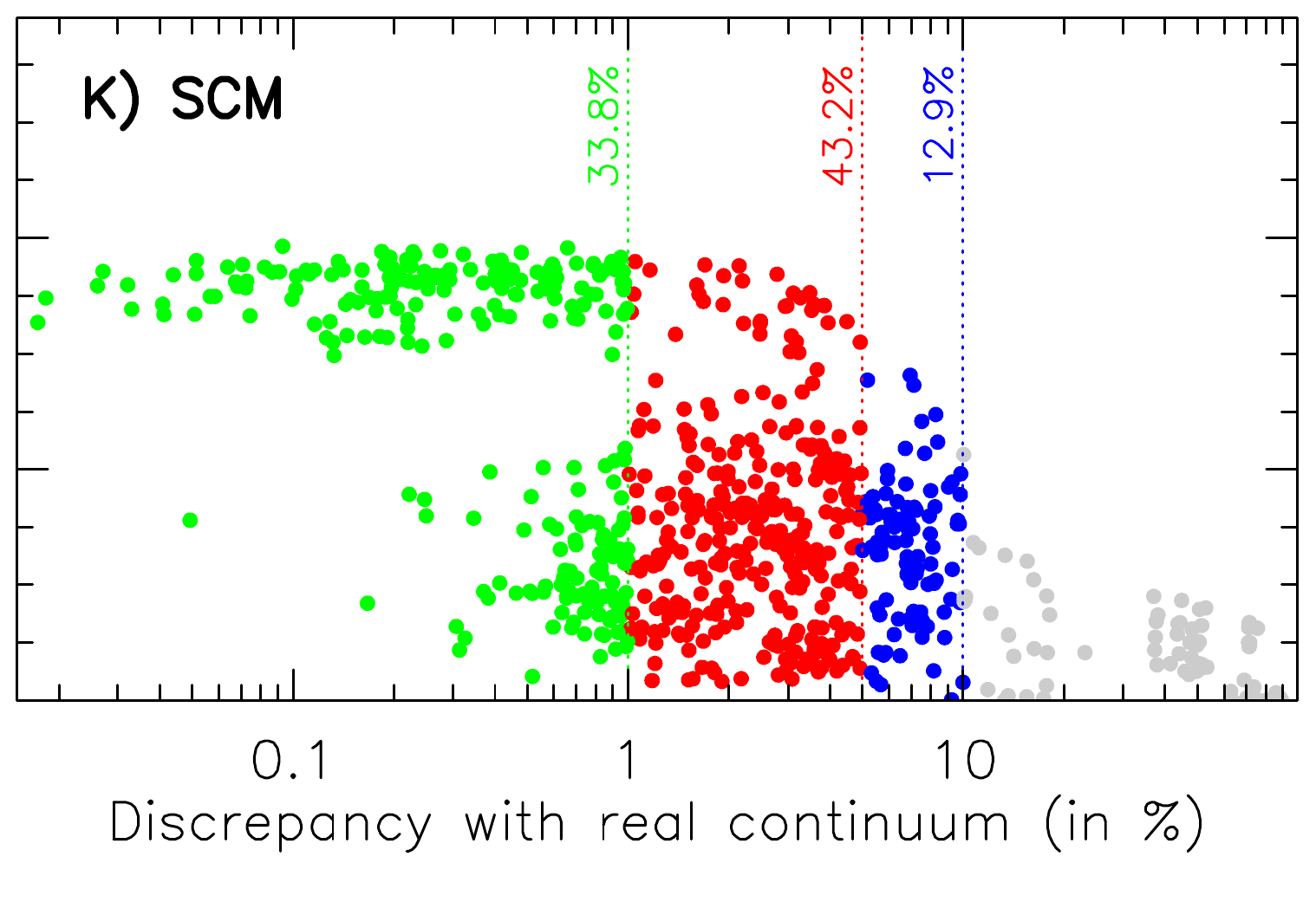} &
        \includegraphics[height=0.115\textheight]{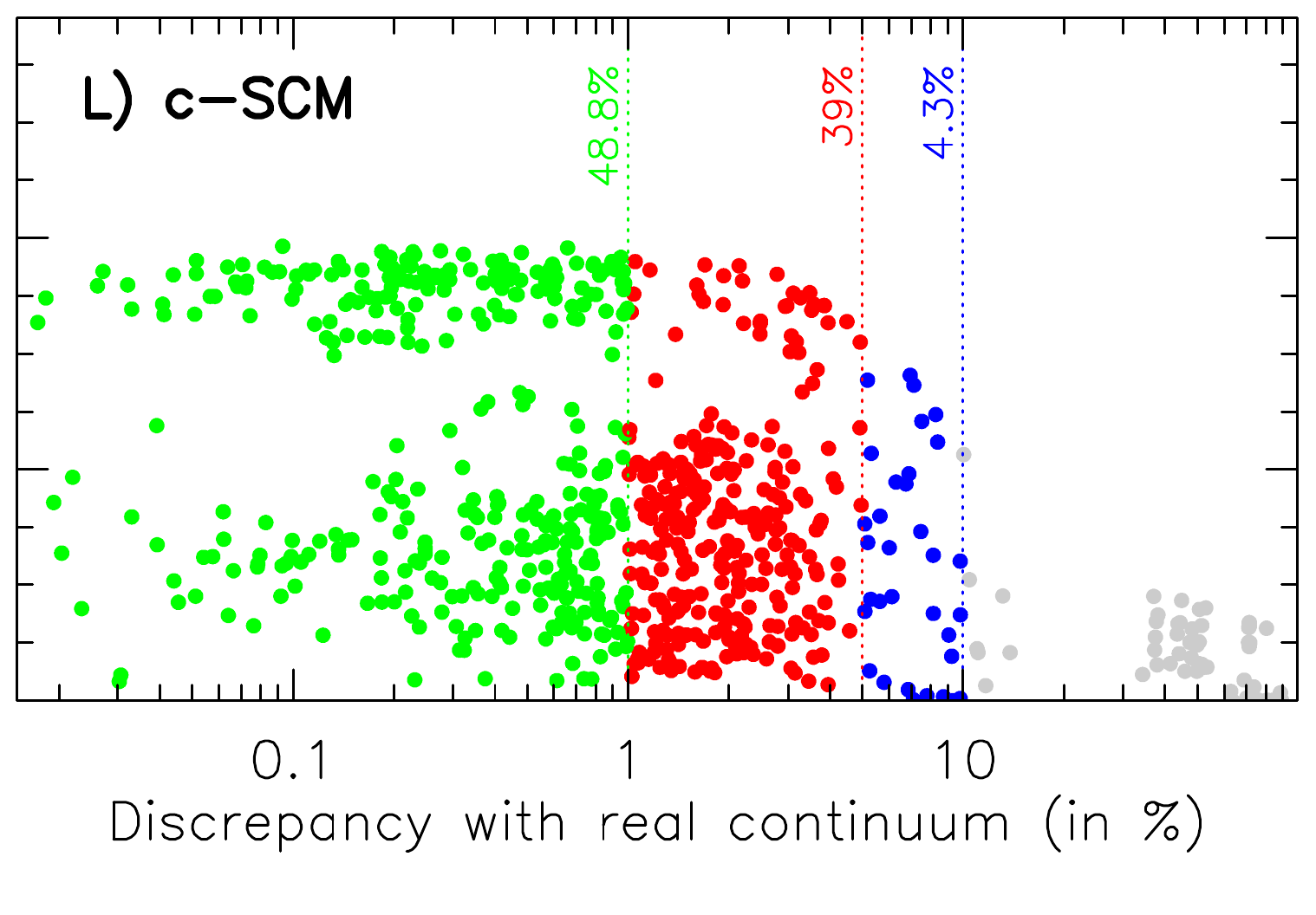} \\
\end{tabular}
\caption{Scatter plots of the percentage of channels with intensity in the range [$\mathrm{continuum}-\mathrm{noise}, \mathrm{continuum}+\mathrm{noise}$] against the discrepancy between the determined continuum level and the real one (in percentage). The different panels correspond to the different methods described in Sect.~\ref{s:contspectra}, as in Fig.~\ref{f:statistics}. In each panel, green dots correspond to synthetic spectra for which the continuum level has a discrepancy $<1$\%, red dots are those spectra with discrepancies in the range 1--5\%, blue dots are those spectra with discrepancies 5--10\%, and gray dots have discrepancies $>10$\%. The success rate (or percentage of spectra with a given discrepancy) are indicated next to the vertical, coloured lines (see also Table~\ref{t:statistics}).}
\label{f:accuracy}
\end{center}
\end{figure*}
%----------------------------------------------------------------------

%----------------------------------------------------------------------
\begin{figure}[t]
\begin{center}
\begin{tabular}[b]{c}
        \includegraphics[width=0.8\columnwidth]{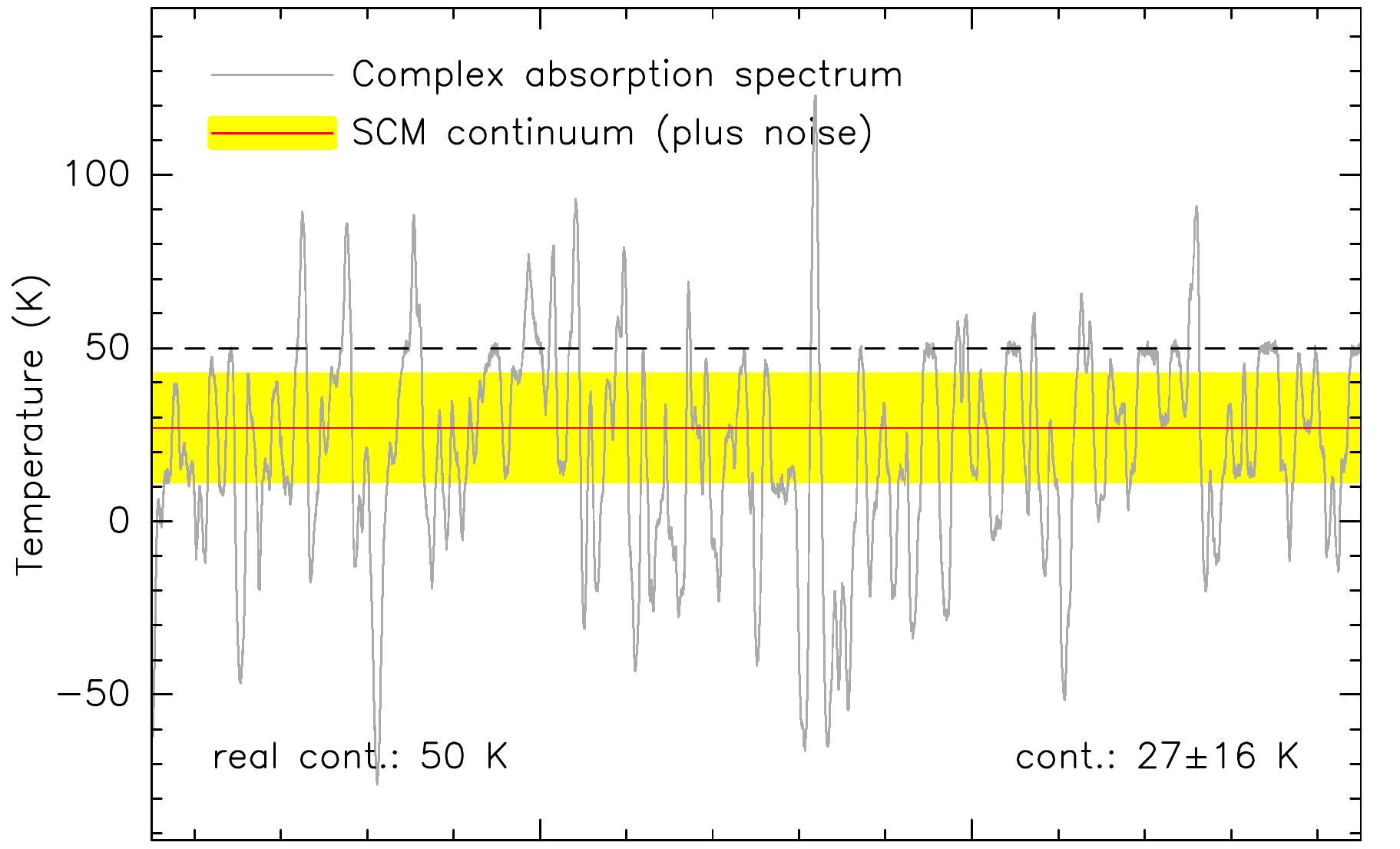} \\
        \includegraphics[width=0.8\columnwidth]{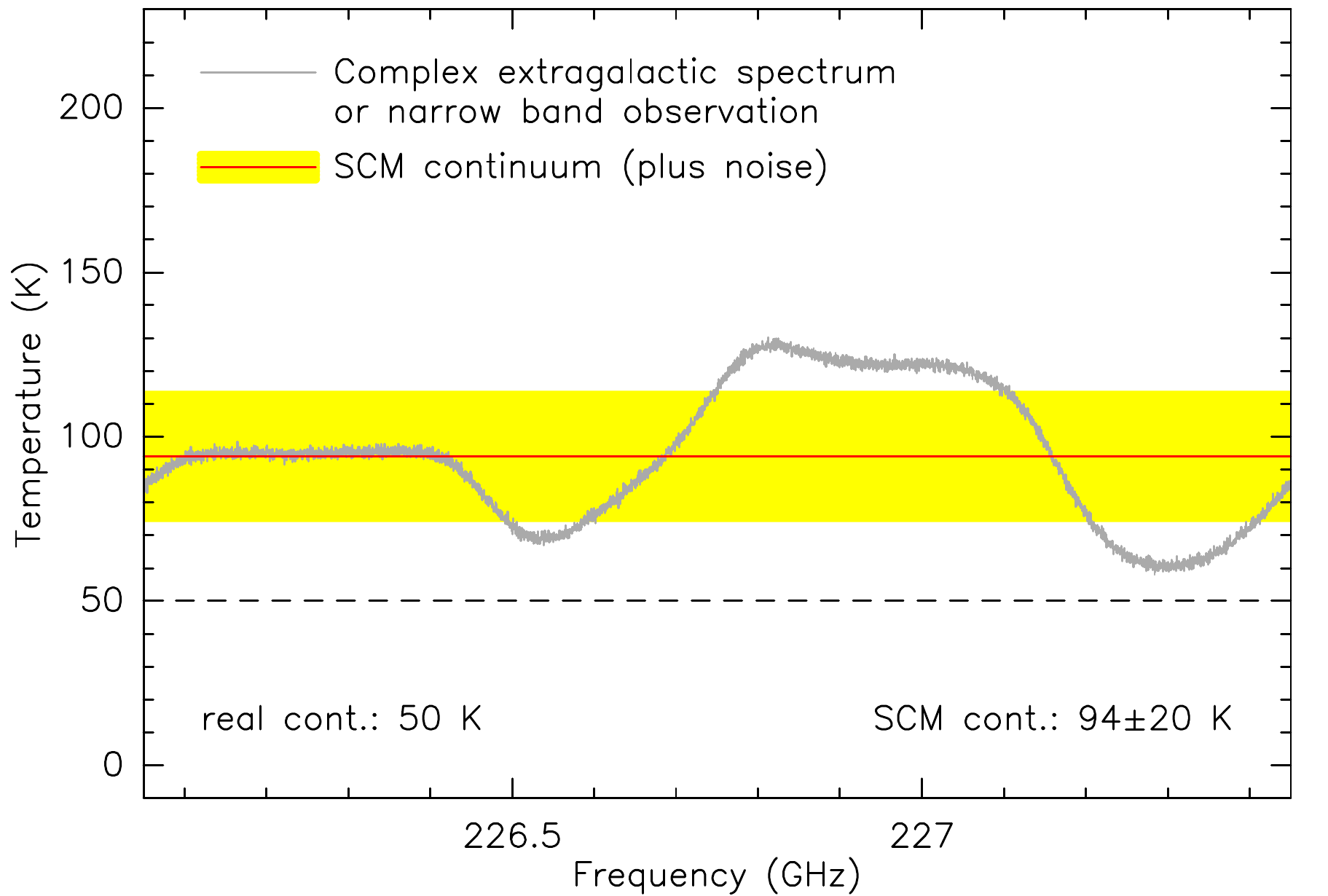} \\
\end{tabular}
\caption{Examples of complex spectra where none of the methods can determine the continuum level. The real synthetic continuum level is 50~K, while the best continuum determination is indicated in the bottom-right corner. See more details in Sect.~\ref{s:statistics}.}
\label{f:special}
\end{center}
\end{figure}
%----------------------------------------------------------------------

%________________________________________________________________
\subsection{Statistical comparison of continuum methods}\label{s:statistics}

The different methods to determine the continuum level have been presented (see Sect.~\ref{s:contspectra}) by analysing five specific synthetic spectra that reproduce cases of common astronomical sources. With the aim of testing, in a statistically significant way, the different methods we produce 50 different synthetic spectra for each one of the five groups depicted in Fig.~\ref{f:contmethod} and Table~\ref{t:contspectra}. Each spectrum is generated with XCLASS: we randomly modify the number of molecules and components as well as the molecular abundances, temperatures and velocities of each component. This results in 250 different synthetic spectra with different spectral features both in emission and absorption. The molecular abundances are in general lower than those of the spectra of Fig.~\ref{f:contmethod}, which results in spectra that are still chemically rich but contain weaker lines, i.e., more similar to average astronomical sources. In a second step, we introduce different levels of noise (1~K, 3~K, and 5~K) to each one of the spectra. The final statistically significant sample consists of 750 different synthetic spectra.

We determine the continuum level for all the spectra using all the methods described in Section~\ref{s:contspectra}. In Fig.~\ref{f:statistics} we present the distribution of the continuum levels for the 750 cases. For all the methods, the distribution peaks around a continuum level of 50~K, however, not all of them have the same level of accuracy. In each panel we indicate the mean, median and standard deviation of the distributions (see also columns 2 to 4 in Table~\ref{t:statistics}). In columns (5) to (9) of Table~\ref{t:statistics} (see also Fig.~\ref{f:accuracy}) we list the success rate determined as the number (in percentage) of synthetic spectra for which the continuum level has been determined with a certain level of accuracy: discrepancies $<1$\%, $<5$\%, and in between 5 and 10\% considered to be good, and discrepancies between 10 and 25\%, and $>25$\%.

Out of all the continuum determination methods, the percentiles (panels \textit{F} and \textit{G}) are clearly skewed towards low values when using the 25$^\mathrm{th}$ percentile, and towards high values when using the 75$^\mathrm{th}$ percentile. These are the most inaccurate methods: less than 60\% of the cases have very good or good continuum level determinations. The mean, median and Gaussian methods have a large standard deviation when applied to the whole spectral histogram, i.e., broad distributions resulting in low success rates ($\approx35$--75\%). The success rate increases up to $\approx85$\% when applying these methods to the central bins of the spectral histogram (cases \textit{C}, \textit{E}, and \textit{I}). The methods with success rates $\approx$90\% are the maximum (panel \textit{A}), the KDE maximum (panel \textit{J}), and the sigma-clipping method (panel \textit{K}), together with the median and Gaussian when applied to the central bins of the histogram. The smallest standard deviation and better mean and median are obtained for the sigma-clipping method, which has the largest success rate for very good continuum determination cases ($\approx77$\%). In Sect.~\ref{s:correction} we introduce a correction to the sigma-clipping method that increases the success rate up to 92.3\%, with about 88\% of the cases with accuracies better than 5\%.

In the distributions of Fig.~\ref{f:statistics}, there are a number of cases where the continuum level is always determined to be around 20~K or 80~K in all the methods considered. These correspond to complex spectra like the ones shown in Fig.~\ref{f:special}. In the top panel, the spectrum is highly dominated by absorption features with less than 10\% of the channels having an intensity close to the real continuum value. The spectrum in the bottom panel is a more extreme example, where no single channel has the intensity of the continuum level (corresponding to 50~K). This situation may occur in extragalactic sources with a large number of blended broad spectral line features, or in sources where, despite having narrow lines, the frequency bandwidth covered is not large enough to include channels free of line emission. The determination of the continuum level in these particular cases requires an accurate modeling of both lines and continuum simultaneously or, in cases of data-cubes, a more complex method that determines the continuum level using also the spectra of the neighbouring, better-behaved pixels. After inspecting these special cases, we investigate the number of channels that are required for an accurate determination of the continuum level. In Fig.~\ref{f:accuracy} we compare the accuracy in the determination of the continuum level with the percentage of channels with intensity in the range [continuum$-$noise/2, continuum$+$noise/2], where the continuum is equal to 50~K and the noise is 1~K, 3~K or 5~K. The accuracy is measured as the discrepancy between the measured continuum and the real one: lower values (i.e., lower discrepancies) correspond to higher accuracies. In general, all the methods but the percentiles and the mean can determine the continuum level with discrepancies $<5$\% in spectra with 30--40\% of channels close to the continuum level. When the number of these channels is reduced to 10--20\%, only some methods find the continuum level with a discrepancy $<5$--10\%. In particular, the sigma-clipping method (and its corrected version, see Sect.~\ref{s:correction}) are able to determine the continuum level for most of the cases (about 90\% of success rate), and only fail in extreme cases like those shown in Fig.~\ref{f:special}.

In summary, the continuum emission level of chemically-rich, line-crowded spectra can be well determined statistically by applying the sigma-clipping algorithm to the distribution of intensities. As shown in Figs.~\ref{f:statistics} and \ref{f:accuracy}, and Table~\ref{t:statistics}, the discrepancy between the derived continuum level and the real one is $<5$\% in most of the cases, and better than 1\% in up to 34\% of the cases.

%----------------------------------------------------------------------
\begin{figure}[t]
\begin{center}
\begin{tabular}[b]{c}
        \includegraphics[width=0.88\columnwidth]{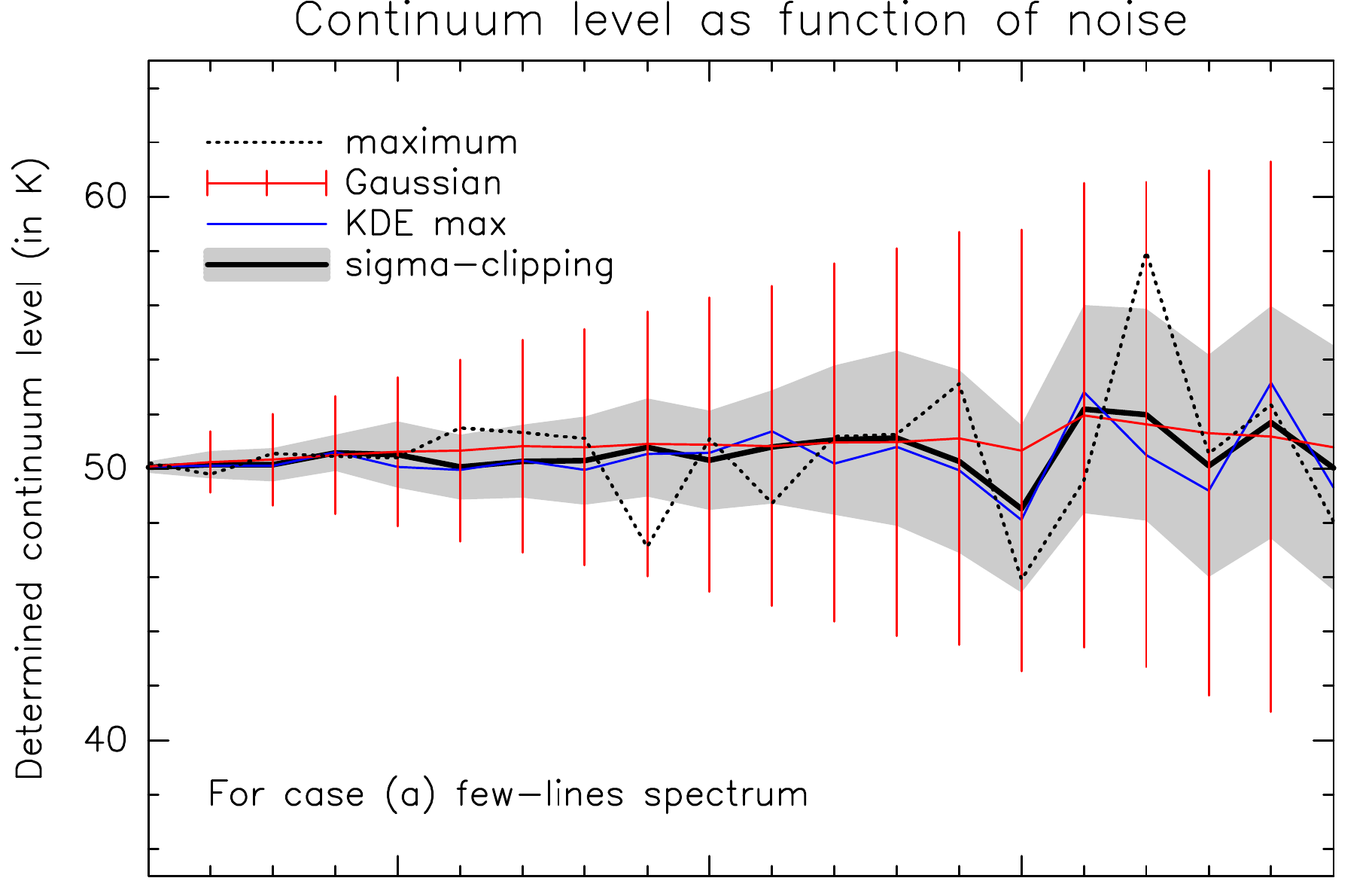} \\
        \includegraphics[width=0.88\columnwidth]{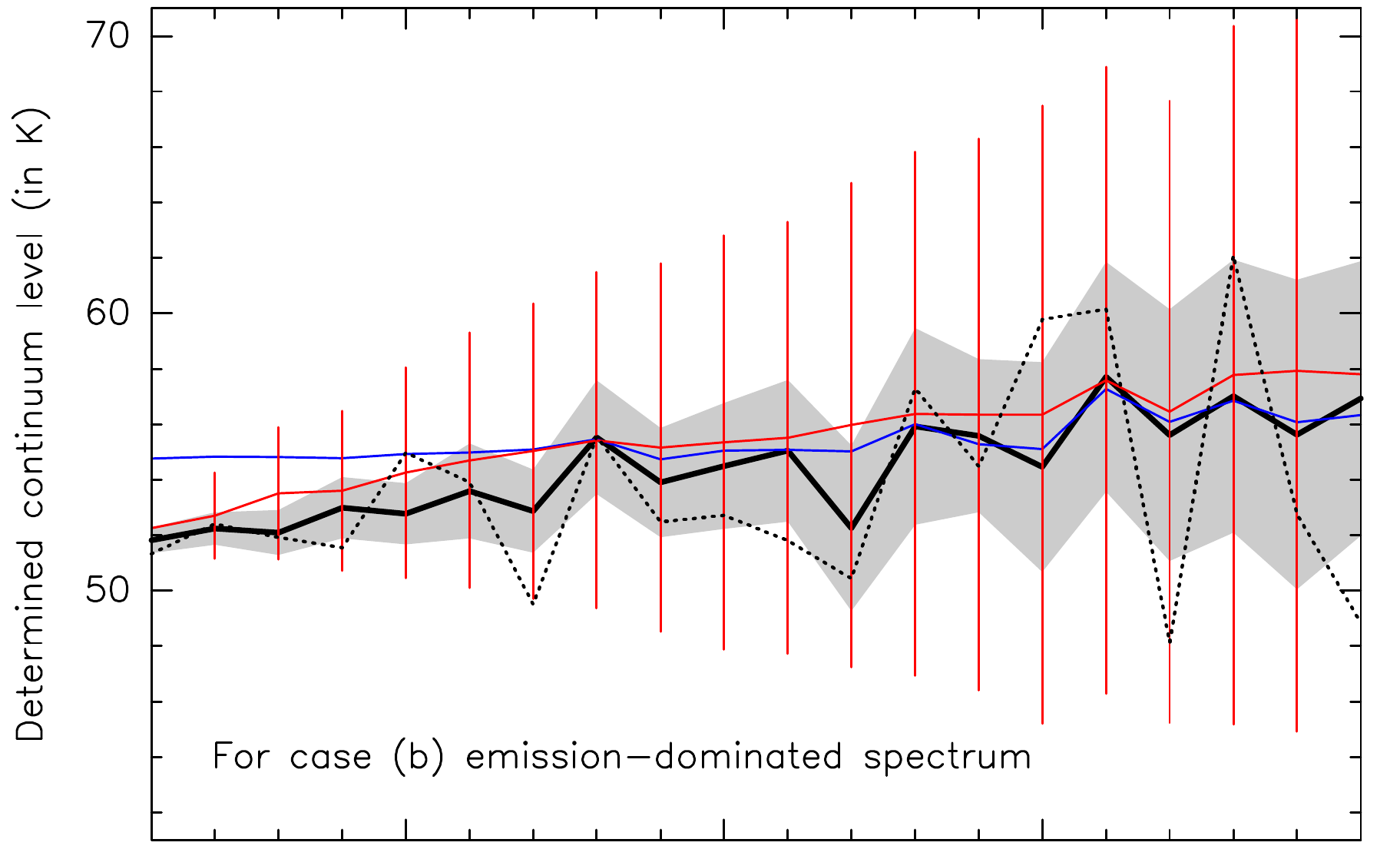} \\
        \includegraphics[width=0.88\columnwidth]{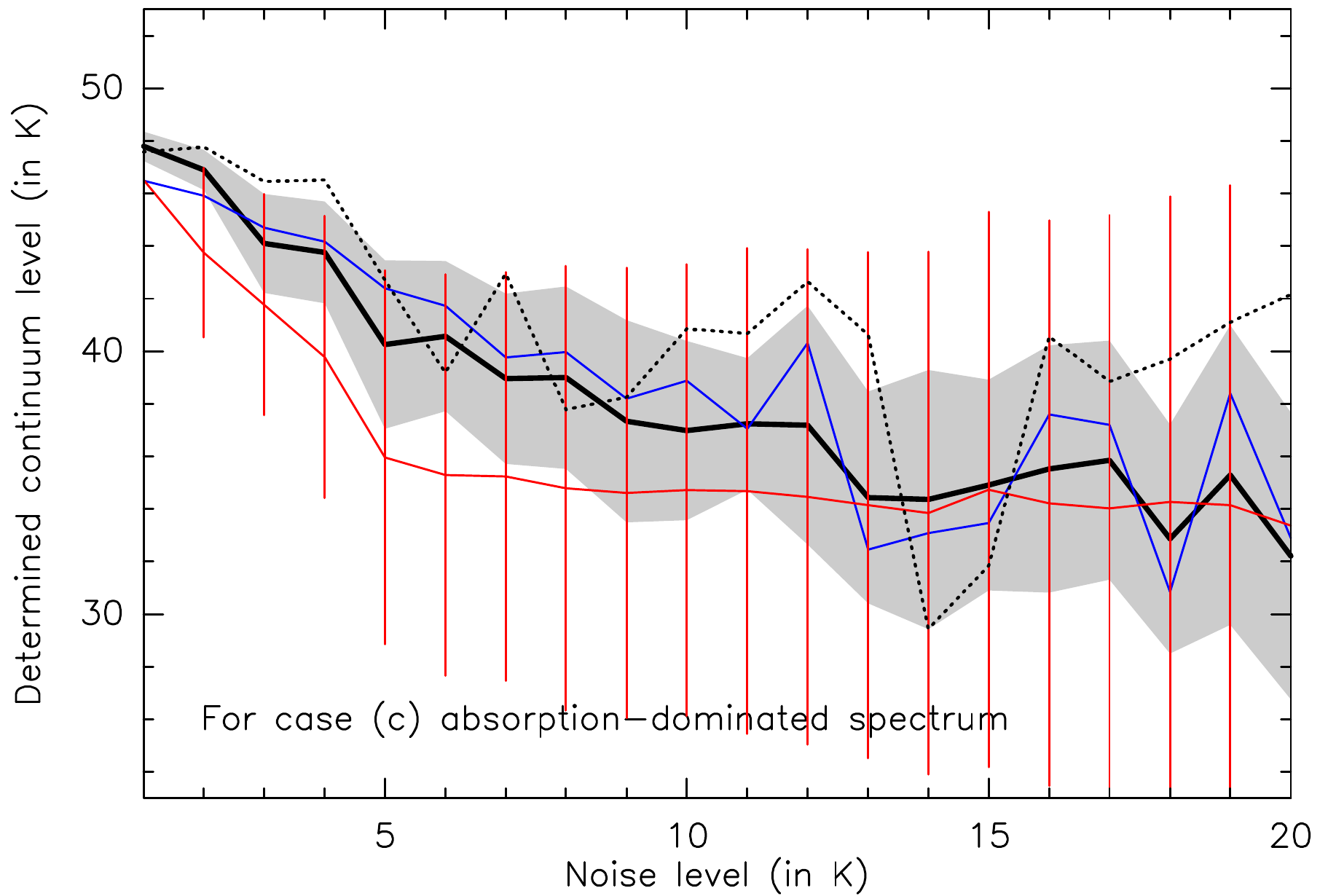} \\
\end{tabular}
\caption{Variation of the continuum emission level as a function of the noise artificially introduced in the synthetic spectra. The dotted line shows the continuum as the maximum of the histogram, the Gaussian continuum level together with its uncertainty are shown with red lines, the KDE maximum continuum level determination is shown in blue, and the sigma-clipping continuum level is shown as a thick black line with its uncertainty depicted as a gray area. The different panels correspond to (\textit{top}) few-lines spectrum, (\textit{middle}) emission-dominated spectrum and (\textit{bottom}) absorption-dominated spectrum.}
\label{f:contnoise}
\end{center}
\end{figure}
%----------------------------------------------------------------------

%----------------------------------------------------------------------
\begin{figure}[t]
\begin{center}
\begin{tabular}[b]{c}
        \includegraphics[width=0.88\columnwidth]{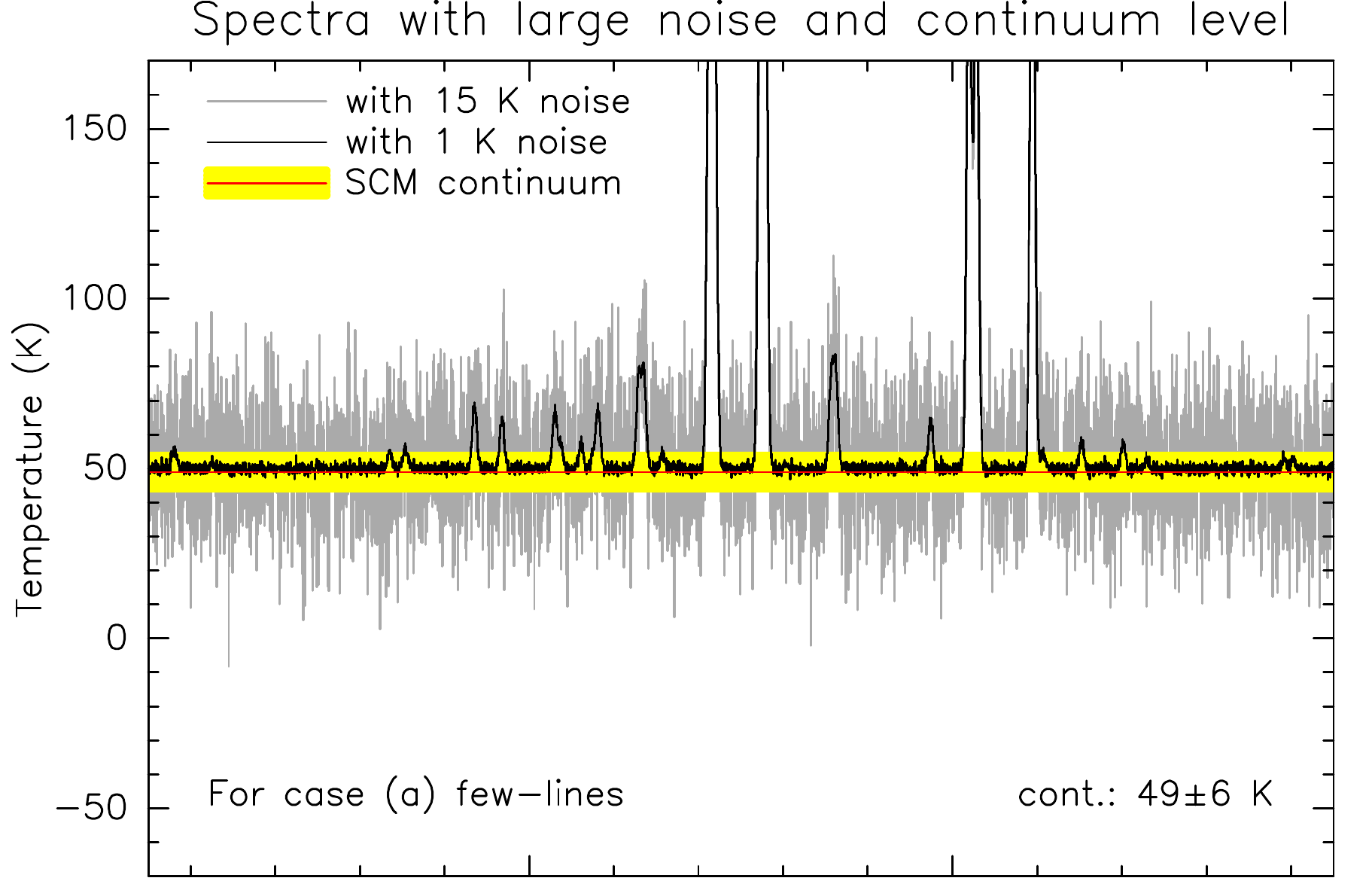} \\
        \includegraphics[width=0.88\columnwidth]{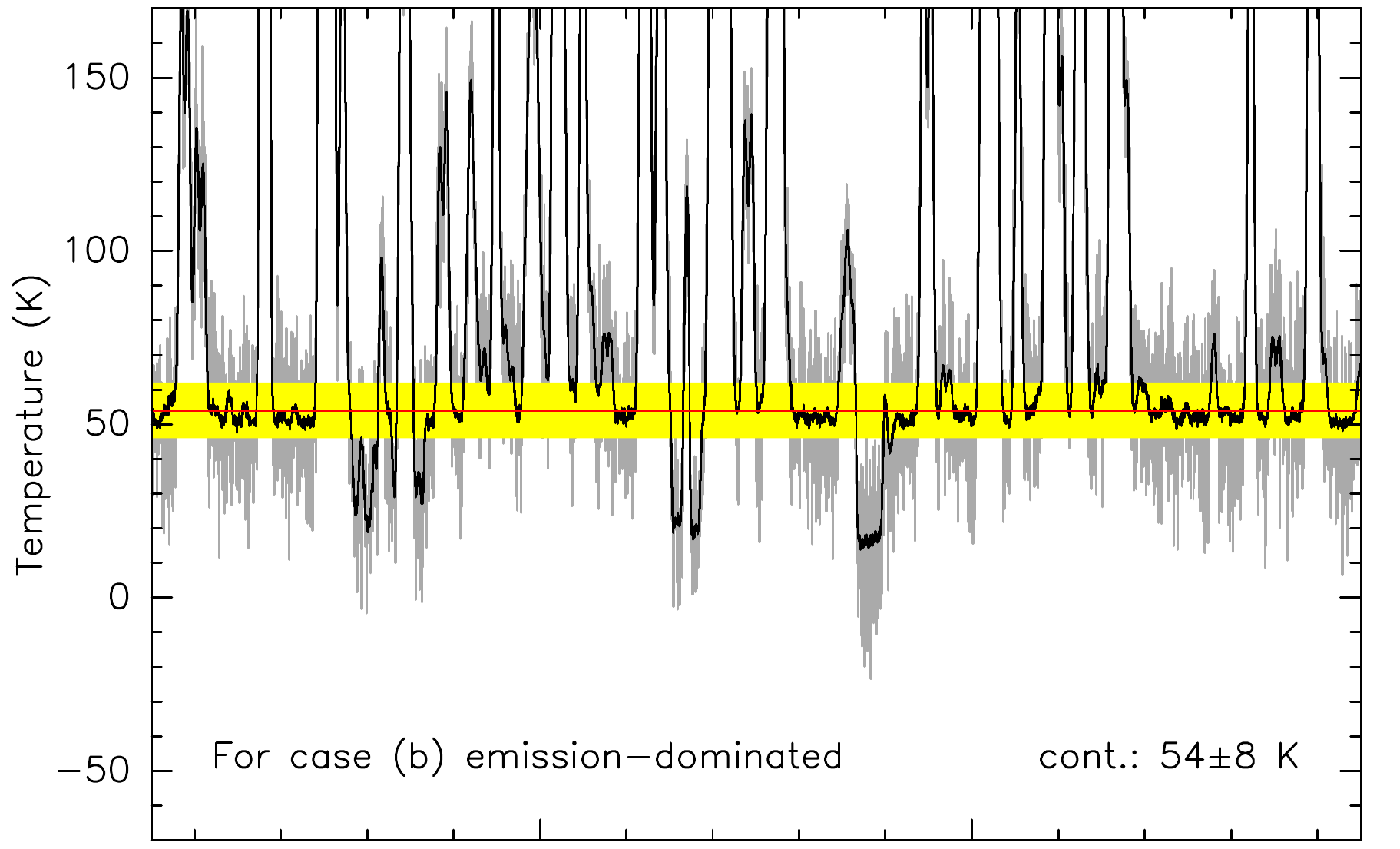} \\
        \includegraphics[width=0.88\columnwidth]{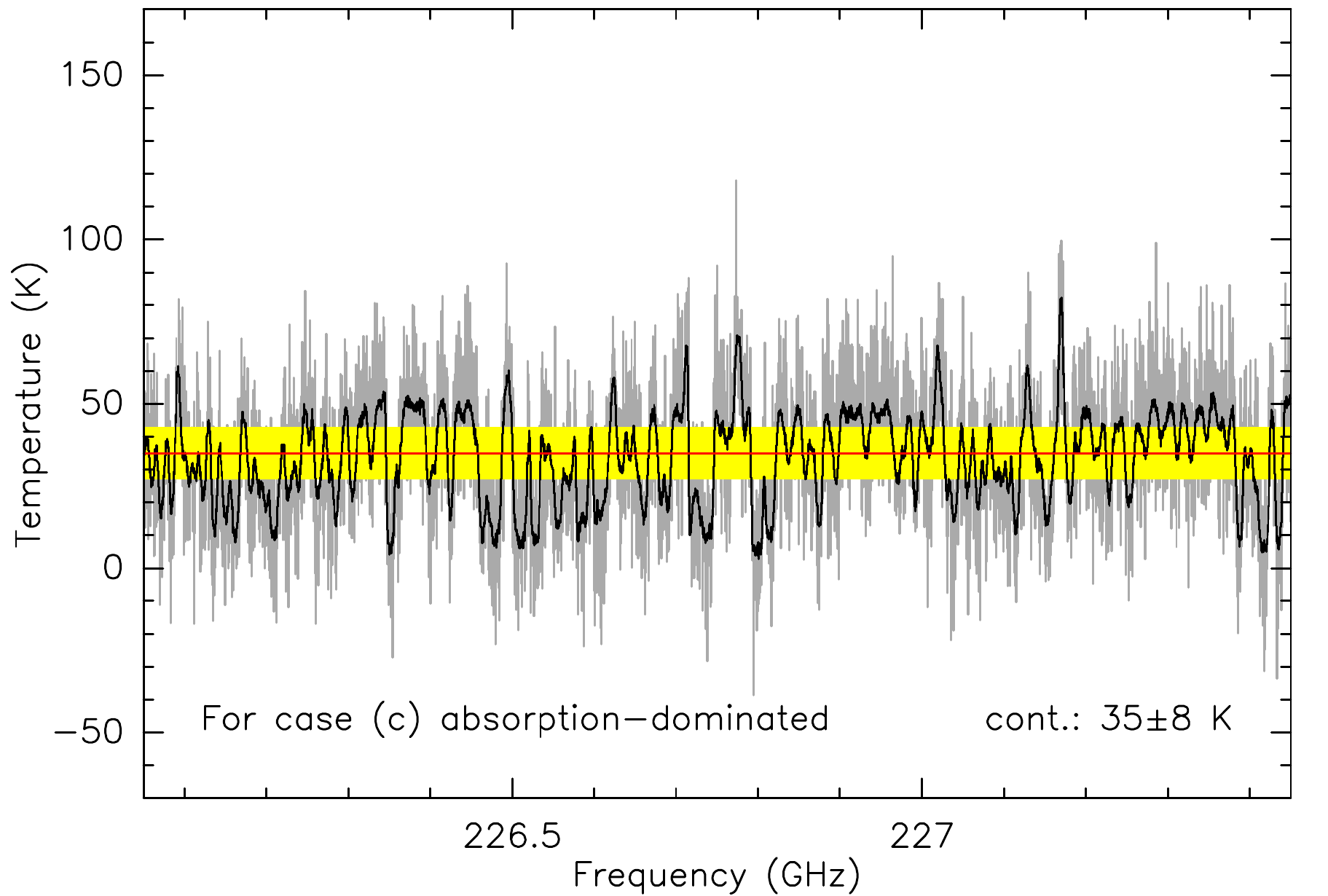} \\
\end{tabular}
\caption{Synthetic spectra for the three cases considered in Sect.~\ref{s:contnoise} corresponding to (\textit{top}) few-lines spectrum, (\textit{middle}) emission-dominated spectrum, and (\textit{bottom}) absorption-dominated spectrum. The gray lines show the spectra with a 15~K noise, while the black lines show the spectra with 1 1~K noise. The red line together with the shaded yellow area correspond to the continuum level (plus noise) as determined with the sigma-clipping method.}
\label{f:specnoise}
\end{center}
\end{figure}
%----------------------------------------------------------------------

%________________________________________________________________
\subsection{Effect of noise}\label{s:contnoise}

We investigate the effects of noise in the continuum level determination. From the methods presented in the previous sections, we consider the four methods that give the best results, i.e., the maximum (\textit{A}), the maximum of the KDE (\textit{J}), the Gaussian fit to the histogram bins close to the maximum of the distribution (\textit{H}), and the sigma-clipping method (\textit{K}). In order to evaluate only the effects of noise, we consider the simplest spectrum corresponding to case (a) in Fig.~\ref{f:contmethod}, i.e., the spectrum with only a few spectral line features, all of them in emission. We have added random Gaussian noise to the synthetic spectrum of different levels ranging from 1~K to 20~K, and determined the continuum level in each case. In a second step, we evaluate the effect of noise in emission- and absorption-dominated spectra.

In the top panel of Fig.~\ref{f:contnoise} we show the continuum level emission determined for the four methods as a function of noise. The four methods are consistent with each other and close to the real continuum level (50~K) for low noise levels (corresponding to 1 to 2~K). For larger noises, the maximum (dotted line), KDE maximum (blue line), Gaussian fit (red line) and sigma-clipping (thick black line) result in continuum levels that are only within 10\%, even for the highest noise levels. The shaded gray area in the figure shows the error computed from the sigma-clipping approach (as explained in Sect.~\ref{s:contspectra}). In this example, the noise in the determination is directly related to the noise added to the spectrum. In the top panel of Fig.~\ref{f:specnoise} we show the spectrum with a 15~K noise in gray together with a spectrum with 1~K noise in black. The sigma-clipping continuum level is shown with a red line plus a shaded yellow area that indicates the error. The high noise introduced in the spectrum hinders an accurate determination of the continuum level; however, the estimated continuum is consistent with a zeroth-order baseline fit to the line-free channels.

In the other panels of Fig.~\ref{f:contnoise} we show the continuum level determined for the four selected methods in the case of an emission-dominated spectrum (central panel) and an absorption-dominated spectrum (bottom panel). In general and as expected, the four methods are consistent with each other and tend to slightly overestimate (by about 5--10\%) the continuum level for emission-dominated spectra, and underestimate the continuum level by 20--25\% in the case of absorption-dominated spectra. A more careful inspection reveals that the KDE maximum has for all the noise levels a constant excess in the continuum determination for the emission-dominated spectrum, while the Gaussian method underestimates the continuum level when compared to the other methods, in the absorption-dominated spectrum. The central and bottom panels of Fig.~\ref{f:specnoise} show the spectra with a large 15~K noise level (in gray) and with 1~K noise (in black) and the continuum as determined with the sigma-clipping approach. We compare these continuum levels with the classical approach (i.e., identifying line-free channels by visual inspection). We have imported the synthetic spectra with 15~K noise into the CLASS program of the GILDAS\footnote{GILDAS: Grenoble Image and Line Data Analysis System, available at \url{http://www.iram.fr/IRAMFR/GILDAS/}} software package, and fit a 0$^\mathrm{th}$-order polynomial baseline to the line-free channels. The continuum levels determined in this way are $50.6\pm15.2$~K, $53.9\pm14.9$~K, and $34.5\pm19.7$~K for the cases shown Fig.~\ref{f:specnoise}, consistent with the values determined using \statcont.

In summary, the effect of noise in the data affects less the process of the continuum level determination when using the sigma-clipping algorithm and when taking into account the uncertainty obtained with that method (see Section~\ref{s:correction}).

%----------------------------------------------------------------------
\begin{figure}[t]
\begin{center}
\begin{tabular}[b]{c}
        \includegraphics[width=0.95\columnwidth]{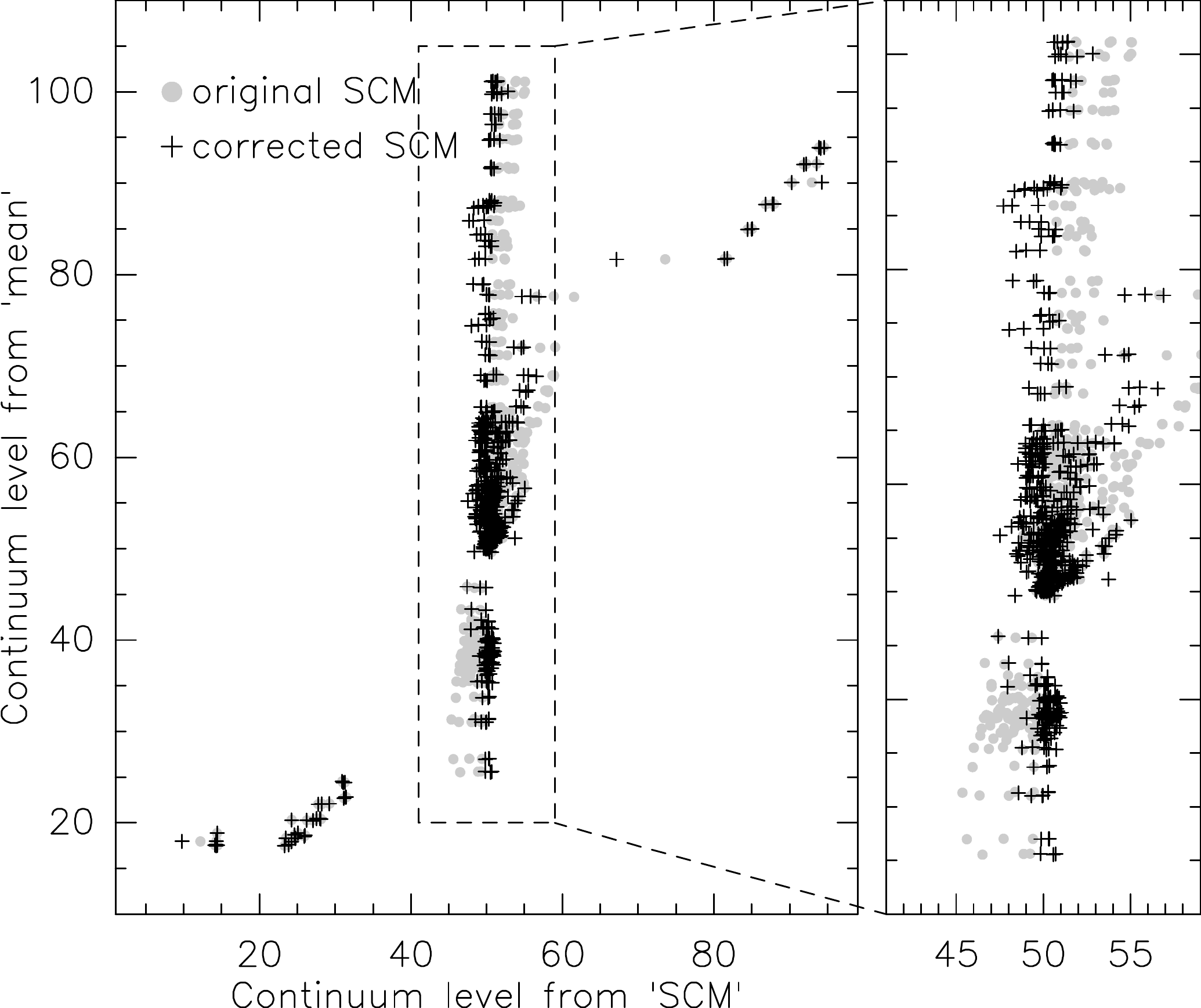} \\
\end{tabular}
\caption{Comparison of the continuum level determined with the original sigma-clipping method (gray dots) and the corrected version (black crosses). In order to better show the differences between the two methods, the y-axis has been selected to show the continuum level determined with the mean. The right panel shows a zoom in of the central region. The black crosses appear clustered closer to the real continuum level, 50~K, compared to the gray dots.}
\label{f:correction}
\end{center}
\end{figure}
%----------------------------------------------------------------------

%________________________________________________________________
\subsection{Corrected sigma-clipping continuum level}\label{s:correction}

In most of the cases shown in Figs.~\ref{f:contnoise} and \ref{f:specnoise}, the real continuum level is only off by $<5$\% with respect to the sigma-clipping continuum value if the uncertainty, that is obtained in the process of determination of the continuum level, is taken into account. Therefore, a more accurate continuum level can be determined if the value of the uncertainty is used to correct for the trends seen in emission- and absorption-dominated spectra (see Sect.~\ref{s:contnoise}). In short, the continuum emission level is corrected by adding or subtracting the value of the uncertainty to the continuum level depending if the analyzed spectrum is an absorption- or emission-dominated spectrum, respectively. This correction improves the continuum determination not only in spectra with large noise, but in all cases. For example, see the corrected continuum levels for the examples shown in Fig.~\ref{f:contmethod}, which are listed in column (12) of Table~\ref{t:contspectra}. All the values are closer to 50~K, with the largest discrepancy occurring for the absorption-dominated spectra with a value of only 2.2\%.

The addition or subtraction of the value of the uncertainty to the continuum level is done automatically in \statcont. Each spectrum is evaluated individually, and \statcont\ compares the number of emission channels with the number of absorption channels. A channel is considered to be dominated by spectral line emission if its intensity is above the value of the continuum as determined with the sigma-clipping method (original sigma-clipping level, or oSCL) plus the rms noise level of the observations: $\mathrm{oSCL}+\mathrm{rms}$. Similarly, a channel is considered to be dominated by absorption if its intensity is below the value $\mathrm{oSCL}-\mathrm{rms}$. Following this approach we can determine the percentage of emission-dominated and absorption-dominated channels in a spectrum. \statcont\ uses this information to add or subtract the uncertainty $\sigma_\mathrm{SCM}$ obtained with the sigma-clipping method depending on the following criteria:
\begin{itemize}
\item \textit{Case A, $+1\times\sigma_\mathrm{SCM}$}: \\
The uncertainty is added to the original continuum level if the spectrum is highly dominated by absorptions. This occurs when (\textit{a}) the fraction of emission channels is $<33$\% and that of absorption channels is $>33$\%, and the difference in channels $\mathrm{absorption}-\mathrm{emission}$ is $>25$\%; or (\textit{b}) when both the fraction of emission and absorption channels is $>33$\% but the difference $\mathrm{absorption}-\mathrm{emission}$ is $>25$\%.
\item \textit{Case B, $+0.5\times\sigma_\mathrm{SCM}$}: \\
For spectra only slightly dominated by absorption, the correction is applied only at a level of half of the uncertainty. This occurs when the fraction of emission is $<33$\% and that of absorption is $>33$\%, but the difference $\mathrm{absorption}-\mathrm{emission}$ is $\le25$\%.
\item \textit{Case C, $+0\times\sigma_\mathrm{SCM}$}: \\
No correction is applied if no clear emission or absorption features are found. This happens if (\textit{a}) emission and absorption fractions are $<33$\%, and (\textit{b}) if emission and absorption fractions are $>33$\% but the absolute difference $\mathrm{abs}(\mathrm{absorption}-\mathrm{emission})$ is $\le25$\%. One example of this last case is a line-rich, but extremely noisy spectrum.
\item \textit{Case D, $-0.5\times\sigma_\mathrm{SCM}$}: \\
When the spectrum is slightly dominated by emission, we partially subtract the uncertainty to the original continuum level. This is the case if the fraction of emission is $>33$\% and the fraction of absorption is $<33$\%, but the difference $\mathrm{emission}-\mathrm{absorption}$ is only $\le25$\%.
\item \textit{Case E, $-1\times\sigma_\mathrm{SCM}$}: \\
For emission-dominated spectra we subtract the uncertainty if (\textit{a}) emission $>33$\% and absorption $<33$\% together with the difference $\mathrm{emission}-\mathrm{absorption}$ $>25$\%, or (\textit{b}) emission $>33$\% and absorption $>33$\% but $\mathrm{emission}-\mathrm{absorption}$ $>25$\%.
\end{itemize}

%with absorptionoriginal sigma-clipping continuum level (oSCL) to the mean value ($\mu$) of the intensities of the spectrum\footnote{This value corresponds to the mean continuum level of the entire sample as described in Sect.~\ref{s:contspectra}.}, taking into account the rms noise level of the observations (rms). There are three possible situations: (1) a spectrum is considered to be emission dominated, if the difference between the mean and the original sigma-clipping continuum is positive and larger, in absolute value, than the rms noise level of the observations, or $\mu-\mathrm{oSCL}>+\mathrm{rms}$. In this case, the uncertainty is subtracted to calculate the final corrected continuum level; (2) a spectrum is absorption-dominated if the difference is negative and larger than the rms noise level, in absolute value, or $\mu-\mathrm{oSCL}<-\mathrm{rms}$. In this case the uncertainty is added; (3) a spectrum is considered to be noise-dominated or with no prominent emission or absorption features dominating the spectrum if the difference is within the rms noise level. In this case, no correction is applied to the original continuum level.

Panel (\textit{L}) in Figs.~\ref{f:statistics} and \ref{f:accuracy} shows the results of the continuum level determination after applying the correction. See also the last row in Table~\ref{t:statistics}. The success rate is $>90$\% and in almost 50\% of the cases the continuum level is determined with a discrepancy $<1$\%. In the last panel of Fig.~\ref{f:accuracy} we show that the corrected method is able to determine the continuum with discrepancies $<10$\% when there are more than 10\% of channels free of line emission. The main failures are due to special spectra like those shown in Fig.~\ref{f:special}. In Fig.~\ref{f:correction} we compare the original sigma-clipping continuum level (gray dots) with the corrected continuum (black crosses). This correction improves the final continuum determination, and reduces the discrepancies with the synthetic real continuum level considerable. Hereafter, this method is called `corrected sigma-clipping method' or c-SCM and is the default method used in \statcont\ to determine the continuum emission level.

%----------------------------------------------------------------------
\begin{figure}[t]
\begin{center}
\begin{tabular}[b]{c}
        \includegraphics[width=0.9\columnwidth]{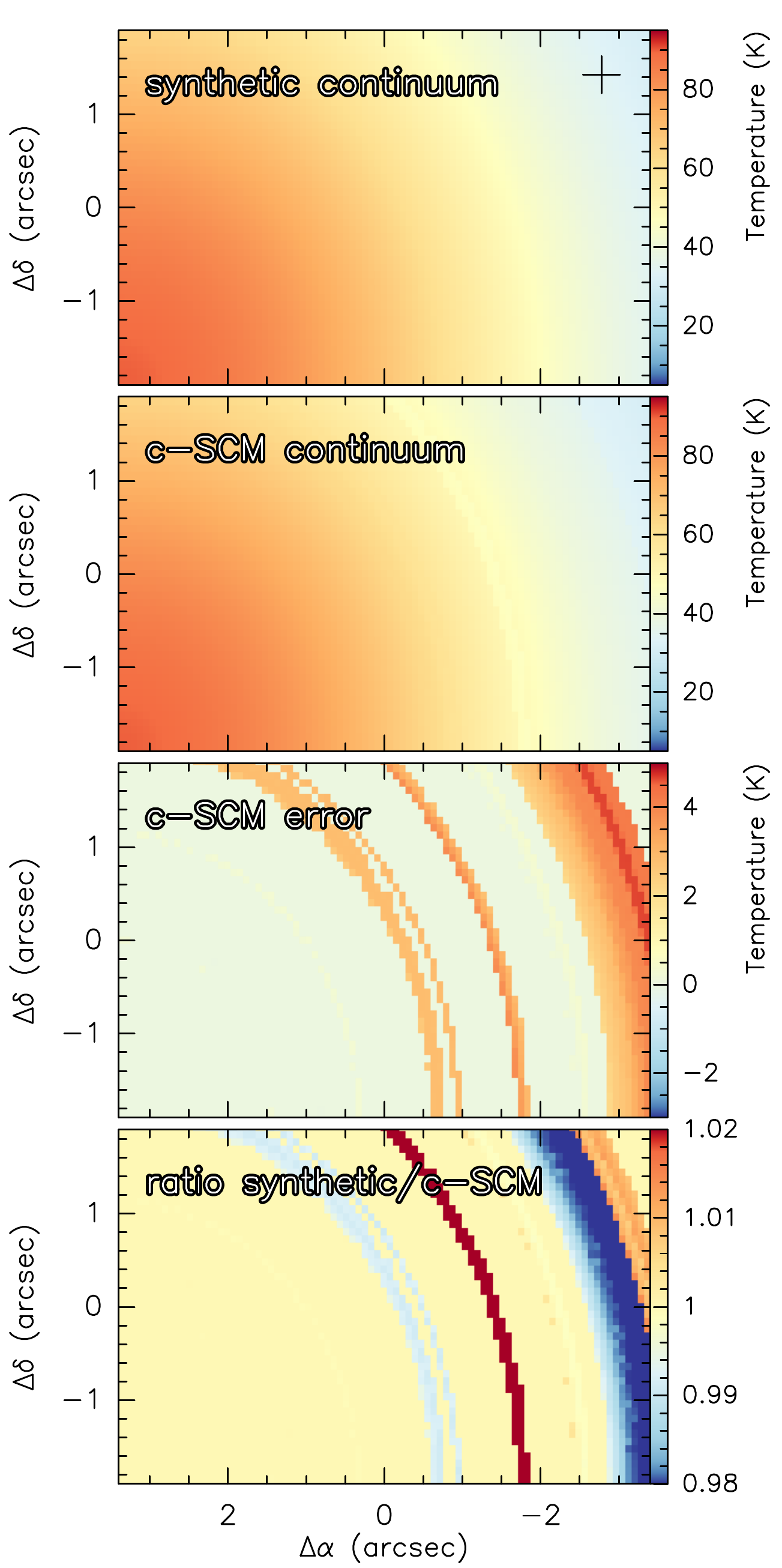} \\
\end{tabular}
\caption{Continuum level determination of a full datacube. The different panels show from top to bottom: (i) Synthetic continuum map created with XCLASS. Each pixel of the map contains a crowded-line spectrum with both emission and absorption lines. The black cross marks the position for which the spectrum is shown in Fig.~\ref{f:contexamplesspec}; (ii) Continuum map determined by using the c-SCM (see Sects.~\ref{s:contspectra} and \ref{s:correction}); (iii) Uncertainty of the continuum level determination obtained with the c-SCM; (iv) Ratio of the synthetic continuum map over the continuum image produced with the c-SCM.}
\label{f:contexamples}
\end{center}
\end{figure}
%----------------------------------------------------------------------

%----------------------------------------------------------------------
\begin{figure}[t]
\begin{center}
\begin{tabular}[b]{c}
        \includegraphics[width=0.9\columnwidth]{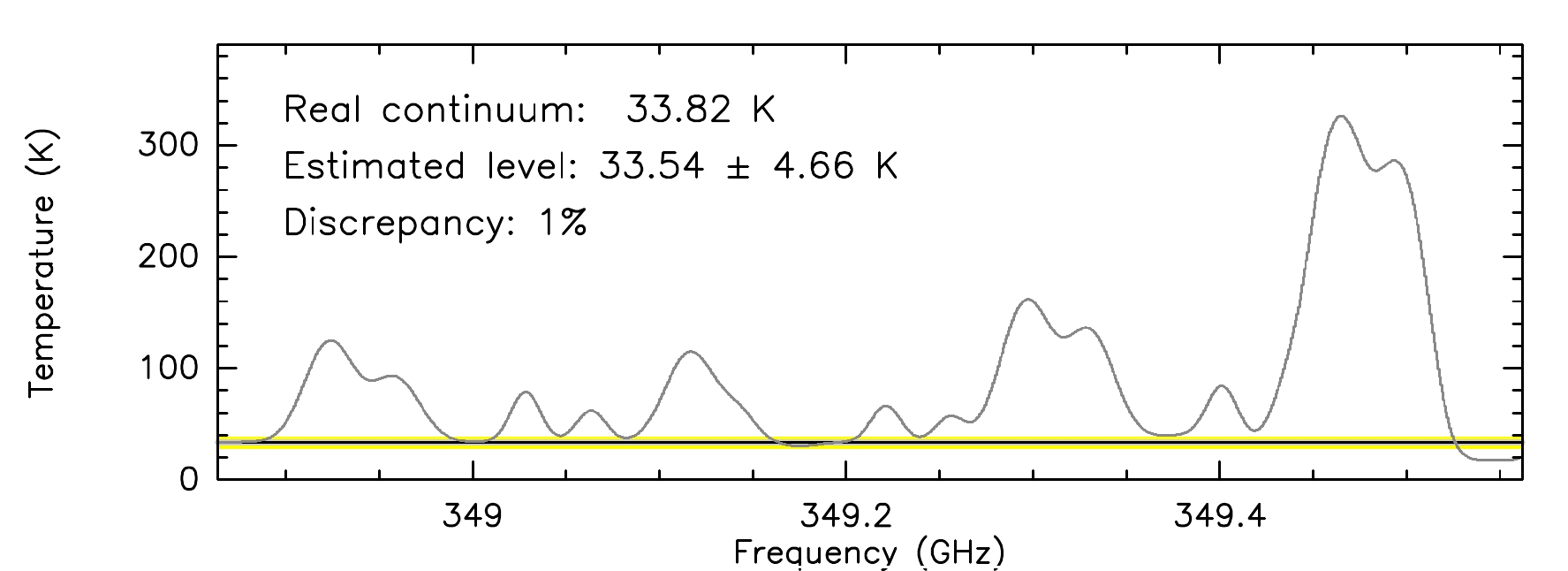} \\
\end{tabular}
\caption{Spectrum extracted at the position depicted in the top panel of Fig.~\ref{f:contexamples}. The black solid line indicates the continuum level determined with c-SCM, while the yellow band shows the uncertainty in the determination of the continuum level. The real continuum level as well as the discrepancy between the real and the determined one are listed in the panel.}
\label{f:contexamplesspec}
\end{center}
\end{figure}
%----------------------------------------------------------------------

%----------------------------------------------------------------------
\begin{figure*}[t]
\begin{center}
\begin{tabular}[b]{c}
        \includegraphics[width=0.84\textwidth]{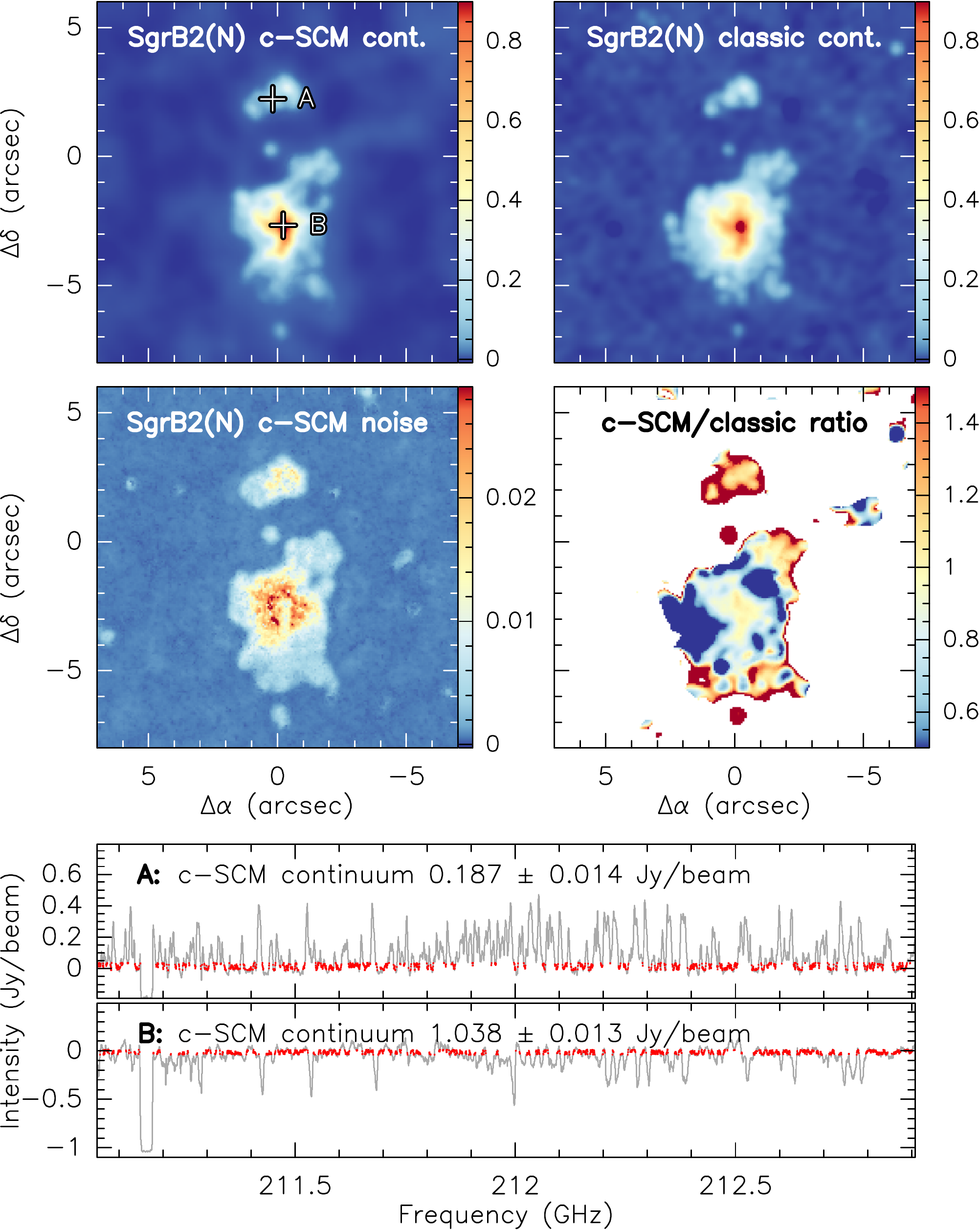} \\
\end{tabular}
\caption{Example case Sgr\,B2(N). (\textit{top-left}) Continuum emission map as determined with the c-SCM method of \statcont. The synthesized beam is 0\farcs4. (\textit{top-right}) Continuum emission map determined after searching and selecting line-free channels. (\textit{middle-left}) Noise map obtained with c-SCM. (\textit{middle-right}) Ratio of the c-SCM to classical-approach continuum maps. (\textit{bottom panels}) Continuum-subtracted spectra using the c-SCM value towards two selected positions A and B, shown in the top-left panel. The channels used for the continuum determination after clipping and converging are depicted in red. The c-SCM continuum emission level subtracted to the original spectra, together with its uncertainty, are listed in the upper part of each panel. See more details of these data in \citet{SanchezMonge2017}.}
\label{f:SgrB2}
\end{center}
\end{figure*}
%----------------------------------------------------------------------

%----------------------------------------------------------------------
\begin{figure*}[t]
\begin{center}
\begin{tabular}[b]{c}
        \includegraphics[width=0.84\textwidth]{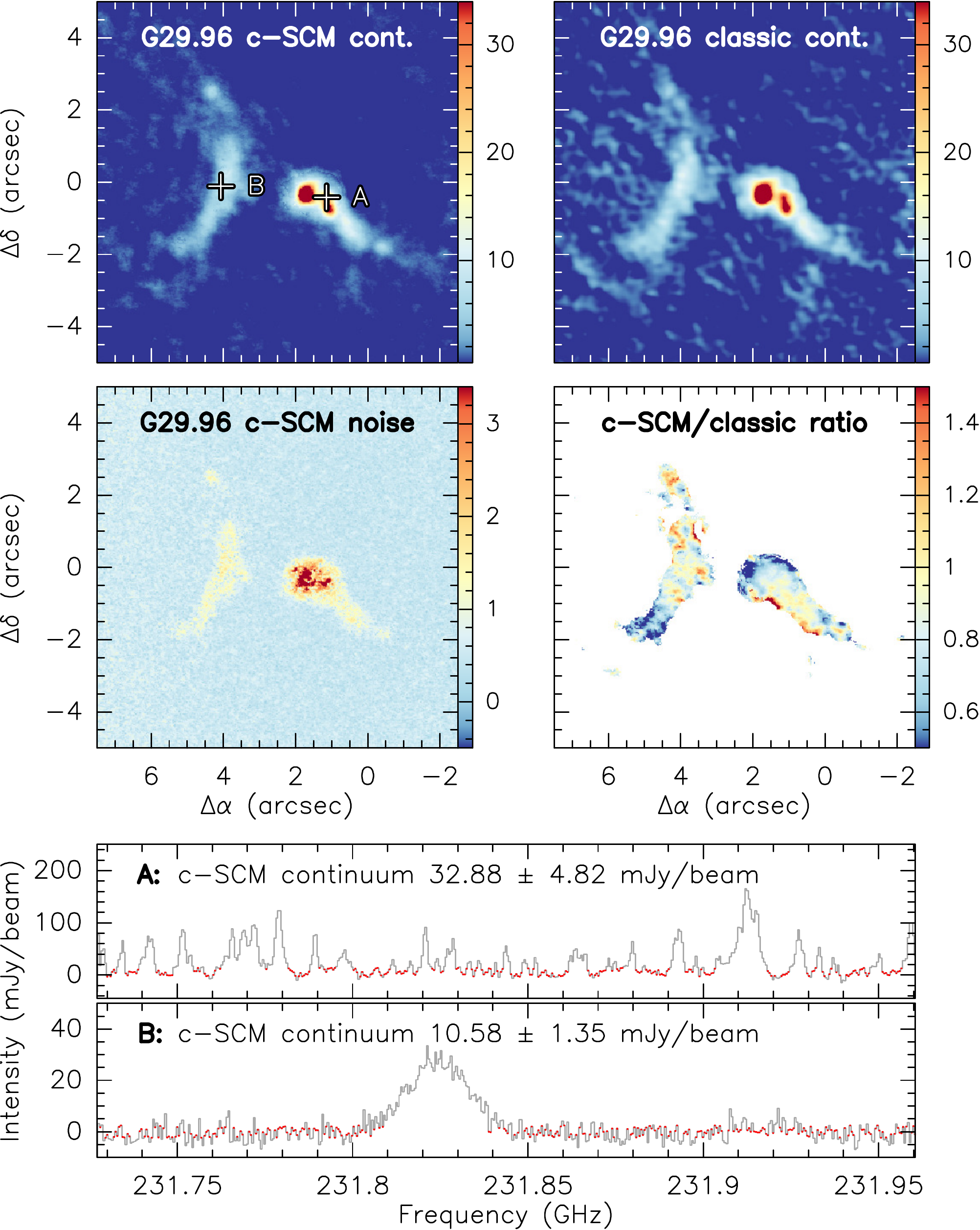} \\
\end{tabular}
\caption{Example case Sgr\,B2(N). (\textit{top-left}) Continuum emission map as determined with the c-SCM method of \statcont. The synthesized beam is 0\farcs2. (\textit{top-right}) Continuum emission map determined after searching and selecting line-free channels. (\textit{middle-left}) Noise map obtained with c-SCM. (\textit{middle-right}) Ratio of the c-SCM to classical-approach continuum maps. (\textit{bottom panels}) Continuum-subtracted spectra using the c-SCM value towards two selected positions A and B, shown in the top-left panel. The channels used for the continuum determination after clipping and converging are depicted in red. The c-SCM continuum emission level subtracted to the original spectra, together with its uncertainty, are listed in the upper part of each panel. See more details of these data in \citet{Cesaroni2017}.}
\label{f:G29}
\end{center}
\end{figure*}
%----------------------------------------------------------------------

%________________________________________________________________
\subsection{Continuum maps from datacubes}\label{s:contdatacube}

In this section we move from the analysis of single spectra (see Fig.~\ref{f:contmethod}) to the analysis of a full datacube. As an example of the analysis of datacubes, the top panel of Fig.~\ref{f:contexamples} shows the continuum emission of a synthetic datacube created with XCLASS. Each pixel contains a line-crowded spectrum, with both emission and absorption lines, and a pixel-dependent continuum level, similar to what is commonly found in chemically-rich star-forming regions observed at millimeter wavelengths and containing bright continuum sources. In Fig.~\ref{f:contexamplesspec} we show an example of the spectra from the datacube corresponding to the position indicated with a black cross in Fig.~\ref{f:contexamples} (top panel). The second panel in Fig.~\ref{f:contexamples} shows the map of the continuum emission level determined with c-SCM (explained in Sect.~\ref{s:correction}). The 2D continuum image is created after applying the method described for a single spectrum on a pixel-to-pixel basis. The method extracts the spectrum at each pixel and determines its continuum level. With this information, an image is constructed by putting back together the information of all pixels. As described above, the c-SCM also provides information on the standard deviation of each continuum emission level, that we relate to the uncertainty of the continuum determination. This information is shown in the third panel of Fig.~\ref{f:contexamples}. Finally, in the last panel we show the ratio between the original synthetic continuum emission and the one derived using the c-SCM. The continuum emission is well determined ($<1$\%) for most pixels in the map, with only few of them showing a discrepancy of about 5\%. The c-SCM in \statcont\ can be used to construct continuum emission maps of datacubes obtained with instruments such as ALMA and the VLA.

%________________________________________________________________
\section{Application to real datasets}\label{s:realcases}

We apply the c-SCM in \statcont\ to determine the continuum level in real observational data. For this we have selected two ALMA datasets focused on the study of sources with a rich chemistry: project 2013.1.00332.S (PI: P.\ Schilke) targeting the well-known hot molecular cores Sgr\,B2(M) and Sgr\,B2(N), and project 2013.1.00489.S (PI: R.\ Cesaroni) targeting a list of six high-mass star forming regions including hot cores like G29.96$-$0.02 and G31.41$+$0.31. The observations were conducted in spectral line mode, and observed different frequency ranges of the ALMA band~6. Details of the observations can be found in \citet{Cesaroni2017}, and in \citet{SanchezMonge2017}.

In order to show the feasibility of \statcont\ applied to real data, we will focus on one spectral window of the Sgr\,B2(N) observations, and another of the G29.96$-$0.02 observations. A brief summary of the observation details are in the following. The selected spectral window of Sgr\,B2(N) covers the frequency range 211~GHz to 213~GHz, with an effective bandwidth of 1875~MHz and 3480~channels, providing a spectral resolution of 0.7~km~s$^{-1}$. On the other hand, for G29.96$-$0.02 we use a spectral window with a narrower effective bandwidth of 256~MHz, centered at the frequency 231.85~GHz and with a spectral resolution of 0.6~km~s$^{-1}$. Figures~\ref{f:SgrB2} and \ref{f:G29} show the continuum emission map as determined with the c-SCM method (top-left panels). In the top-right panel, we compare with the continuum map that is created after searching for line-free channels following the classical approach. The middle panels show uncertainty in the determination of the continuum level with c-SCM, and the ratio of the two continuum images. Finally, the bottom panels show the continuum-subtracted spectra obtained from c-SCM.
%These spectra are obtained from line-only datacubes that \statcont\ produces after subtracting the continuum contribution.
%the original spectra, i.e., line plus continuum emission, for two representative positions in each case. The continuum level is listed in each panel and depicted with a solid horizontal black line.

The classical continuum emission maps for Sgr\,B2(N) and G29.96$-$0.02 are produced after identifying line-free channels in an average spectrum extracted around position~A (see Figs.~\ref{f:SgrB2} and \ref{f:G29}). These positions correspond to the chemically richest ones in both regions (see \citealt{SanchezMonge2017, Cesaroni2017}), and therefore constitute the obvious candidates to search for line-free channels. For Sgr\,B2(N), the brightest source towards position~B is less chemically rich, but has a slightly different line-of-sight velocity that would results in a shift of a few channels if the line-free channels would have been identified at this position. Therefore, our selected channels contain no major line contamination towards position~A, but are partially contaminated towards position~B. This introduces some structure to the continuum emission: the spherical-like emission surrounding the brightest source comes from different molecular lines (S\'anchez-Monge et al.\ in prep.), and not only from continuum emission. The continuum emission image produced with c-SCM in \statcont\ does not contain this structure, but only structures mainly dominated by continuum-only emission. The noise map obtained with c-SCM corresponds to uncertainties of about 1--5\%, reaching values up to 10\% towards positions with complex, chemically-rich spectra. The spectra of the two selected positions correspond to a line-rich source with an emission-dominated spectrum (position~A), and a spectrum with most of the lines in absorption against a bright background continuum (position~B). The difference in line-of-sight velocity between sources in the region is not as relevant in G29.96$-$0.02 as in Sgr\,B2(N). This results in two similar continuum maps (top panels in Fig.~\ref{f:G29}) produced both with the c-SCM and the classical approach. For this region we have selected two representative spectra: position~A contains a chemically-rich source with narrow, emission line features, while position~B corresponds to a line-poor spectrum with only continuum emission from an \hii\ region and emission from the H31$\alpha$ recombination line.

In summary, \statcont\ is able to automatically determine the continuum level in sources with complex line spectra. It provides (a) continuum emission maps that are necessary to study and understand e.g., the number of sources or cores in a region, and (b) continuum-subtracted line datacubes or images that are useful to characterize the origin of each spectral line.

%________________________________________________________________
\section{Conditions for \statcont\ usage on real data}\label{s:conditions}

In the following we list and summarize the conditions under which \statcont\ can be used to determine the continuum level of astronomical sources.
\begin{itemize}
\item \statcont\ can be applied to any data set (observations or simulations) that has been observed (or produced) in spectral line observing mode, i.e., with enough channels to resolve spectral line features.
\item Single spectra or 3D datacubes (i.e., two spatial axis and one spectral axis) can be used as input to determine the continuum level.
\item The continuum level must not vary significantly with frequency across the selected frequency range. The variation is expected to be within the noise level of the observations. For example, for observations of a star-forming region with ALMA at millimeter wavelengths, reaching a sensitivity of about 5~mJy, we recommend to use a frequency range $<5$~GHz.
\item A minimum fraction of 10\% for line-free channels, or channels with no significant contamination of bright emission/absorption line features is advisable to determine the continuum level with accuracies better than 5\%. For a typical ALMA observation of a line-rich source, we recommend a minimum frequency range of about 500~MHz, or 1~GHz in the case of complex (i.e., combination of bright emission and absorption features).
\item Despite \statcont\ being developed to analyze data in the radio regime (i.e.,\ centimeter to submillimeter wavelengths), there is no software limitation in applying it to different regimes of the electromagnetic spectrum (e.g., infrared, optical, ultraviolet).
\end{itemize}

%________________________________________________________________
\section{Description of the \statcont\ procedures}\label{s:procedures}

The \statcont\ python-based tool determines the continuum level of spectral line data for either a single spectrum or a 3D datacube. The code is freely available for download\footnote{\url{http://www.astro.uni-koeln.de/~sanchez/statcont}} from GitHub or as a \texttt{STATCONT.tar.gz} file. \statcont\ uses the \texttt{ASTROPY}\footnote{\url{http://www.astropy.org}} package-template and is fully compatible with the \texttt{ASTROPY} ecosystem. In the following we describe the input files that have to be provided by the user, as well as the output products and the main procedures that can be used. The general structure of the working directory consists of two main subdirectories: \texttt{data} containing the data to be analyzed and provided by the user, and \texttt{products} containing the final products.

\subsection{Input files}\label{s:input}

The input files are those that will be analyzed to determine the continuum emission level and that are stored in the directory \texttt{data}. They can be (i) standard ASCII files with two columns corresponding to the frequency or velocity and the intensity, or (ii) FITS images with at least three axis, two of them corresponding to the spatial axis and a third one with the frequency or velocity information. The input files can be stored in specific subdirectories within the \texttt{data} directory. A similar specific directory will be created in \texttt{products} to contain the final products. 

\subsection{Output products}\label{s:products}

\statcont\ generates two main final products: (i) the continuum emission level, which for a single spectrum file is stored as an ASCII file saved in the \texttt{products} directory, while for spectral line datacubes, the product is a FITS image containing only the continuum emission and labeled as \texttt{\_continuum}; and (ii) a continuum-subtracted line-only ASCII spectrum or FITS datacube labeled as \texttt{\_line}. For those continuum determination methods with an estimation of the uncertainty (i.e., Gaussian and sigma-clipping, see Sect.~\ref{s:cont}), \statcont\ also produces a third product that contains information on the error estimation. This is stored as a value in an ASCII file, or as a new FITS image with the label \texttt{\_noise}. Finally, the tool can also generate plots similar to those shown in Fig.~\ref{f:contmethod} containing the spectrum and intensity distribution with markers of the continuum level estimate. These plots are produced pixel by pixel when the input parameter is a FITS datacube.

\subsection{Main procedures}\label{s:tasks}

The main procedure within \statcont\ is the determination of the continuum level. The c-CSM is the default, but the user can easily select one or more of the continuum determination methods described in Sect.~\ref{s:contspectra} to estimate the continuum level, and if necessary compare between them. In addition to the continuum level determination it is worth mentioning three options that the user can select.

The \texttt{----cutout} option allows the user to select a smaller rectangular area (or cutout) of the original FITS datacube, that will be used to determine the continuum level. This can be useful if the extension of the source of interest is small compared to the total coverage of the FITS input image.

The \texttt{----merge} option combines two or more input files in order to increase the total frequency coverage and better determine the continuum level. As shown in Sect.~\ref{s:cont}, narrow frequency ranges result in less reliable continuum level estimates. In order to avoid unrealistic results, it is important that all the files that will be merged have the same intensity scale and have been obtained with the same instrument, as well as the same dimensions (number of pixels and pixel size) in the two spatial axis.

The \texttt{----spindex} option is intended to take into account variations of the intensity of the continuum level with frequency, i.e., the spectral index $\alpha$ defined as $S_\nu\propto\nu^\alpha$ where $S_\nu$ is the intensity and $\nu$ the frequency. In order to determine the spectral index, the user has to provide a number of input files covering different frequency ranges. As a product, \statcont\ generates an ASCII file or a FITS image containing the spectral index in each pixel. The spectral index is determined by performing a linear fit to the variation of the logarithm of the continuum emission against the logarithm of the frequency. The frequency corresponds to the central frequency of each file for which the continuum level has been determined. An example of the usage of this option is shown in \citet[their Fig.~7]{SanchezMonge2017} where the authors determine spectral index maps of the two star forming regions Sgr\,B2(N) and Sgr\,B2(M) by using 40 different data-cubes with bandwidths of $\approx2$~GHz each one and covering the whole frequency range from 211~GHz to 275~GHz.

Finally, the \texttt{----model} option makes use of the spectral index calculated by \statcont\ to produce a new ASCII file or FITS file that contains the line plus continuum emission, where the line is directly obtained after subtracting the continuum emission (see Sect.~\ref{s:products}) and the continuum is a model determined from the spectral index. Therefore, this continuum will not be constant over a selected frequency range, but will change according to the spectral index. It is worth noting that the continuum level that is originally subtracted to the original data is flat (i.e., constant over frequency), while the inclusion of the model continuum has a spectral index. This is a good approximation under the assumption that the (flat) continuum level was originally determined within a frequency range where the continuum can be assumed to be constant. For typical astronomical sources the spectral index varies between $-1$ and $+4$. Therefore, in the millimeter/submillimeter wavelength regime, this is a good approximation if we consider original frequency ranges to determine the continuum level of about $<5$~GHz.

%________________________________________________________________
\section{Summary}\label{s:summary}

We have developed \statcont, a python-based tool, to automatically determine the continuum level in line-rich sources. The method works both in single spectrum observations and in datacubes, and inspects in a statistical way the intensity information contained in a spectrum searching for the continuum emission level. We have tested different algorithms against synthetic observations. The methods that more accurately estimate the continuum level are the position of the maximum of a histogram distribution of intensities of the spectrum, the Gaussian fit to this histogram distribution, the maximum of a Kernel Density Estimation of the intensities, and the sigma-clipping method that iteratively searches for the median of the sample by excluding outlier values. The method that provides accurate results and is less affected by user inputs is the sigma-clipping method. Furthermore, it provides together with the continuum emission level, information on the uncertainty (or noise level) in its determination. We use this noise level to automatically correct the determined continuum level in emission-dominated and absorption-dominated spectra. This approach is introduced in \statcont\ as corrected sigma-clipping method, or c-SCM. We have applied c-SCM to more than 750 different synthetic spectra obtaining accuracies in the determination of the continuum level better than 1\% in almost 50\% of the cases, and better than 5\% in almost 90\% of the cases. If the spectrum to be analyzed contains more than 10\% of line-free channels, c-SCM is accurate in the determination of the continuum level.

For cases with less than 10\% line-free channels, c-SCM may fail if the spectrum is highly complex combining broad emission and absorption line features. The continuum level of these complex spectra can be obtained by simultaneously fitting the line and continuum emission. This is beyond the scope of the \statcont\ tool that is not designed to fit spectral lines, but is currently being tested using the software XCLASS \citep{Moeller2017}. Schw\"orer et al.\ (in prep) uses XCLASS to simultaneously fit the main spectral lines and the continuum level of the hot cores in Sgr\,B2(N) and Sgr\,B2(M). This can improve the determination of the continuum level with respect to \statcont. A current limitation of \statcont\ is that it assumes that the continuum level, or baseline of the spectrum, is flat, i.e., 0th order polynomial. Improvements in the continuum determination method can go in the direction of considering more flexible baselines (e.g., 1st order polynomials). Other improvements, in the case of interferometric observations, is the application of the \statcont\ methods directly to the uv-data, instead of in the images. If possible, this can be used to produce higher-dynamic range continuum images, but may introduces other difficulties that go beyond the scope of this paper.

The main products of \statcont\ are the continuum emission level, together with its uncertainty, and datacubes containing only spectral line emission, i.e., continuum-subtracted datacubes. In addition, the user can provide a number of input files at different frequencies to compute the spectral index (i.e., variation of the continuum emission with frequency).

In summary, the continuum emission level of complex sources with a rich chemistry can be well determined (discrepancies $<5$\% for 90\% of the cases) applying c-SCM in \statcont, and therefore can be used to construct continuum emission maps of datacubes obtained with instruments such as ALMA and the VLA. This method has been applied successfully to complex sources like Sgr\,B2(N) and Sgr\,B2(M) and other hot molecular cores like G31.41$+$0.31 (e.g., \citealt{SanchezMonge2017, Cesaroni2017}).

\begin{acknowledgements}
The authors thank the anonymous referee for his/her thorough review and comments that improved the manuscript. This work was supported by Deutsche Forschungsgemeinschaft through grant SFB\,956 (subproject A6). This paper makes use of the following ALMA data: ADS/JAO.ALMA\#2013.1.00332.S and ADS/JAO.ALMA\#2013.1.00489.S. ALMA is a partnership of ESO (representing its member states), NSF (USA) and NINS (Japan), together with NRC (Canada) and NSC and ASIAA (Taiwan), in cooperation with the Republic of Chile. The Joint ALMA Observatory is operated by ESO, AUI/NRAO and NAOJ. \'A.\ S.-M.\ is grateful to Th.\ M\"oller for providing the synthetic datacube, and to E.\ Bergin for the suggestion of the package name. The figures of this paper have been done using the program GREG of the GILDAS software package.
\end{acknowledgements}

\end{document}